\RequirePackage{fix-cm}
\documentclass[twocolumn]{svjour3}          
\smartqed  
\usepackage{graphicx}
\usepackage{amsmath}  
\usepackage{amssymb}  
\usepackage{color}  
\newcommand{\flux}{erg~cm$^{-2}$~s$^{-1}$\,}
\newcommand{\lum}{erg~s$^{-1}$\,}
\newcommand{\msun}{{\rm M}_{\odot}}
\newcommand{\rsun}{R_{\odot}}

\begin{document}

\title{Time Domain Astronomy with the THESEUS Satellite
}


\author{S. Mereghetti         \and
        S.~Balman  \and 
      M.~Caballero-Garcia \and 
       M.~Del Santo \and 
           V.~Doroshenko  \and          
        M. H.~Erkut  \and          
        L.~Hanlon  \and 
          P.~Hoeflich  \and          
        A.~Markowitz  \and          
        J.P.~Osborne  \and 
        E.~Pian  \and          
        L.~Rivera Sandoval  \and          
        N.~Webb  \and 
         L.~Amati \and
         E.~Ambrosi \and
          A.P.~Beardmore  \and          
          A.~Blain \and
          E.~Bozzo \and
          L.~Burderi \and
          S.~Campana \and
          P.~Casella \and
          A.~D'A\`i \and
       F.~D'Ammando  \and          
       F.~De~Colle \and
          M.~Della Valle  \and          
      D.~De Martino  \and 
      T.~Di Salvo \and
      M.~Doyle \and
      P.~Esposito \and 
      F.~Frontera \and
      P.~Gandhi \and
              G.~Ghisellini\and 
              D.~Gotz \and
              V.~Grinberg \and
              C.~Guidorzi \and
               R.~Hudec \and
              R.~Iaria \and
              L.~Izzo \and
               G.~K.~Jaisawal \and
              P.~Jonker \and 
              A.K.H.~Kong \and
              M.~Krumpe \and
              P.~Kumar \and
              A.~Manousakis
              A.~Marino \and
              A.~Martin-Carrillo \and
              R.~Mignani \and
              G.~Miniutti \and
              C.G.~Mundell \and
                K.~Mukai \and          
                A.A.~Nucita \and 
                P.T.~O'Brien \and
                M.~Orlandini \and
                M.~Orio \and
                 E.~Palazzi \and
               A.~Papitto \and  
               F.~Pintore \and 
               S.~Piranomonte \and 
               D.~Porquet \and
               C.~Ricci \and
               A.~Riggio \and
                 M.~Rigoselli \and  
                  J.~Rodriguez \and
                  T.~Saha \and
                  A.~Sanna \and
                  A.~Santangelo \and
                  R.~Saxton \and
         L.~Sidoli  \and 
         H.~Stiele \and
         G.~Tagliaferri \and
         F.~Tavecchio \and  
         A.~Tiengo \and
         S.~Tsygankov \and
         S.~Turriziani \and
         R.~Wijnands \and
        S.~Zane  \and
        B.~Zhang
}


\institute{S. Mereghetti, R.~Mignani, M. Rigoselli, L. Sidoli \at
              INAF, IASF-Milano,
              via A.~Corti 12, I-20133 Milano, Italy\\
              \email{sandro.mereghetti@inaf.it}           
           \and
 S. Balman \at
 Istanbul University, Faculty of Science, Dept. of  Astronomy and Space Sciences, Beyazit, Istanbul, 34119, Turkey 
           \and
M. Caballero-Garcia \at
 Instituto de Astrof\'{\i}sica de Andaluc\'{\i}a (IAA-CSIC),
  P.O. Box 03004, E-18080, Granada, Spain
           \and
M. Del Santo, E. Ambrosi, A.~D'A\`i, F. Pintore, A. Marino\at
  INAF -- IASF Palermo,
  via U. La Malfa 153, 90146 Palermo, Italy 
           \and
 V. Doroshenko, A. Santangelo \at
 Institut fuer Astronomie und Astrophysik Tuebingen,
   Sand 1, Tuebingen, Germany\\
   Space Research Institute of the Russian Academy of Sciences,
    Profsoyuznaya Str. 84/32, Moscow 117997, Russia
           \and
 M.H. Erkut \at
Faculty of Engineering and Natural Sciences, Istanbul Bilgi University,
 34060, Istanbul, Turkey
           \and
 L. Hanlon, M. Doyle, A. Martin-Carrillo \at
Space Science Group, School of Physics, University College Dublin, 
Belfield, Dublin 4, Ireland
           \and
 P. Hoeflich \at
 Florida State University,
Tallahassee, Fl 32309, USA 
           \and
 A. Markowitz \at
Centrum Astronomiczne im.\ Miko{\l}aja Kopernika, 
Polskiej Akademii Nauk, ul.\ Bartycka 18, 00-716 Warszawa, Poland\\
University of California, San Diego, Center for Astrophysics and Space Sciences,
 MC 0424, La Jolla, CA, 92093-0424, USA
 \and
  J.P. Osborne, A.P. Beardmore, A. Blain, P.T. O'Brien \at
     Dept of Physics \& Astronomy,  University of Leicester, 
Leicester LE1 7RH, UK    
           \and
 E. Pian, L. Amati,  M.~Orlandini, E. Palazzi \at
   INAF, Astrophysics and Space Science Observatory,
    via P. Gobetti 101, 40129 Bologna, Italy
           \and
 L. Rivera Sandoval \at
University of Alberta, Canada
           \and
 N. Webb \at
Institut de Recherche en Astrophysique et Plan\'etologie,
9 Avenue du Colonel Roche, BP 44346,
31028 Toulouse Cedex 4,         France     
           \and
 E. Bozzo \at
    Department of Astronomy, University of Geneva, 
Ch. d'Ecogia 16, 1290, Versoix (Geneva), Switzerland  
           \and
 L. Burderi, A. Riggio, A. Sanna \at
Università degli Studi di Cagliari, Dipartimento di Fisica, 
SP Monserrato-Sestu, KM 0.7, I-09042 Monserrato, Italy
           \and
 S. Campana, G. Ghisellini, G. Tagliaferri, F. Tavecchio \at
    INAF - Osservatorio Astronomico di Brera,
     via Bianchi 46, I-23807 Merate, Italy
                      \and
 P.~Casella, A. Papitto, S.~Piranomonte \at
    INAF -- Osservatorio Astronomico di Roma,
via di Frascati 33,  I-00044, Monte Porzio Catone, Italy  
           \and
 F.D'Ammando \at
    INAF - Istituto di Radioastronomia,
via P. Gobetti 101, I-40129 Bologna, Italy  
           \and
 De Colle \at
Instituto de Ciencias Nucleares, Universidad Nacional Autonoma de Mexico,
 A. P. 70-543 04510 D. F. Mexico
           \and
 M. Della Valle, D. De Martino\at
INAF -- Capodimonte Observatory,
Salita Moiariello 16, 80131, Napoli, Italy
           \and
 T. Di Salvo, R. Iaria \at
Universit\`a di Palermo, via Archirafi 36 - 90123 Palermo, Italy
           \and
 P. Esposito, A. Tiengo \at
 Scuola Universitaria Superiore IUSS Pavia,
 piazza della Vittoria 15, 27100 Pavia, Italy
           \and
 F. Frontera, C. Guidorzi \at
University of Ferrara,
via Saragat  1,  44100 Ferrara, Italy
           \and
 P. Gandhi \at
School of Physics \& Astronomy, University of Southampton,
 SO17 1BJ, UK
           \and
 D. Gotz, J. Rodriguez \at
AIM-CEA/DRF/Irfu/D\'epartement d’Astrophysique, CNRS, 
Universit\'e Paris-Saclay, Universit\'e de Paris, 
	 Orme des Merisiers, F-91191 Gif-sur-Yvette, France
           \and
 V.~Grinberg,  A. Santangelo \at
 Institut fuer Astronomie und Astrophysik Tuebingen,
   Sand 1, Tuebingen, Germany
           \and
 R. Hudec \at
Czech Technical University in Prague, Faculty of Electrical Engineering, Prague, Czech Republic \\
Astronomical Institute, Czech Academy of Sciences,
 Ondrejov, Czech Republic \\
Kazan Federal University, Kazan, Russian Federation
           \and
 L. Izzo \at
DARK, Niels Bohr Institute, University of Copenhagen,
 Lyngbyvej 2,DK-2100 Copenhagen, Denmark 
           \and
 G. K. Jaisawal \at
National Space Institute, Technical University of Denmark,
 Elektrovej, Kgs. Lyngby 2800, Denmark
           \and
 P. Jonker \at
SRON, Netherlands Institute for Space Research, Sorbonnelaan 2, 3584 CA Utrecht, The Netherlands \\
Department of Astrophysics/IMAPP, Radboud University, P.O. Box 9010, 6500 GL, Nijmegen, The Netherlands
  \and
  A.K.H. Kong, H. Stiele \at
  Institute of Astronomy, National Tsing Hua University, 
   Guangfu Road 101, Sect. 2, 30013 Hsinchu, Taiwan
           \and
 M. Krumpe \at
    Leibniz-Institut f\"ur Astrophysik Potsdam, 
An der Sternwarte 16, 14482 Potsdam, Germany
           \and
 P. Kumar \at
 University of Texas, Austin, TX , USA
           \and
 A. Manousakis \at
Department of Applied Physics \& Astronomy, College of Sciences 
\& Sharjah Academy for Astronomy, Space Sciences, and Technology, 
University of Sharjah, P.O.Box 27272, Sharjah, UAE
           \and
 G. Miniutti \at
Centro de Astrobiolog\'ia (CSIC-INTA), Camino Bajo del Castillo s/n, 
E-28692 Villanueva de la Ca\~nada, Spain
           \and
 C. Mundell \at
Department of Physics, University of Bath,
Claverton Down, Bath, BA2 7AY, UK
           \and
 K. Mukai \at
CRESST II and X-ray Astrophysics Laboratory, NASA/GSFC, Greenbelt, MD 20771, USA \\
Dept. of Physics, University of Maryland Baltimore County, 1000 Hilltop Circle, Baltimore MD 21250, USA
 \and
 A.A. Nucita \at
Department of Mathematics and Physics "Ennio De Giorgi",
University of Salento, 
Via per Arnesano,  73100, Lecce, Italy
           \and
 M. Orio \at
INAF, Osservatorio di Padova, vicolo Osservatorio 5, 35122 Padova, Italy \\
Dept. of Astronomy, University of Wisconsin, 475 N, Charter Str., Madison, WI 53706, USA       
    \and
    D. Porquet \at
Aix Marseille Univ, CNRS, CNES, LAM, Marseille, France
           \and
 C. Ricci \at
    N\'ucleo de Astronom\'ia, Facultad de Ingenier\'ia y Ciencias,
Universidad Diego Portales , Santiago de Chile, Chile
           \and
 T. Saha \at
Centrum Astronomiczne im.\ Miko{\l}aja Kopernika,
Polskiej Akademii Nauk, ul.\ Bartycka 18, 00-716 Warszawa, Poland
           \and
 R.  Saxton \at
Telespazio UK Ltd., 350 Capability Green, 
Luton, Bedfordshire LU1 3LU - UK 
          \and
 S. Tsygankov  \at
Department of Physics and Astronomy,  FI-20014 University of Turku, Finland \\ 
Space Research Institute of the Russian Academy of Sciences,
Profsoyuznaya Str. 84/32, Moscow 117997, Russia 
           \and
 S. Turriziani \at
Physics Department, Gubkin Russian State University (National Research University),
 65 Leninsky Prospekt, Moscow, 119991, Russian Federation
           \and
 R. Wijnands \at
Anton Pannekoek Institute for Astronomy, University of Amsterdam,
 Postbus 94249, 1090 GE Amsterdam, The Netherlands
           \and
 S. Zane \at
Mullard Space Science Laboratory,
University College London, UK 
           \and
 B. Zhang \at
Department of Physics and Astronomy,
University of Nevada, Las Vegas, USA
%
%
%
%
}

\date{Received: date / Accepted: date}

\maketitle

\begin{abstract}
THESEUS is a medium size space mission of the European Space Agency, currently under evaluation for a possible launch in 2032.   Its main objectives are to investigate the early Universe through the observation of gamma-ray bursts and to study  the  gravitational waves electromagnetic counterparts and neutrino events. 
On the other hand, its instruments, which include a wide field of view X-ray (0.3-5 keV) telescope based on  lobster-eye focussing optics and a  gamma-ray spectrometer with imaging capabilities in the 2-150 keV range, are also ideal for carrying out unprecedented studies in time domain astrophysics. In addition, the presence onboard of a 70 cm near infrared telescope will allow simultaneous multiwavelegth studies.
Here we present the THESEUS capabilities for studying the time variability of different classes of sources in parallel to, and without affecting, the gamma-ray bursts hunt.
\end{abstract}


\section{Introduction}
 \label{intro}
 
THESEUS (Transient High Energy Sky and Early Universe Surveyor, \cite{2018AdSpR..62..191A}) is a  space mission devoted to the study of Gamma-ray bursts (GRB)   which has been selected by the European Space Agency for an assessment study in response  to the call for a satellite of medium size class (M5) to be launched in 2032. THESEUS has been conceived with the two main objectives of {\it a)}  exploiting GRBs to investigate the early Universe and {\it b)} providing a great contribution to multi-messenger astrophysics with the observation of the electromagnetic counterparts of gravitational waves (GW) and neutrino events. 

To fulfill these scientific objectives, THESEUS will carry wide field monitors operating in the X-ray-to-$\gamma$-ray energy range  able to discover and accurately localize GRBs and other transient high-energy sources. These monitors will be complemented by   a near infrared (NIR) telescope for the  characterization of the discovered transients thanks to rapid follow-up observations after  autonomous slews of the satellite.  
 
The wide fields of view, high sensitivity and broad energy coverage of the X-ray and gamma-ray monitors, coupled with the high cadence observational strategy of THESEUS, are also ideal to carry out unprecedented studies of the variability properties of different classes of Galactic and extra-Galactic high-energy sources.  These studies will be done in parallel with the main THESEUS scientific objectives   without requiring additional satellite resources.  
In Section \ref{sec-THESEUS} we give a brief description of the  instruments on board THESEUS and   the modality in which the satellite will be operated. We then present in the following Sections the  science cases for different classes of sources, ranging from nearby flaring stars to the most distant AGNs.

   \begin{table*}[ht]
\caption{Main characteristics of the THESEUS mission.} 
\label{tab:THESEUS}
\begin{center}       
\begin{tabular}{ | p{4cm} |p{7cm}   |}
\hline
   


\multicolumn{2}{|c|}{   }  \\
\multicolumn{2}{|c|}{ \bf   Mission Profile }  \\
\multicolumn{2}{|c|}{   }  \\

\hline\rule[-1ex]{0pt}{3.5ex} 
Field of regard  (visible      &   $>$50\% of the sky \\
  sky region  at a given time)  &    \\
      
\hline\rule[-1ex]{0pt}{3.5ex} 
Orbit  &   inclination 5.4 deg, altitude 550-640 km,\\
           &  period   96.7 min (at 600 km)  \\
      
\hline\rule[-1ex]{0pt}{3.5ex} 
Autonomous slews   &  $>$7$^{\circ}$/min \\
      
\hline\rule[-1ex]{0pt}{3.5ex} 
Mission duration  &  4 yr (nominal),  $>$2 yr (possible extension) \\

\hline
\multicolumn{2}{|c|}{   }  \\
\multicolumn{2}{|c|}{\bf SXI }  \\
\multicolumn{2}{|c|}{   }  \\
\hline\rule[-1ex]{0pt}{3.5ex} 
Energy range &  0.2-5 keV \\

\hline\rule[-1ex]{0pt}{3.5ex} 
Field of view  &    31$\times$61 deg$^2$       \\

\hline\rule[-1ex]{0pt}{3.5ex} 
Sensitivity  (3$\sigma$ in 1000 s)  &  2$\times$10$^{-11}$ erg cm$^{-2}$ s$^{-1}$ (0.2-5 keV)   \\

\hline\rule[-1ex]{0pt}{3.5ex} 
Location accuracy  (99\% c.l.)  &  $<$2 arcmin    \\

\hline
\multicolumn{2}{|c|}{   }  \\
\multicolumn{2}{|c|}{\bf XGIS }  \\
\multicolumn{2}{|c|}{   }  \\
\hline\rule[-1ex]{0pt}{3.5ex} 
 Energy range       & 2 keV  - 10 MeV  (total),  2-150 keV (imaging) \\
 
 \hline\rule[-1ex]{0pt}{3.5ex} 
Imaging field of view         & $\sim$30$\times$60 deg$^2$   (full sensitivity)     \\
of 2 combined cameras          & 117$\times$77 deg$^2$   (zero sensitivity)     \\
                    
\hline\rule[-1ex]{0pt}{3.5ex} 
  Sensitivity  (5$\sigma$ in 1000 s) & 5$\times$10$^{-10}$ erg cm$^{-2}$ s$^{-1}$ (2-30 keV)  \\ 
                                              & 2$\times$10$^{-9}$ erg cm$^{-2}$ s$^{-1}$  (30-150 keV)    \\ 
\hline\rule[-1ex]{0pt}{3.5ex} 
Location accuracy  (90\% c.l.)  &  $<$5 arcmin (for a source at SNR$>$20)    \\
                                              &  $\sim$15 arcmin (for  SNR=7)    \\
 
\hline
 \multicolumn{2}{|c|}{   }  \\
\multicolumn{2}{|c|}{\bf IRT }  \\
\multicolumn{2}{|c|}{   }  \\

\hline\rule[-1ex]{0pt}{3.5ex} 
Wavelength range &  700-1800 nm  \\

\hline\rule[-1ex]{0pt}{3.5ex} 
Field of view  &     15$\times$15 arcmin$^2$  \\

\hline\rule[-1ex]{0pt}{3.5ex} 
Sensitivity    &  20.9 (I), 20.7 (Z), 20.4 (Y), 20.7 (J), 20.8 (H) \\
                         &  (SNR=5 in 150 s) \\
      
\hline\rule[-1ex]{0pt}{3.5ex} 
Location   accuracy  (99\% c.l.)  &  $<$1 arcsec    \\

\hline

\end{tabular}
\end{center}
\end{table*}

\section{The THESEUS Mission}
\label{sec-THESEUS}

The THESEUS satellite will carry two high-energy instruments, the Soft X-ray Imager (SXI, 0.3-5 keV, \cite{2020SPIE11444E..2LO}) and the  X and Gamma Imager and Spectrometer 
(XGIS, 2 keV-10 MeV, \cite{2020SPIE11444E..2KL}), and an Infrared Telescope (IRT, 0.7-1.8 $\mu$m, \cite{2020SPIE11444E..2MG}). The main properties of these instruments, described in more detail below, are summarized in Table \ref{tab:THESEUS}. All the instruments will operate simultaneously. GRBs and other transient sources occurring in the large imaging fields of view of the SXI and XGIS will be detected and localized in real time by the on-board software. Their coordinates and other relevant information will be immediately down-linked thanks to a network of VHF ground stations, and the satellite will autonomously reorient itself in order to place the  discovered source in the IRT field of view.

\begin{figure*}
\center
\includegraphics[width=16cm]{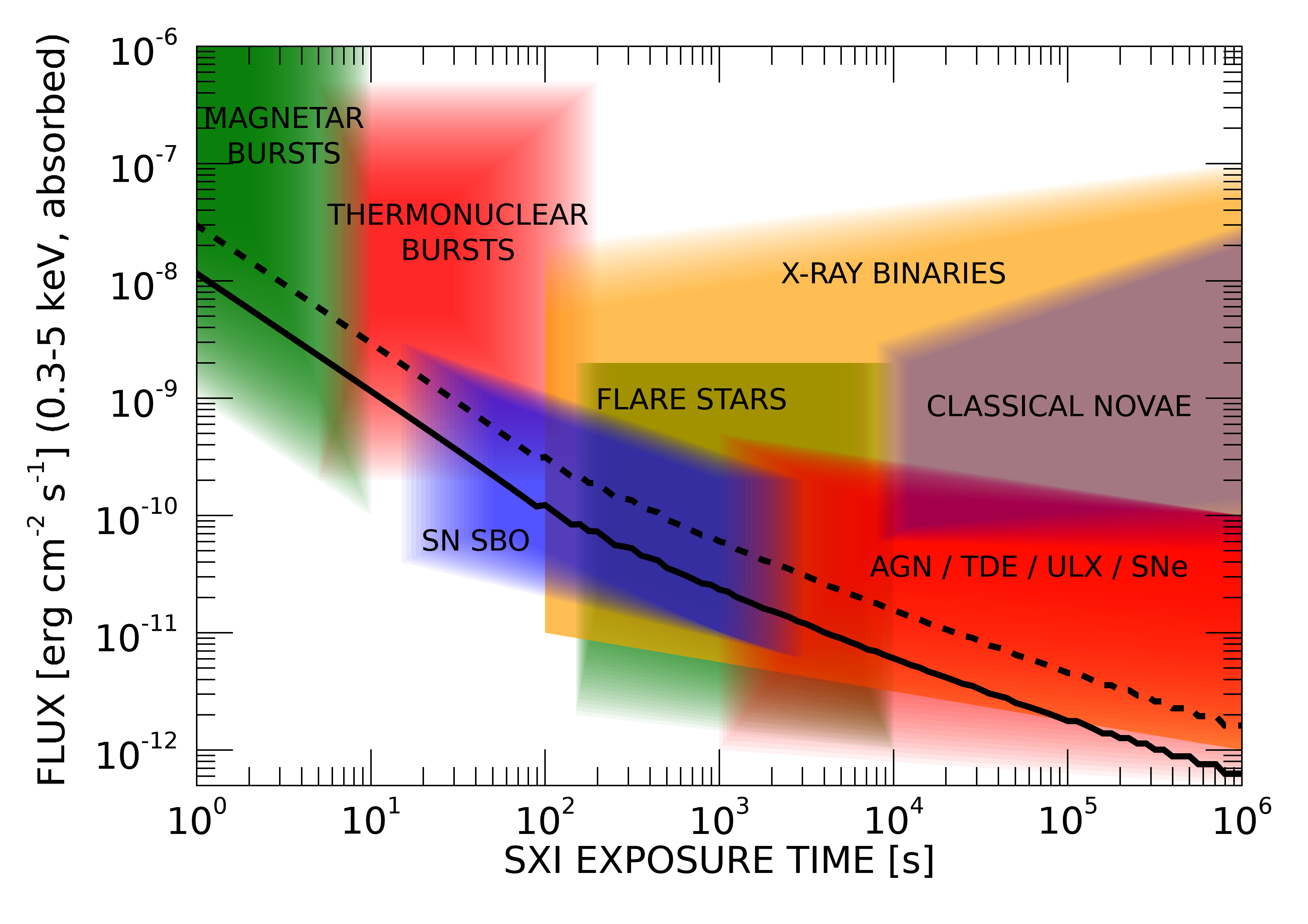}
   \caption{SXI sensitivity as a function of exposure time for sources with an absorbed power law spectrum with photon index $\Gamma$=2 and  N$_H$=5$\times$10$^{20}$ (solid line) or   N$_H$=10$^{22}$   cm$^{-2}$ (dashed line). 
Both curves refer to absorbed source fluxes (0.3-5 keV) with a 95\% probability to exceed a detection  threshold of 3$\sigma$ (see Appendix for details). 
Approximate regions of the fluxes and variability time scales expected from different classes of sources are also indicated. 
}  
  \label{fig-sxi-sens}
\end{figure*}

THESEUS will be placed in a low  equatorial orbit (height $\sim$600 km, inclination $<6^{\circ}$), to guarantee a low and stable  background, and will operate with a pointing strategy optimized to maximise the number of detected GRBs and, at the same time,  facilitate  their follow-up  with large ground based telescopes (see Sect. \ref{sub-pointing}). 
To fully exploit the THESEUS capabilities,   in particular those offered by the IRT, small deviations from the nominal pointing directions (up to a few degrees) are foreseen to place interesting targets, selected through a Guest Observer Program, in the IRT field of view.  THESEUS will have a nominal lifetime of four years,  with possible extensions not limited by onboard consumables.

 \subsection{SXI}
 \label{sub-SXI}

The SXI is a focussing X-ray telescope composed of two identical units providing high-sensitivity over a very large field of view thanks to  a micro-pore  mirror in lobster-eye configuration  coupled to wide area CMOS detectors \cite{2020SPIE11444E..2LO}. It will operate in the 0.3-5 keV energy range providing a uniform sensitivity,  of the order of few milliCrabs in one ks, across the whole field of view of 31$\times$61  deg$^2$.   The source location accuracy will be $<$1-2 arcmin. 

One of the main advantages of the lobster-eye optics is that they can focus X-ray photons over a very wide field of view, with uniform efficiency and point spread function, independent of the off-axis angle of the source. The SXI point spread function (PSF) has a characteristic shape consisting of a narrow core produced by the source photons undergoing two reflections in the micro pores (50\% of the total) and two cross arms produced by single-reflected photons.  The narrow PSF core has a FWHM of 6 arcmin (at 1 keV) across the whole field of view.  The SXI background is dominated by diffuse X-ray photons of cosmic origin and, therefore, it depends on the pointing direction, due to the variations in the soft X-ray component of Galactic origin. A sky-averaged value of  1.23$\times$10$^{-5}$ cts s$^{-1}$ arcmin$^{-2}$     (this includes also the particle component)  has been adopted to compute the  sensitivity   shown in Fig.~$\ref{fig-sxi-sens}$.  
Note that the flux sensitivity depends  on the spectral shape and interstellar absorption of the source. 
The curves in Fig.~$\ref{fig-sxi-sens}$ refer to a source with power law spectrum of photon index $\Gamma$=2 and two differerent values of absorption,  N$_H$=5$\times$10$^{20}$  and  N$_H$=10$^{22}$  cm$^{-2}$. 
We describe in the Appendix how to compute the SXI sensitivity for sources of different spectral shapes.

 \subsection{XGIS}
   \label{sub-XGIS}

\begin{figure}
\center
\includegraphics[width=8cm]{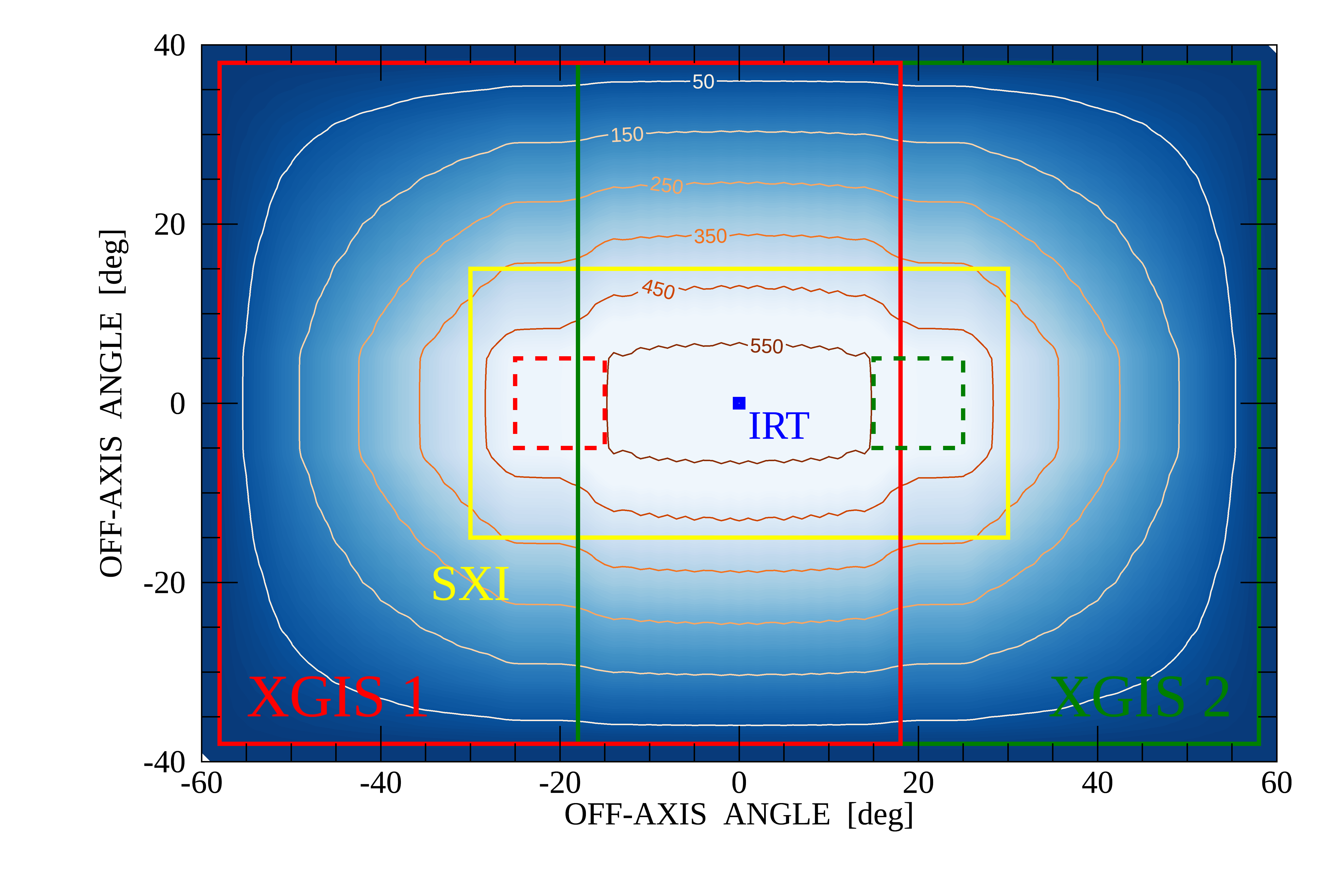}
  \caption{Fields of view of the THESEUS instruments. The  red (left) and green (right) squares indicate the  partially coded FoV (77$\times$77 deg$^2$, solid lines) and fully coded FoV (11$\times$11 deg$^2$, dashed lines) of the two XGIS units. The yellow rectangle is the SXI FoV (61$\times$31 deg$^2$). The blue square indicates the pointing direction of the IRT.   The  contour lines indicate the effective area at 10 keV provided by the sum of the two XGIS units, assuming a 50\% open fraction for the coded masks.  
  }  
  \label{fig-fov}
\end{figure}

The XGIS consists of two identical units employing coded masks and  position sensitive detection planes  based on  arrays of  Silicon Drift Detectors (SDD) coupled to CsI crystal scintillator bars \cite{2020SPIE11444E..2KL}. The SDD are used both for direct  photon detection in the lower energy range (2-30 keV) and as readout of the light signals recorded in the CsI scintillators,  that operate in the higher energy range (30 keV - 10 MeV). In the following we will refer to the SDD and CsI detectors as XGIS-X and XGIS-S, respectively. The tungsten coded mask (1.5 mm thickness) is placed at a distance of 63 cm from the detector and provides imaging capability up to $\sim$150 keV over a square field of view of 77$\times$77 deg$^2$ for each XGIS unit \cite{2020SPIE11444E..8SG}.  The two XGIS units are pointed at directions offset by $\pm$20$^{\circ}$ with respect to the satellite boresight (defined by the IRT pointing direction). Thus the FoV of the two units partially overlap to give a total XGIS field of view of 117$\times$77 deg$^2$ (at zero sensitivity), as shown in Fig.~\ref{fig-fov}. 
  
\begin{figure}
\center
\includegraphics[width=8cm]{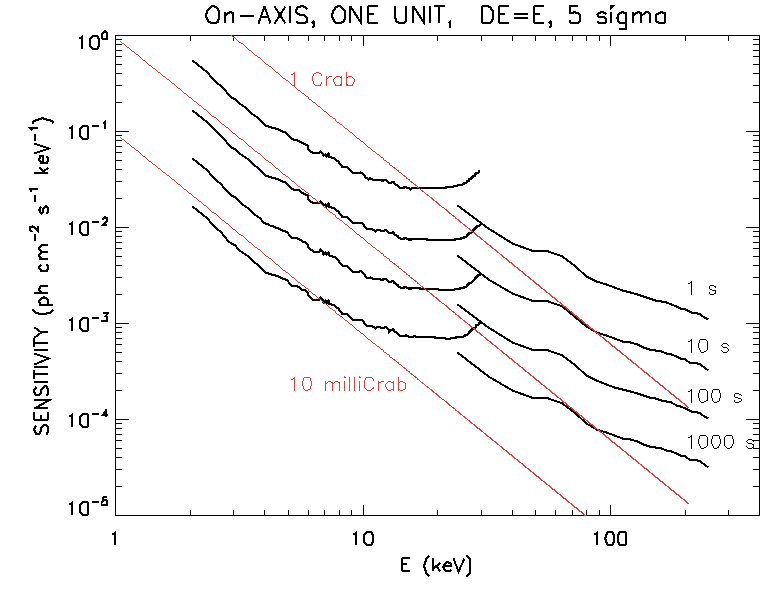}
  \caption{XGIS imaging sensitivity (5$\sigma$ c.l., $\Delta$E/E=1)  for different exposure times. The curves refer to a single unit and sources in the fully coded field of view ($\sim10\times10$ deg$^2$). One Crab corresponds to $3\times10^{-8}$ erg cm$^{-2}$ s$^{-1}$ in the 2-30 keV range and to $1.6\times10^{-8}$ erg cm$^{-2}$ s$^{-1}$ in the 30-150 keV range.  }  
  \label{fig-sens-ene}
\end{figure}

The XGIS-X and XGIS-S   sensitivities are shown in Fig.~\ref{fig-sens-ene}.  Note that this is the sensitivity that can be achieved in the images, i.e. it is valid up to approximately $\sim$150 keV. For higher energies the coded mask, and the collimating structure that support it, become progressively transparent. This results in a sensitivity to high-energy transient events over a wider field of view, approaching 2$\pi$ sr, without imaging   and with only rough localization capabilities. More details on the XGIS expected performances can be found in \cite{2020SPIE11444E..8QM} and \cite{2020SPIE11444E..8PC}.

   \subsection{IRT}
  \label{sub-IRT}

The IR telescope on board THESEUS (IRT, \cite{2020SPIE11444E..2MG}) has been designed with the main scope of detecting the GRB counterparts and measuring their redshift through multiband photometry and moderate resolution spectroscopy.  The IRT consists of a 70 cm aperture telescope with an IR camera at its focus, sensitive in the 0.7-1.8 $\mu$m range.  Five filters (I, Z, Y, J, and H) will be available to acquire images over a field of view of 15$\times$15 arcmin$^2$, reaching  magnitude limits  of $\sim$21    (5$\sigma$) in 150 s exposures.
Slitless spectroscopy with resolution R$\sim$400 over the 0.8-1.6 $\mu$m range will be provided on a separate field of view of 2$\times$2 arcmin$^2$.

   \subsection{Pointing strategy and Guest Observers Program}
  \label{sub-pointing}

\begin{figure}
\center
\includegraphics[width=8cm]{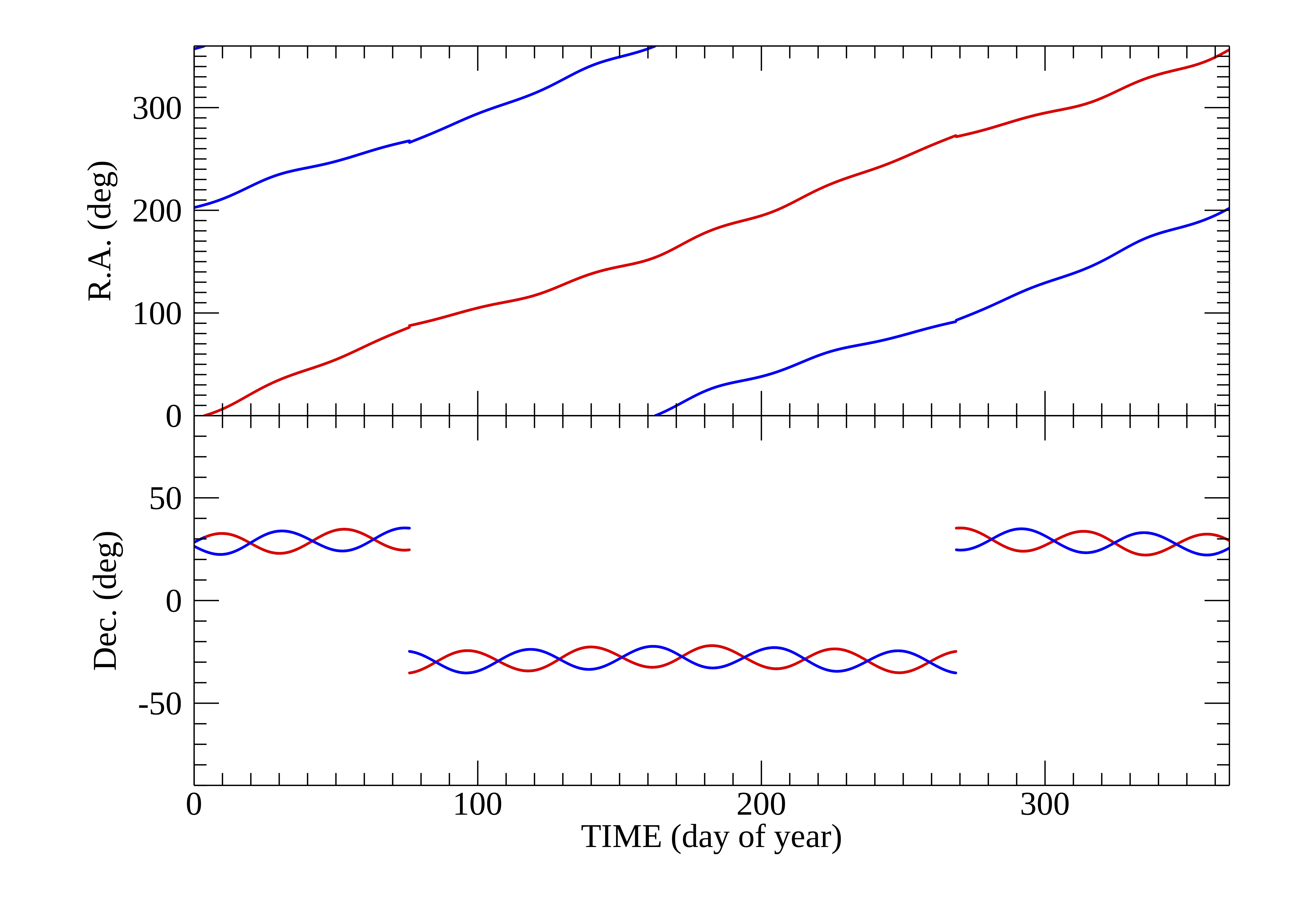}
  \caption{Right ascension (top panel) and declination (bottom panel) of the Survey Mode baseline pointing directions for one year of THESEUS observations.  Twice per orbit the satellite alternates between pointing directions  on the blue and red lines.}  
  \label{fig-point}
\end{figure}

\begin{figure}
\center
\includegraphics[width=8cm]{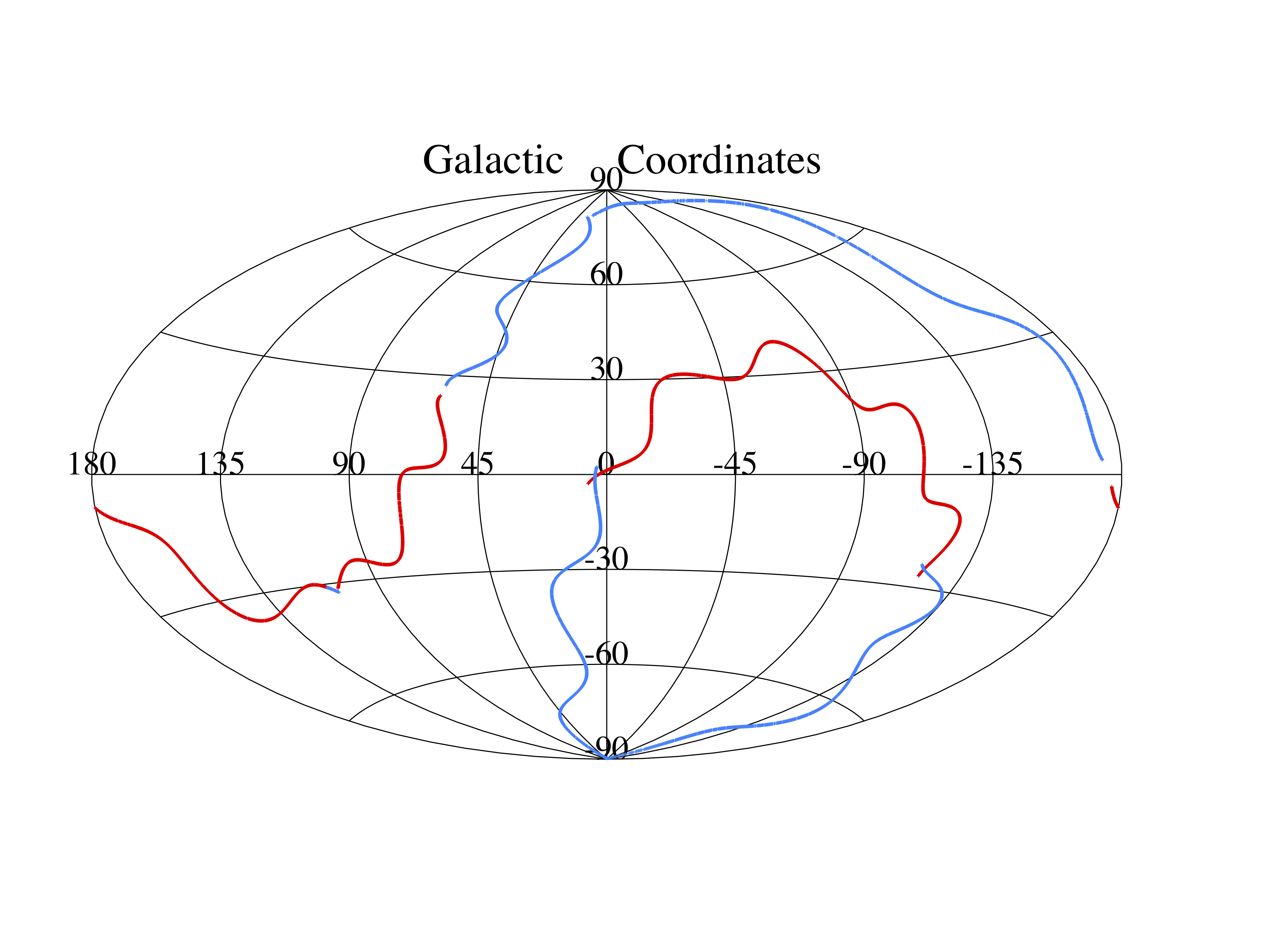}
  \caption{Survey Mode pointing directions for one year of THESEUS observations plotted on a sky map in Galactic coodinates.}  
  \label{fig-point_lb}
\end{figure}

\begin{figure}
\center
\includegraphics[width=8cm]{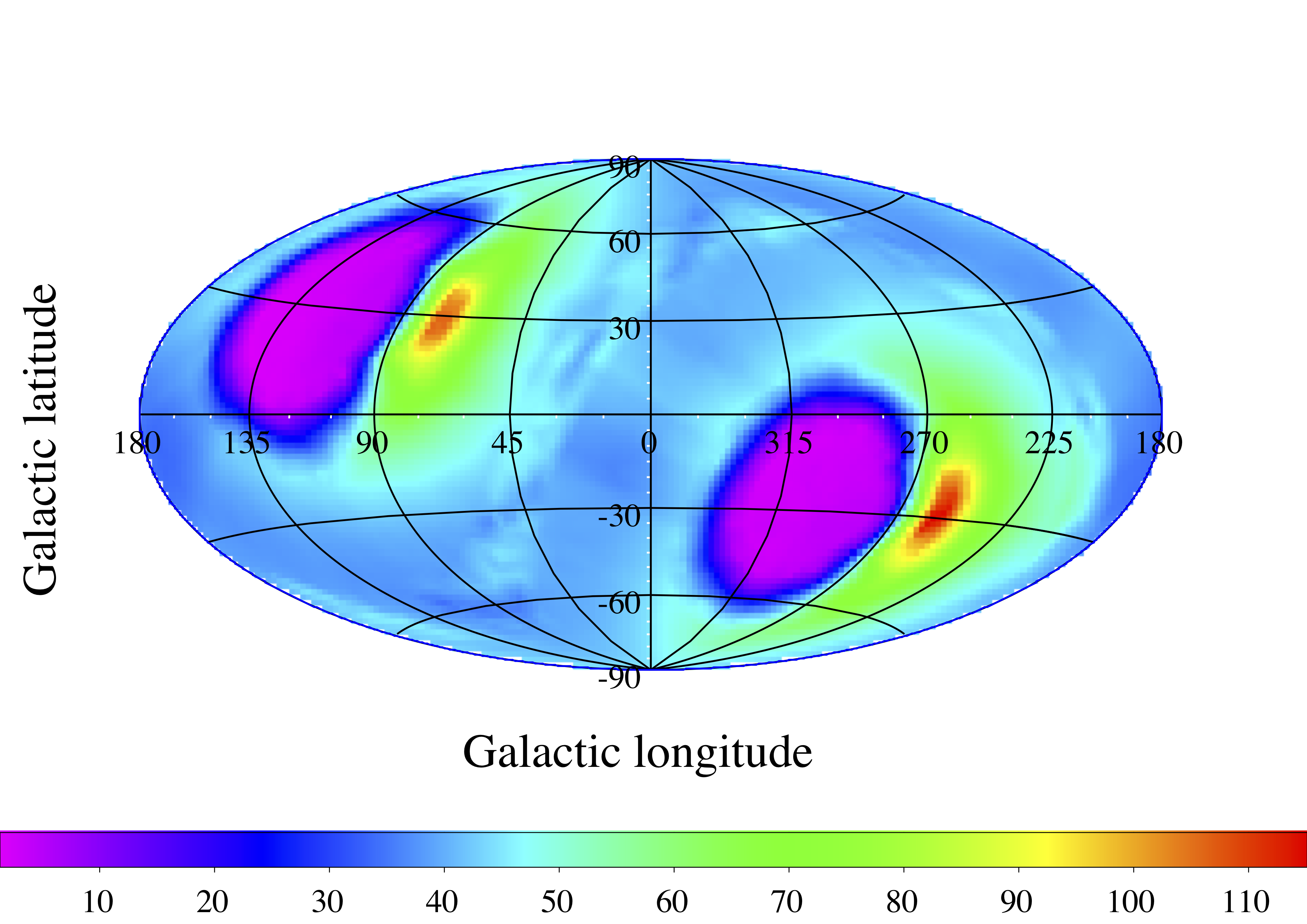}
  \caption{Expected sky exposure for  the SXI instrument  computed from  simulations of a nominal mission duration of 4 yrs.  The color scale is in units of days. }  
  \label{fig-expo}
\end{figure}

THESEUS will spend most of the observing time in Survey Mode, during which the SXI and XGIS instruments will be able to detect in real time GRBs and other transient sources occurring in their fields of view.  When a GRB is detected and localized by the on-board triggering software of one (or both) of these instruments,  the satellite will autonomously  slew to place the source in the field of view of the IRT.  After a sequence of photometric and spectroscopic IRT observations, driven by the properties of the candidate NIR counterpart, the satellite will resume the predefined Survey Mode observing plan. 

During the THESEUS Assessment Study phase, several possible pointing strategies for the Survey Mode have been evaluated, in order to demonstrate the feasibility of the main scientific objectives of the mission. Extensive simulations of the THESEUS operations were done based on a state-of-the-art modelisation of the GRB populations and taking into account all the satellite operational and technical constraints. 

This resulted in the selection of a baseline strategy that provides an optimal trade-off between the total rate of detected GRBs and a distribution of their sky positions easily accessible to large ground based facilities.  In this baseline, in order to minimize the Earth occultation of the field of view, two exposures are foreseen for each orbit of the satellite.  They will be pointed at directions with  opposite right ascension values (i.e. RA$_i$ and RA$_i$ +12 h) and declination of about $+30^{\circ}$ or  $-30^{\circ}$, depending on the season of the year.  The pointing directions will gradually change in  right ascension, covering the whole range during the year, as illustrated in Fig.~\ref{fig-point}. In Fig. \ref{fig-point_lb} we show the pointing directions in a sky projection in Galactic coordinates.

Considering the wide fields of view of the SXI and XGIS instruments, small deviations (less than a few degrees) from the nominal pointing  directions will have little effect on the rate of detectable GRBs, but they  can be very useful to place interesting sources inside the IRT field of view. Considering the large sky density of potentially interesting sources (e.g. AGNs, active stars, X-ray binaries)   it should be always possible to find IRT targets not too far from the nominal pointing directions. It is foreseen that such targets can be selected from      proposals submitted by the general astronomical community in the framework of a Guest Observing program. Some examples of interesting science cases are given in Sect.~\ref{sec-GO}, while we refer to the accompanying paper \cite{} for a more detailed discussion. 

This observing strategy will result in a non-uniform sky coverage (Fig.~\ref{fig-expo}). Sources in  regions close to the  pointing directions will be observed with a high cadence,  $\sim$15 times per day, for periods of visibility lasting from a  few weeks to a few months,  depending on their position. 

Most of the Survey Mode pointings will have a duration of about 2.3 ks. 
The satellite pointings resulting from GRB triggers will typically have a longer duration and   a more uniform sky distribution in the sky. 
In many of the simulations reported in the following sections we will use as a reference exposure times of $\sim$2 ks (corresponding to the single individual pointings).


\section{Stars}
\label{sec-stars}

Stellar flares will undoubtedly constitute the most numerous class of transient events discovered by the SXI, due to its wide field of view and great sensitivity in the soft X-ray range. Therefore, contrary to the case of previous GRB missions which operated at higher energies, for  THESEUS there is the need to recognize these events on-board in order to distinguish them from GRBs. For this reason, a catalog containing the sky coordinates of known and candidate flaring  stars will be used by the on-board software to avoid too many undesired automatic satellite slews.

However, after such triggers, both the SXI and XGIS will continue to operate normally, thus acquiring useful data  on the observed flares. 
These will provide an unprecedentedly large database for the statistical characterization of flares from all the classes of active stars. Most studies of this  kind carried out up to now at X-ray energies were based on relatively small samples and/or concentrated on specific observations of stellar clusters and star forming regions. 
THESEUS  will instead permit a thorough statistical characterization of the X-ray flare properties, and, in particular, of their rate of occurrence as a function of stellar type and age, based on a large and unbiased sample. Note that the level of X-ray flaring activity has important implications for the habitability zone of exoplanets \cite{2017ApJ...848...41L}.

As discussed below,  stars also emit ``super-flares'', which are  characterized by larger luminosities and longer durations than those of normal flares. The study   of these more rare events with THESEUS is obviously of  extreme interest. For this reason, the on-board  software will  be able to distinguish them from ``normal'' flares, thus enabling autonomous satellite slews to place the star in the  IRT field of view, and  rapid alert broadcasting  to permit their follow-up by ground-based observatories.

\subsection{Stellar flares and super-flares}

Stellar flares form in a similar way as those in the Sun. They occur in  close proximity of active regions, which are confined areas with 
magnetic fields of 1-2\,kG. Magnetic loops from these regions extend far away from the chromosphere into the stellar corona, where magnetic reconnection events 
occur \cite{1988ApJ...330..474P,2010ARA&A..48..241B}. These events are accompanied by a sudden release of energy across the electromagnetic spectrum, from 
radio (gyrosynchroton process) to ${\gamma}$-rays (with X-rays originating at the base of the loop), passing through the optical/IR.

Young stars and stars in close binary systems rotate much more rapidly than the Sun, and,  as a consequence, their magnetic fields are stronger. This translates  into a greater coverage of starspots and/or active regions, stronger chromospheric and coronal emission, and more frequent and powerful  flares \cite{2007ApJ...665L.155M,2016AJ....151..114M}.

The largest solar flares have radiated (total integrated) energies exceeding 10$^{32}$ erg, with maximum coronal temperatures of a 
few $10^6$\,K (MK hereafter) \cite{2015AstL...41...53S}. Large stellar flares can be 10$^6$ times more energetic than in the Sun, with 
coronal temperatures around 100 MK 
and energy releases up to 10$^{38}$ erg \cite{1996A&A...311..211K,2007ApJ...654.1052O}. 
A  flare in 2008 of the nearby 30-300 Myr-old M dwarf flare star EV Lac  had a lower limit on energetic release 
of $6\times10^{34}$ erg \cite{2010ApJ...721..785O}.
X-ray flares energies were found up to  $2\times10^{35}$ erg in very young low mass stars \cite{2007A&A...471..645C} and up to $10^{38}$ erg in
active binary systems  \cite{2014efxu.conf..138T}.
A very large flare, detected up to 100 keV and with released energy of about 10$^{37}$ erg, was also observed with BeppoSAX from the eclipsing binary star Algol \cite{1999A&A...350..900F}.
The interpretation of these  flaring events  assumes that they involve the same physical processes   at work  in  the Sun, as confirmed by multi-wavelength observations of plasma heating and particle  acceleration in stellar flares \cite{2010ARA&A..48..241B}.

An X-ray survey with the Monitor of All sky X-ray Image (MAXI) instrument showed that   the number of stars  emitting extremely large flares is very limited:  in two years of data only ten out of the 256 active binaries within 100 pc were detected, with four of them  showing multiple flares  \cite{2016PASJ...68...90T}. No  flares were detected from solar-type stars, despite the fact that fifteen G-type  main-sequence stars lie within 10 pc. This implies that the frequency of the hard X-ray super-flares from solar-type stars which have ${\rm L}_{\rm X}>10^{30}$ \lum is very small.

Events of this type (long and highly energetic) in late spectral types (K-M) stars have been seen to occur in a dozen stars.
A total number of 23 giant flares from 13 active stars (eight RS CVn systems, one Algol system, three dMe stars and 
a young stellar object) were detected during the first two years of the all-sky X-ray monitoring with   MAXI \cite{2016PASJ...68...90T}. 
This number can be significantly   increased by THESEUS thanks to its continuous surveying capabilities in soft and hard X-rays. 
It is  expected that hard X-ray events like the one that happened in DG CVn (see below) are very rare, but when they happen they give enormous insights onto the  physics of the formation of stellar flares. 

DG CVn is a very young and nearby (18 pc)  low-mass M-dwarf,  and also one of the brightest nearby stellar radio emitters \cite{1999AJ....117.1568H}. 
It  is in a binary system consisting of two M dwarfs,  one of wich rotates very rapidly, with a period of 8 hr \cite{2003ApJ...583..451M}. 
An exceptional event occurred on 23rd  April 2014, when one of the two stars flared to a level bright enough ($\sim3.4\times10^{-9}$ \flux in the 15-100 keV band)  to trigger the Swift 
Burst Alert Telescope (BAT)  \cite{2014ATel.6121....1D}. Two minutes later, after Swift had slewed to point in the direction of this 
source, the Swift X-ray Telescope  and the Ultraviolet Optical Telescope  started to  observe this flare. These observations,   
supported by ground-based optical and radio campaigns, continued (intermittently) for about 20 days and 
provided a unique case history of such a rare event. Its decay lasted more than two weeks in soft X-rays, and it included a number of smaller superimposed secondary flares. Other  studies  reported additional data indicating radio and optical bursts from this system during this time period \cite{2015MNRAS.446L..66F,2015MNRAS.452.4195C,2016ApJ...832..174O}.

The most powerful solar flares had energies of about 10$^{32}$ erg.
Up-scaling  of solar flares models would require enormous starspots (up to 48$^{\circ}$ across) to match stellar super-flares, thus much bigger than any sunspots seen in the last four centuries of solar observations. 
A possible explanation  to produce super-flares in   Sun-like stars  is that they host a dynamo     much stronger than that of the Sun \cite{2013A&A...549A..66A}. 
 Recent studies using data  from the Kepler satellite, 
point out to both the few instances and the possibility of solar-type stars undergoing super-flares with luminosity as high as 10$^{35-37}$ \lum under certain conditions \cite{2014MNRAS.442.3769K,2014ApJ...792...67C,2015ApJ...798...92W}.

This indicates the potential of such events as powerful releases of energy.  Planets around these stars would be  exposed to enourmous releases of energy that may be harmful to any  presence of life. 
Multi-wavelength and high-cadence observations of super-flares are necessary to understand their influence in the evolution of exo-planet 
atmospheres. 
So far only a handful of M-dwarf superflares have been recorded with multi-wavelength high-cadence observations. 
Recently, the sample of M-dwarf stars emitting super-flares has been doubled (to a total of 44, out of a sample of 300 flaring stars) using optical data from the  TESS satellite \cite{2020ApJ...902..115H},   confirming  the trend (already observed in X-rays \cite{2014efxu.conf..138T}) that  the number of super-flares decreases with  luminosity/maximum temperature. 
It was found that   43\%, 23\% and 5\% of the flares emit at temperatures above 14\,000 K, 20\,000 K 
and 30\,000 K, respectively. The largest and hottest flares briefly reach 42\,000 K. Note that all these measurements  have been
made in the optical \cite{2020ApJ...902..115H}, so typically cooler (by ${\times}10^3$) than the corresponding coronal temperatures measured in X-rays. It is found that exo-planets orbiting young ($\le200$\,Myr) 
M-type stars typically receive X-ray and UV fluxes 100--1000 times larger than those  from Proxima Centauri.

G-type main-sequence and Sun-like stars are also believed to experience super-flares, though much less powerful than in the case of K and M dwarfs \cite{2019ApJ...876...58N}. The effect of  star spots can be very disturbing in deriving the real luminosity of such energetic events. A recent study  has found 2341 and 26 super-flares from 265 and 15 solar-type 
and Sun-like stars, respectively \cite{2021ApJ...906...72O}. This was based on   Kepler  and   GAIA data, removing the rotational variations due to star-spots in the light curves (plus
other effects). This study confirmed that both the peak energy and frequency of the super-flares
decrease as the rotational period of the stars increase (as already suggested for K-M dwarf stars).  The maximum energy realeased in these events is of $4{\times}10^{34}$\,erg in Sun-like stars, and   events as energetic as $(7-10){\times}10^{33}$\,erg occur once every 3000-6000 years \cite{2021ApJ...906...72O}.

With THESEUS, we will be able to catch flares with luminosities ${\gtrsim}10^{32}$ \lum that can be detected up to 200 pc. In this volume we estimate a number of ${\lesssim}10\,000$ stars of G-K-M spectral type (see Chap. 2 from \cite{2008PhDT.......342C}). 
In two years of MAXI mission only ten out of the 256 active binaries within 100 pc have been detected. This means 0.02 super-flaring stars detected per year and per hundred parsec. Doubling  the length of the survey (4 yr of nominal THESEUS mission), the 
distance to which super-flares   can be detected, and the whole sample of G-K-M spectral type stars as input, gives   that up to $\sim$3000 flaring stars could be detected\footnote{This  is an upper limit, given the sky coverage  and time constraints of  THESEUS that would limit its capacity
to detect all the flares. Also, in this calculation it has been assumed that a normal G-K-M star has the same probability to undergo a (detectable) flare as a super-flare 
for an active binary system.}.
If we consider that only the sample of known and most nearby K-M stars are capable to produce such powerful super-flares as detected with
previous X-ray missions, this gives a number of 300 stars. If only one tenth of these flaring stars might produce super-flares  \cite{2020ApJ...902..115H} then the estimated  upper limit of detected super-flares during   THESEUS  nominal life is of ${\lesssim}30$.

\subsection{The Fe K$_{\alpha}$ line in super-flares}

Modelling  Fe K lines produced by fluorescence during flares  we can infer the properties of  the illuminating source, i.e. the flare-loop (see, e.g., \cite{2010ApJ...721..785O}  for EV Lac). EV Lac is a nearby ($d=5.06$\,pc) low mass (${\approx}0.35\,{\rm M}_{\odot}$ \cite{2000A&A...353..987F}) late spectral-type (M3) star that has been observed to emit super-flares in X- and ${\gamma}$-rays with the Swift and   Konus satellites. During its largest observed flare in 2008, its 0.3-100\,keV peak flux reached a value 
of ${\approx}5{\times}10^{-8}\,{\rm erg}\,{\rm cm}^{-2}\,{\rm s}^{-1}$. This provided a unique occasion to detect, observe and study the evolution of the Fe K lines.  The Fe K$_{\alpha}$ emission feature showed variability on time-scales of $\sim$200\,s, difficult to explain using only  the fluorescence hypothesis \cite{2010ApJ...721..785O}.
A proposed alternative scenario is that this emission was produced by collisional ionization from accelerated 
particles. It was found that the spectrum of accelerated particles can explain the  Fe K$_{\alpha}$    flux as well as  the absence of  non-thermal emission in the 20–50\,keV range.  The observation of similar events with THESEUS can give valuable insights into the fluorescence vs. collisional hypothesis for the origin of the observed Fe K$_{\alpha}$  line.

For young stars, the possible presence of a disk complicates the geometry, with reflection off the surrounding disk likely being a dominant contributor \cite{2005ApJS..160..503T}.  The detection of Fe K$_{\alpha}$ line emission in the X-ray spectra of normal stars without discs, such as EV Lac \cite{2006A&A...460..733L},  gives an independent constraint on the height of flaring coronal loops. These results can be compared with results from purely theoretical flare hydrodynamic calculations and provide  crucial inputs to the models.

\begin{figure}
\center
\includegraphics[width=8.5cm]{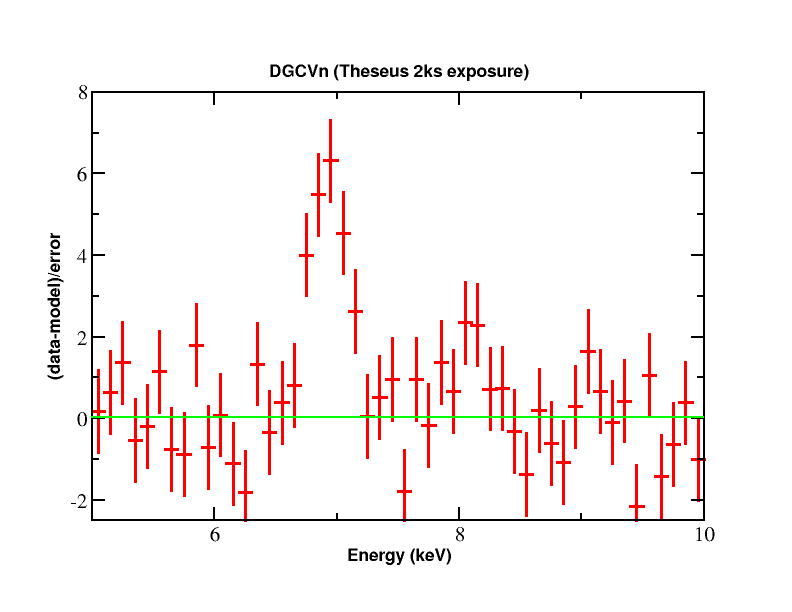}
  \caption{Simulated residuals of a 2\,ks XGIS-X spectrum showing Fe lines from the 280\,MK optically thin thermal emission model fit to the Swift DG CVn super-flare 
\cite{2016ApJ...832..174O} when fitted with a non-thermal bremsstrahlung model which fits the Swift data equally well. THESEUS is easily able to distinguish these models, constraining
the plasma evolution on short timescales. The normalization of the model was chosen so that the fluxes in the 0.3-10 keV and 14-100 keV energy ranges are 
${\sim}4{\times}10^{-9}\,{\rm erg}\,{\rm cm}^{-2}\,{\rm s}^{-1}$  (i.e. similar to the fluxes of the first super-flare of DG CVn during 2014). }
  \label{stars_residuals}
\end{figure}

The Fe line  in EV Lac is the only one observed so far in a main-sequence star (apart from the Sun).  
Observation of super-flares with THESEUS  will lead to  Fe line detections in many more cases.  
 The simulation of a super-flare similar to that of DG CVn  shows that THESEUS will be able to significantly detect these fluorescent emission lines  in a single 2\,ks pointing   (see Fig.~\ref{stars_residuals}), thus allowing  to  study the line flux variability at high time resolution. Constraints on the hard X-ray flux level will be provided as well.

\subsection{The Neupert effect and the timing of stellar flares}

The Neupert effect \cite{1968ApJ...153L..59N} is a time delay between different energy ranges observed in flares from the Sun, as well as in just a few radio and X-ray stellar flares, e.g. from UV Ceti  and Proxima Centauri \cite{1996ApJ...471.1002G,2002ApJ...580L..73G,2004A&A...416..713G}.  

It was found  that the X-ray light curve approximately follows the time integral of the V-band emission (or of the radio emission in the case of the Sun).  
In the framework of the chromospheric evaporation scenario \cite{1968ApJ...153L..59N}, this is explained by the fact that the  high-energy electrons travel along magnetic fields, where the high-pitch  angle population emits prompt gyrosynchrotron emission and the low pitch angle population impacts in the chromosphere to produce prompt radio/V(IR) band emission. The hot thermal  plasma (soft X-rays) evolves as a consequence of the accumulated energy deposition, hence the integral relation. 

\begin{figure}
\includegraphics[width=8.5cm]{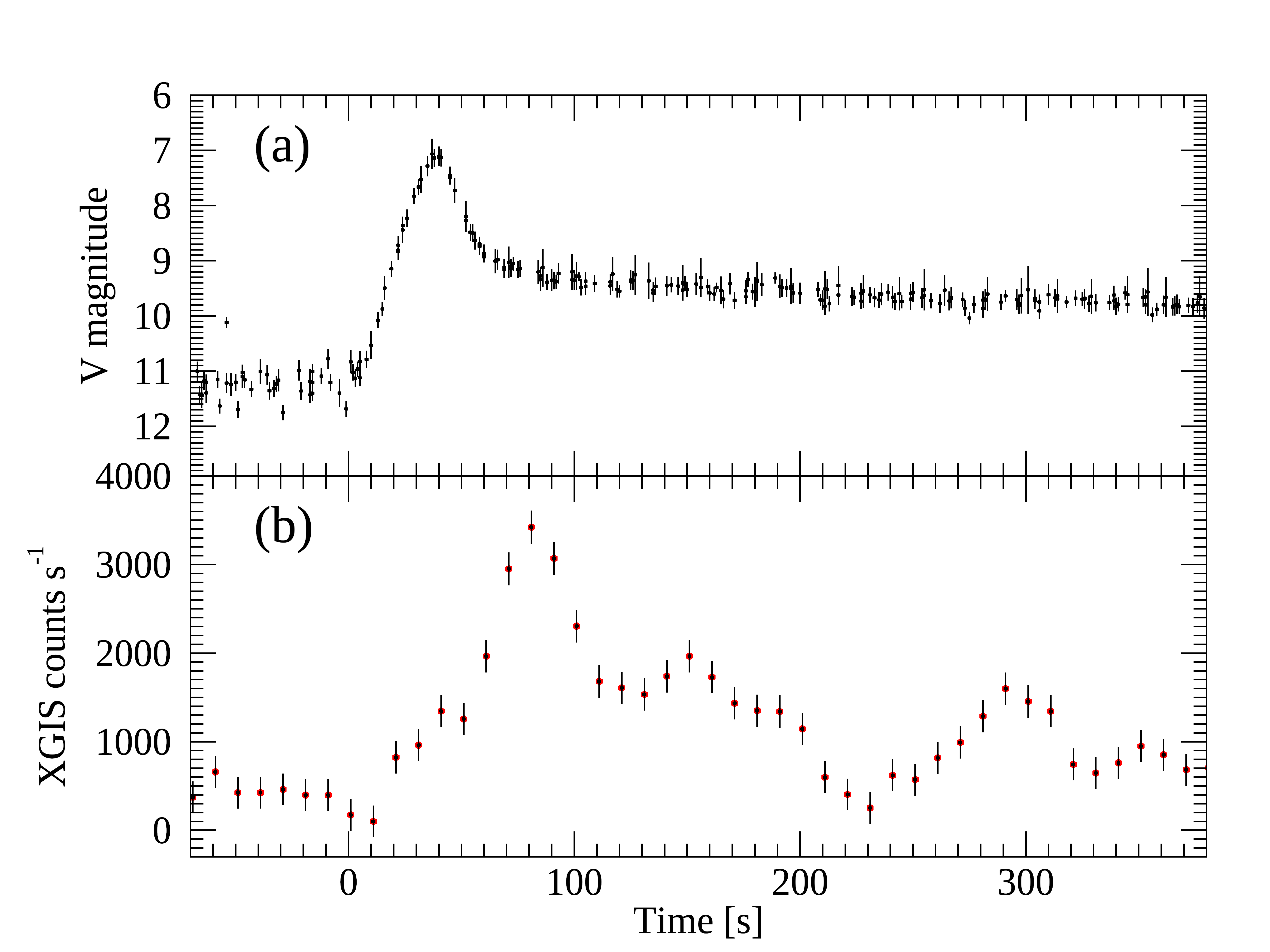}
\caption{{\bf a)} Observed light curve of the optical emission of the initial part of the DG CVn super-flare \cite{2015MNRAS.452.4195C}. 
{\bf b)}  Simulated XGIS light curve in the 2-30 keV energy range. 
 }
\label{neupert_delay}
\end{figure}

Recently,  this relation has been tentatively observed also in the hard X-ray band in powerful flares from DG CVn detected with Swift/BAT 
\cite{2015MNRAS.452.4195C}. 
In the chromospheric evaporation scenario, the soft X-ray emission is a signature of the thermal emission from the   plasma  heated by the impact of  the accelerated particles, that are responsible for the early optical/radio-emission. 
Therefore, the detection of hard X-ray emission following the integral of  the impulsive optical emission is something unexpected. This indicates that, contrary to what was previously understood, either the plasma heats up to E $>$ 15 keV or   the 
particles emit radiation following a non-thermal kinetic distribution. 
The X-ray data of the  initial part of the super-flare were well fitted by a thermal plasma model with    temperature   ${\rm T}_{\rm X}=280$\,MK and peak flux of $\sim9\times10^{-9}$  erg cm$^{-2}$ s$^{-1}$ in the 0.01--100 keV range \cite{2016ApJ...832..174O}.
Based on these spectral properties and on  the Swift/BAT light curve \cite{2015MNRAS.452.4195C},   we simulated the  2-30 keV light curve expected in the XGIS for a similar event (lower panel of Fig.~\ref{neupert_delay}).  The brightness of the flare in the visual band (top panel of Fig.~\ref{neupert_delay}) implies that the  IRT will be able to obtain significant detections with very small integration times. 
These simulations indicate that observations of  optical/X-ray delays of the light curves, as well as X-ray spectroscopy, of similar events with THESEUS are feasible and will provide definite insights into the nature of  these exceptional and rare events. 
 
\section{White Dwarf X-ray Binaries}
\label{sec-WD}

X-ray binaries containing white dwarfs (WD)  form a large and heterogeneous population. They consist of several classes of sources with different X-ray properties, depending on the type of companion star, magnetization of the WD, and interplay between the main powering mechanisms, i.e. accretion and thermonuclear burning (see, \cite{1995Warner,2017Mukai,2020Balman}, for rewiews).

In  Cataclysmic Variables (CVs) the  donor star is generally a Roche-lobe overflowing late-type main sequence star,  or sometimes a slightly evolved star. CVs have orbital periods of 1.4-13 hr,  with a few exceptions up to 2-2.5 days, and  can be divided into two main classes: non-magnetic and magnetic. 
When the   magnetic field of the WD is weak  ($B$ $<$ 0.01 MG) an accretion disk can form and reach all the way down to the WD.  
Sub-types of non-magnetic CVs are dwarf novae (DNs) that  show state changes and nova-likes (NLs) consisting of CVs that are almost always in high state (except for a few low states). There are  several subclasses among nonmagnetic CVs and  there has been suggestions on some subclasses to host suspected low magnetic field WDs.  
The other class is that of magnetic CVs (MCVs), comprising about 25\% of the CV population, and it is in turn divided into two sub-classes according to the degree of synchronization of the binary. Polars have  $B\sim20-230$ MG, which prevents the formation of a disk and channels the accretion flow  directly onto the magnetic pole(s) of the WD. 
The magnetic and tidal torques cause the WD rotation to synchronize with the binary orbit. 
The Intermediate Polars (IPs) contain fast asynchronously rotating WDs ($P_{spin}/P_{orb} \sim$0.1) due to their weaker magnetic field, $B\sim$5-30 MG.
They may be disk-fed, stream-fed, or in a hybrid mode in the form of stream-fed disk-overflow,  which may be diagnosed by spin, orbital and sideband periodicities at  different wavelengths \cite{1995ASPC...85..185H,1997MNRAS.289..362N}.

Other related classes of accreting WD binaries (AWBs) comprise the so-called AM CVn systems,  hosting either two WDs  or a He-star plus a WD,  and the Symbiotics, in which the companions are red giants or Mira stars and accretion is usually sustained by stellar winds.
AM CVns, being ultra compact systems with $P_{orb}$ between 5 and 65 min,  are particularly interesting as sources of low frequency gravitational waves for the future LISA mission.

\subsection{Magnetic Cataclysmic Variables}

MCVs, being  the brightest X-ray emitting CVs with luminosities 10$^{30-34}$ \lum ,   are widely studied in  X-rays and readily detected  in surveys, thus playing a crucial role in our understanding of the  Galactic X-ray binary populations \cite{2020deMartino}.

In MCVs, the accretion flow close to the WD is channeled along the magnetic field lines reaching supersonic velocities and produces a stand-off shock above the WD surface. The post-shock region is hot (kT$\sim$ 10-50  keV) and cools via thermal bremsstrahlung, producing hard X-rays 
and cyclotron radiation emerging in the optical/NIR band. Both emissions are partially thermalized by the WD surface and re-emitted in the soft X-rays (blackbody kT$\sim$ 30-50 eV) and/or EUV/UV. 
The relative proportion of the two cooling mechanisms strongly depends on the field strength and the local mass accretion rate. Cyclotron radiation dominates for  high field polars and suppresses bremsstrahlung cooling and high shock temperatures.
The  post-shock region has been diagnosed by spectral, temporal and spectro-polarimetric analysis in the optical, NIR and   X-ray regimes, which have shown  complex field topology  with different emission regions of several accretion spots, as in quadrupole effects \cite{2020Mason} and differences between the primary and secondary pole geometries and emissions \cite{2015SSRv..191..111F,2018MNRAS.481.2384P}. The complex geometry and emission properties of MCVs makes them ideal laboratories to study the accretion processes in moderate magnetic field environments, also helping to  understand the role of magnetic fields in close-binary evolution.

Two aspects of interest for THESEUS observations are  a)  variability properties and  b) search for 
reflection humps in  a selected number of bright known systems:

a) High and low states in MCVs  consist of luminosity variations up to two  orders of magnitude, occurring on timescales of weeks to several months
with the systems lingering in one or the other state for months or years. Whether the occurrence of 
low states is due to magnetic spots temporarily located at the L1 point or the
donor star underfilling its Roche lobe is still debated \cite{1994ApJ...427..956L}. They are poorly 
explored in  X-rays, as instead largely done in the optical band.  X-ray monitoring of state changes  
over the THESEUS lifetime will be crucial to assess the re-shaping of the accretion geometry close to the WD 
surface in response to changes in the mass transfer rate from the donor star \cite{2020ApJ...896..116L}. 
The unique long-term coverage of THESEUS will thus allow us to trace the time history of the mass  accretion rate and accretion geometry which can only efficiently diagnosed in the X-ray band. 

\begin{figure}
    \centering
    \includegraphics[width=\columnwidth]{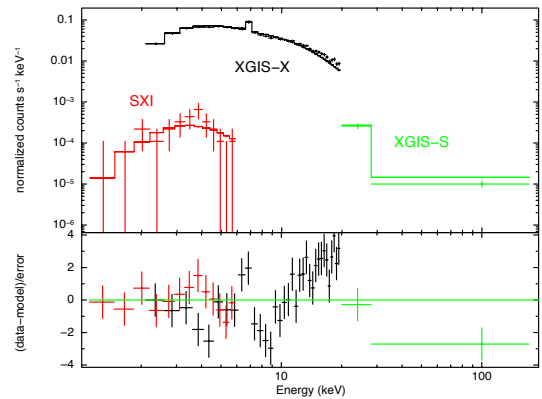}
    \caption{A simulation of a 25 ks exposure of a MCV with THESEUS. The spectrum was simulated with 
a multi-temperature plasma emission model (with T$_{\rm max}$=45 keV) and a Compton  reflection  component. 
An average intrinsic N$_H$ 
of 5$\times$10$^{22}$ cm$^{-2}$  is assumed.  The count rates 
are 0.0013 cts s$^{-1}$ (0.3-6 keV),   0.66 cts s$^{-1}$ (2-30 keV)  and   0.0033 cts s$^{-1}$ (30-150 keV).  
The bottom panel shows the 
residuals obtained when the reflection component is not included in the fit. 
}
    \label{fig:3-D-mcvrefl}
\end{figure}

In addition, and  on much shorter timescales,  besides the aforementioned periodic variabilities,  
the power spectra of IPs also display 
frequency breaks, that can be explained by fluctuations propagating in a truncated optically thick accretion disk \cite{2010Revnivtsev,2011Revnivtsev,1997MNRAS.292..679L}.
This model has been recently applied to study the differences of spectral and geometrical characteristics  of the accretion columns, as well  as  for  the WD mass determinations \cite{2019Suleimanov}. 
THESEUS, with its scanning capability in 4 year time-span,  can be used as a tool to combine 
timing and spectral analysis to derive constraints on WD masses in   these systems. 
 
b) In MCVs, X-ray emission is located close to the polar regions of the WD surface, which is expected to reflect a significant fraction of intrinsic X-rays above 10 keV, producing a Compton reflection hump. Improved hard X-ray sensitivity with the imaging  NuSTAR satellite has provided  the first robust detection of a Compton hump in three objects \cite{2015Mukai}.
A reflection hump has been also detected in a symbiotic system (nonmagnetic, \cite{2018Luna}). 
Such reflection humps  can be  detected with the THESEUS  XGIS instrument in the bright CVs.
To demonstrate these capabilities, we show in Fig.~\ref{fig:3-D-mcvrefl}   simulations based on the 
spectral model fitted to NY Lup  (see \cite{2015Mukai}), corresponding to source fluxes of 
8.5$\times$10$^{-11}$ erg  cm$^{-2}$ s$^{-1}$  (2-30 keV) and  (1-2)$\times$10$^{-11}$ erg 
cm$^{-2}$ s$^{-1}$  (30-150 keV).
 We  adopted  a 25 ks exposure,  which is about a one-day coverage of the THESEUS detectors when they 
are on the source. 
The  simulated joint spectra of the three detectors, using an absorbed multi-temperature plasma,  are fitted  with the reflection amplitude  set to 0 to show how well the  the Compton reflection hump can be detected.

\subsection{Nonmagnetic Cataclysmic Variables and Related Systems}

Keplerian accretion disks around nonmagnetic WD   dominate in the UV optical and NIR, while X-rays are emitted from regions within the inner disk 
closer to the WD. 
In the context of standard accretion theory,  these are believed to be boundary layers and  may 
be optically thin, producing bremsstrahlung-like emission in the hard X-rays during low mass accretion rate states, 
$\dot M_{acc}$$<$10$^{-(9-9.5)}$ M$_{\odot}$ yr$^{-1}$, or optically thick, producing 
blackbody-like emission when $\dot M_{acc}$$\ge$10$^{-(9-9.5)}$ M$_{\odot}$ yr$^{-1}$. 

The mass transfer rate from the donor is believed to be constant in the standard accretion disk theory. 
 However, a consistent picture of the aforementionend high/low states in both nonmagnetic and MCVs 
is still missing \cite{2014Zemko,2020AdSpR..66.1090S}. 
In addition, a few systems observed in low states have revealed flare-like variability in X-rays, 
suggesting sporadic mass transfer enhancement events (e.g., \cite{2013Bernardini}). 
 
Many CVs exhibit dwarf nova (DN) outbursts,  lasting days to weeks and with a  wide range of duty 
cycles (months to years), during which the optical brightness increases by 2-6 magnitudes.  
These outbursts are explained by the  disk instability model (DIM \cite{2001NewAR..45..449L}, \cite{2020AdSpR..66.1004H}), similarly to those of low-mass X-ray binaries (LMXBs). 
The disk is usually in a cool, mostly neutral, inactive state, accumulating matter transferred from the 
secondary until it switches  to a hot, mostly ionized, active state, dumping the accumulated mass onto the 
WD. One key question is how, and how quickly, the boundary layer responds to the variable mass accretion 
rate: many DNe (also, in some AWB active states) 
show an anticorrelation between optical and X-rays (e.g., SS Cyg \cite{2003MNRAS.345...49W}, 
see \cite{2020Balman} for a review), while relatively few show direct correlation (e.g., 
GW Lib \cite{2009MNRAS.399.1576B}), the  diversity in the X-ray and optical behaviour is not 
predicted by the standard accretion or DIM theories. Whether it depends on the mass accretion rate and
WD mass has still to be assessed \cite{2011PASP..123.1054F}. 
The DIM also faces a significant challenge in explaining the quiescent behaviour of DNe and 
related objects. The model depends on the disk accumulating mass during the inter-outburst 
period, and as the disk becomes more massive, the accretion rate  onto the central object 
(hence the  luminosity) should increase from the end of one outburst to the beginning of 
the next. However, this has never been confirmed (e.g., VW Hyi \cite{2019MNRAS.488.5104N}, 
SS Cyg \cite{2004ApJ...601.1100M}) and  actually the opposite behavior has   been observed. 
Most DNe will be  easily at reach of the IRT, and the   simultaneous data  obtained for these sources will help in tracing the correlated or anticorrelated behaviour between NIR and X-ray emission. 

To date, only a small group of  CVs have been observed with sufficient cadence and duration in X-rays 
to allow a systematic study of DN outbursts, quiescent variability, and high and low states or active 
states of AWBs.  Although about a thousand of CVs are known to date, not all of them are 
sufficiently bright to allow such study. Those at reach of the THESEUS instruments are selected by  
using the second Swift/XRT source catalogue \cite{2020ApJS..247...54E}, which contains about 
113 CVs at fluxes above  $4\times10^{-12}$ \flux in the 0.3-10\,keV range, of which 47\% are magnetic, 
27\% are DNe, 12\% are nova-likes and 14\% of other types.  
The SXI can detect and study  at least half of these objects in a one-day (25 ks) exposure  at 
a good enough cadence, whereas XGIS-X can detect these at higher count rates and a better cadence 
can be applied when necessary. A one-day exposure time (even shorter in bright cases), is a 
suitably adequate timescale to study DN outbursts or high and low states of  these AWBs, 
\textit{systematically}.  We think this may be a lower limit and a larger number can be studied 
at lower cadence and new ones may be detected owing to the survey quality and the transient  nature of AWBs.

Soft X-ray/EUV emission, with temperatures $\sim$5--25 eV has been detected only in  a handful of DN systems in outburst,
and  also in some symbiotics during active states. The absence of the soft components in most systems is not due to absorption, since they have generally low interstellar extinction \cite{2020Balman}. DNe  in outburst show hard X-ray emission, but at a lower fluxes and temperatures compared to their quiescent hard X-ray emission (T$_{max}$=6-55 keV). The total X-ray luminosity during the outburst is typically in the range 10$^{29}$- a few$\times$10$^{33}$ erg s$^{-1}$ and the range holds for quiescence as well, with the lower level at 10$^{28}$ erg s$^{-1}$. At high mass accretion rates  ($\dot M_{acc}\ge$10$^{-9}$ M$_{\odot}$ yr$^{-1}$),  as opposed to standard steady-state disk model calculations (where soft X-ray emission is expected from BLs), observations of nonmagnetic CVs (namely nova-likes) show a hot optically thin X-ray source with luminosities $\le$ a few $\times$10$^{32}$  erg s$^{-1}$. Most AM CVn systems  exhibit similar hard X-ray spectra at relatively lower luminosities and X-ray temperatures. The symbiotic systems indicate  a brighter range with L$_x$ $\le$ several $\times$10$^{34}$ 
erg s$^{-1}$ in the hard X-rays.

These departures from the expectations of standard disk accretion theory can be interpreted in the context of  advective hot flows \cite{2020Balman}. 
In fact, multi-wavelength observations with moderate or high-resolution spectral and timing data indicate that the matter flow in the inner regions of the disk   differs  from that expected in the standard  picture in quiescence (e.g., DNe) and in high states (e.g., nova-like  and DNe), with the presence of extended regions (vertical and radial) emitting particularly in  X-rays (and to some extent in UV).  
The X-ray  power density spectra of at least 10 DNe in quiescence show frequency breaks at 1-6 mHz, as a result 
of truncation of the optically thick accretion disk (i.e. accretion flow) at radii $\sim$(3-10)$\times$10$^9$\ cm.  
These break frequencies are inversely correlated with the X-ray temperatures and a broken power law-like 
relation exists for X-ray luminosities \cite{2019Balman}. These characteristics  
can be readily explained with the existence of advective hot flows in  the X-ray emitting regions that are not removed in the outbursts. 
The main characteristics of the power density spectra  of nonmagnetic CVs indicate that,  
regardless of the type of DN (and perhaps high state CVs), X-ray broad-band noise in quiescence and 
its evolution in outburst are similar, which is a result of the properties of the inner advective hot 
flow region.

\begin{figure}
    \centering
    \includegraphics[width=\columnwidth]{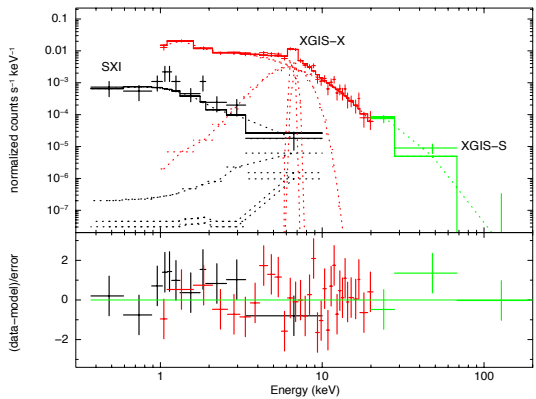}    
    \caption{A 25 ks exposure simulation of a nonmagnetic CV with 2-30 keV flux of 6$\times$10$^{-12}$   erg cm$^{-2}$ s$^{-1}$ (see text for details).  
 The count rates in the three detectors are 0.002 cts s$^{-1}$ (0.3-6 keV),   0.074 cts s$^{-1}$ (2-30 keV)  and   0.0005 cts s$^{-1}$ (30-150 keV).  
    }
    \label{fig:3-D-nCV}
\end{figure}

One of the  main THESUS contributions for our understanding of non-magnetic CVs and related systems, will 
be to study their spectral-timing characteristics comparatively,  to characterize the hysteresis effect 
(i.e., accretion history) and the spectral variations of these sources on different timescales. THESEUS 
will provide the first complete set of energy spectra together with power spectra of a large number of 
AWB.  One can study the structure changes in the disks where the advective hot flows dominate and 
determine its timing and noise properties while studying the energy spectral changes in the sources 
through low and high states, together with   unprecedented  X-ray light curves, systematically, over a four year time-span.
To demonstrate the spectral capabilities of THESEUS, we have simulated a standard nonmagnetic CV spectrum for all detectors  (see Fig.~\ref{fig:3-D-nCV}). 
We used a  bremsstrahlung with  
 10 keV temperature  with 6-7 keV iron line complex including the fluorescence  line at 6.4 keV, H-like and He-like iron (N$_H$ $\sim10^{21}$ cm$^{-2}$) and simulated a 25 ks exposure.
The SXI has the shortest source visibility window ($\sim$7 days), but as long as SXI observes the source also XGIS data will be available. XGIS provides a longer   visibility window  with about 900 ks exposure time.
We note that the total number of source  photons   will be about 2000, which is suitable for power spectra and broadband noise studies.

\subsection{Novae in outburst}

 Nova outbursts have been recorded for many centuries due to their optical luminosity, which allows even naked-eye observations in some cases. Erupting novae are also among the most luminous X-ray sources and many discoveries have been made in this energy band in recent years
\cite{Orio2012,Osborne2015,Ness2015}. The root cause of a nova outburst is the explosive burning in degenerate matter at the bottom of an accreted envelope on a WD. The importance of novae studies is twofold. One aspect of the importance of novae is that they are candidate progenitors of SNe Ia, especially the so called recurrent novae, that host massive WDs and whose outbursts are repeated multiple times over human life timescales of years or tens of years \cite{Starrfield2012a,Starrfield2020}. 
Given the crucial contribution of SNe Ia in determining distances and cosmological parameters, the best possible understanding of these interacting WD binaries is required. 

The other important issue is the contribution of Novae to the Galactic nucleosynthesis \cite{1991A&A...248...62D}. Long ago Novae have been predicted as producers of light elements as Beryllium and Lithium (e.g. \cite{1978ApJ...222..600S}), however only very recently observational evidence has been accumulated in this direction \cite{2015Natur.518..381T,2015ApJ...808L..14I,2020MNRAS.492.4975M}, although this topic is still a matter of lively debates (e.g. \cite{2020A&A...639L..12S}).  

 Novae emit X-rays for three reasons. Some time after optical maximum, the burning-heated photosphere of the WD shrinks back to almost the pre-outburst radius, and becomes a very bright super-soft X-ray source (e.g. \cite{Nelson2008,Rauch2010,Osborne2015}) for periods lasting from few days to 20 years (see also the models' predictions \cite{Starrfield2012b,Wolf2013}). The ejecta of novae are also feature-rich emitters of high energy radiation, with shocks causing hard X-ray and, surprisingly, even GeV emission \cite{Franckowiak2018}. A third mechanism has not been observationally explored yet, but it is predicted by the theoretical models: in the  first few hours after the explosion, an extremely luminous soft X-ray source should be detectable, but it is short-lived and can only be revealed by sensitive, wide-field survey instruments. This is called the ``fireball phase'' and allows monitoring of the shock breakout \cite{Starrfield1998,Kato2015}. 
 
\begin{figure}
    \centering
    \includegraphics[width=\columnwidth]{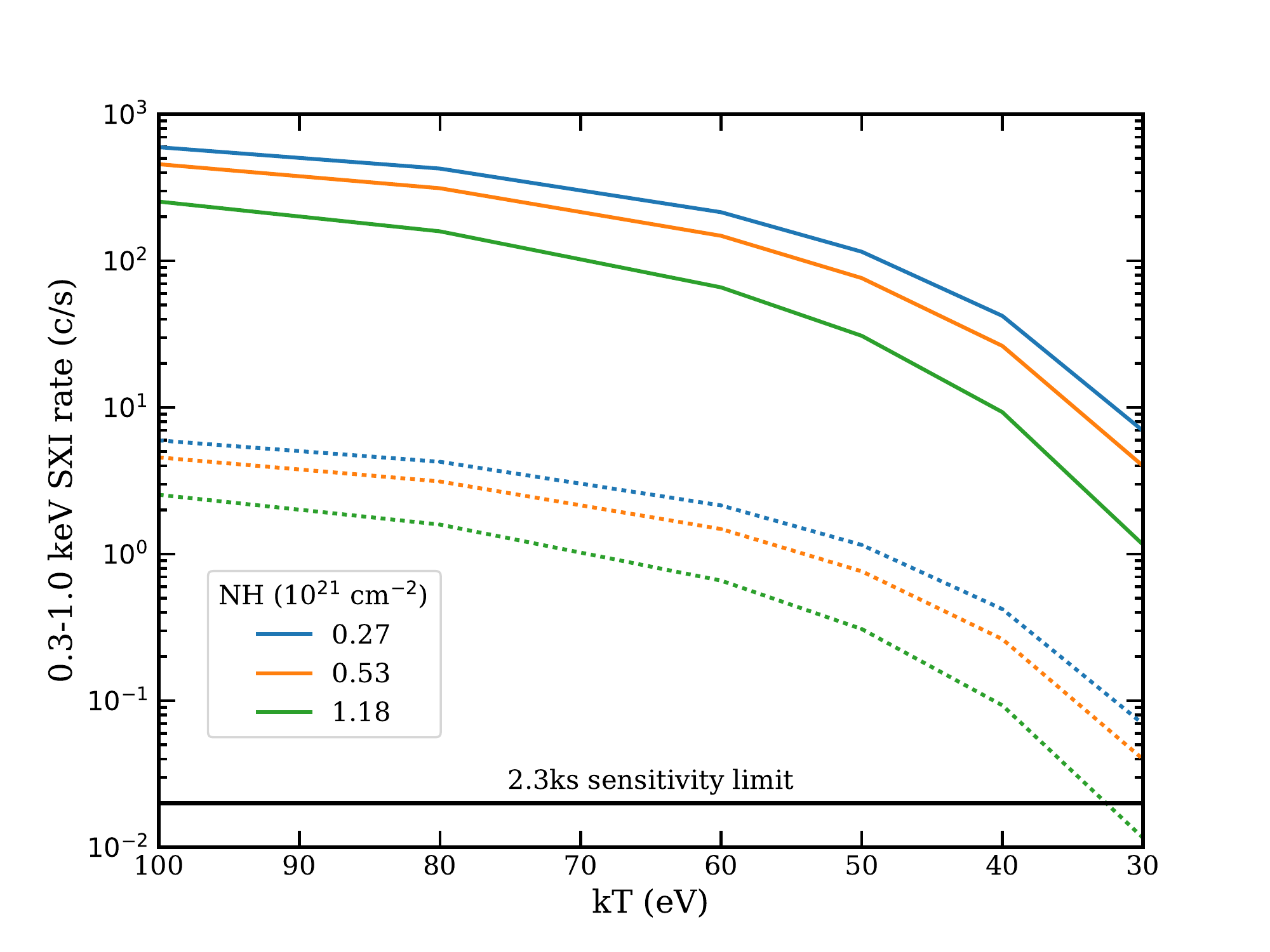}
    \caption{Predicted SXI count rates from the as-yet undiscovered classical nova shock breakout derived from the model of Kato et al. \cite{Kato2016}. The upper and lower sets of curves correspond to distances of 1 and 10 kpc, the typical Galactic range. The colours refer to differing absorbing column densities, corresponding to 25\%, median and 75\% of the cumulative distribution of Galactic HI values.
The SXI detection limit for the standard unbroken observation length in the dynamic pointing scheme is shown. The peak temperature of the shock breakout is a measure of the mass of the WD, the range displayed covers the mass range of $1.3-0.6{\rm\ M_{\odot}}$}.
    \label{fig:sxi-nova-fireball}
\end{figure}

 Despite all the recent progress, there is much about nova physics that will remain unknown up to the THESEUS era. The ultimate goal and promise of THESEUS is actually the discovery of the fireball phase in a number of novae. So far, there has been only one claim of detection of this phase, for a peculiar nova in a high mass system \cite{Morii2013}, but  this X-ray source was hotter and more luminous than the theoretical predictions, so its real nature is still a matter of debate. Otherwise, Swift and MAXI obtained only upper limits on the luminosity and blackbody temperature of the fireball \cite{Kato2015,Morii2016}. THESEUS, with the wide field of view of the SXI is the ideal instrument to characterise for the first time this pre-optical-maximum state during the shock breakout at up to a few times 10$^{38}$ erg s$^{-1}$. Figure~\ref{fig:sxi-nova-fireball} illustrates the typical range of SXI count rates expected from the fireball phase as a function of peak shock breakout temperature, column density and distance, showing it is readily detectable.

\begin{figure}
    \centering
    \includegraphics[width=\columnwidth]{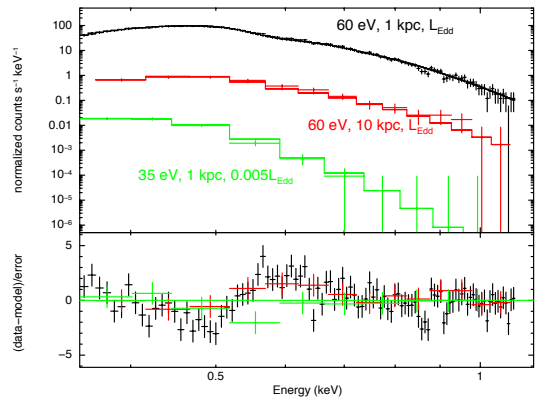}
    \caption{A simulation of soft X-ray component spectrum of a nova for the SXI instrument using a 60 eV blackbody model  with N$_H$=2$\times$10$^{21}$ cm$^{-2}$.
    The black curve assumes a 60 eV blackbody emission with Eddington luminosity at d=1 kpc, giving 21.1 cts $s^{-1}$ in the SXI, whereas the red curve is the same blackbody emission for  d=10 kpc   simulated for the constant bolometric luminosity phase   (2.5 ks, predicted rate 0.22 cts $s^{-1}$). 
    The green curve represents a 0.005L$_{\rm Edd}$ of a 35 eV cooling WD after X-ray turn-off with an exposure of 150ks and d=1 kpc. 
    }
    \label{fig:sxi-nova}
\end{figure}

 THESEUS  will also detect the luminous and hot supersoft phase in novae with effective temperature above $\approx$ 400,000 K at the (0.005-1) times the Eddington luminosity (a few times 10$^{38}$ erg s$^{-1}$), and trace the evolution of the soft X-ray components of novae during the outburst stage \cite{1998Balman,Nelson2008,Orio2018}. With these parameters,  THESEUS will also detect  the most massive WD in novae, those that are close to the Chandrasekhar mass \cite{Starrfield2012b,Wolf2013}, so these serendipitous observations will really constrain the statistics of viable type Ia SNe candidates. In order to show the capabilities of THESEUS SXI on following the evolution of the soft component (blackbody emission) of novae during the outburst stage, we simulated two blackbody spectra using 2.3 ks exposure  (see Fig.~\ref{fig:sxi-nova}), at the Eddington luminosity with 720,000 K effective temperature. The brighter spectrum is at 1 kpc (black) and the dimmer one (red) is at 10 kpc source distance. The green spectrum (at the bottom) is a cooling WD at about 400,000 K effective temperature with 0.005L$_{\rm Edd}$ that could be detected after the X-ray turn-off, once the constant bolometric luminosity evolution ends. We note that for the cooling WDs, the entire 150 ks (about a week) duration of scan data need to be used to achieve this spectrum. Such a low cadence after the X-ray turn-off is still suitable for studies.
  
Last but not least, THESEUS will also constrain the shock physics via at least two methods. The SXI will measure the X-ray temperature and luminosity. In combination with the orders of magnitude improvement in sensitivity of the Cherenkov Telescope Array compared to the Fermi LAT, the acceleration of the highest energy particles can be characterised in close-by (within a few kpc) novae.  Also, by combining  SKA measurements, the ejecta geometry and energetics can be further constrained. In addition, for the few novae detected at higher energies by the XGIS, the buried shock hypothesis \cite{Metzger2015} for the origin of the bright optical emission of novae can be definitively tested.

\begin{figure}
    \centering
    \includegraphics[width=\columnwidth]{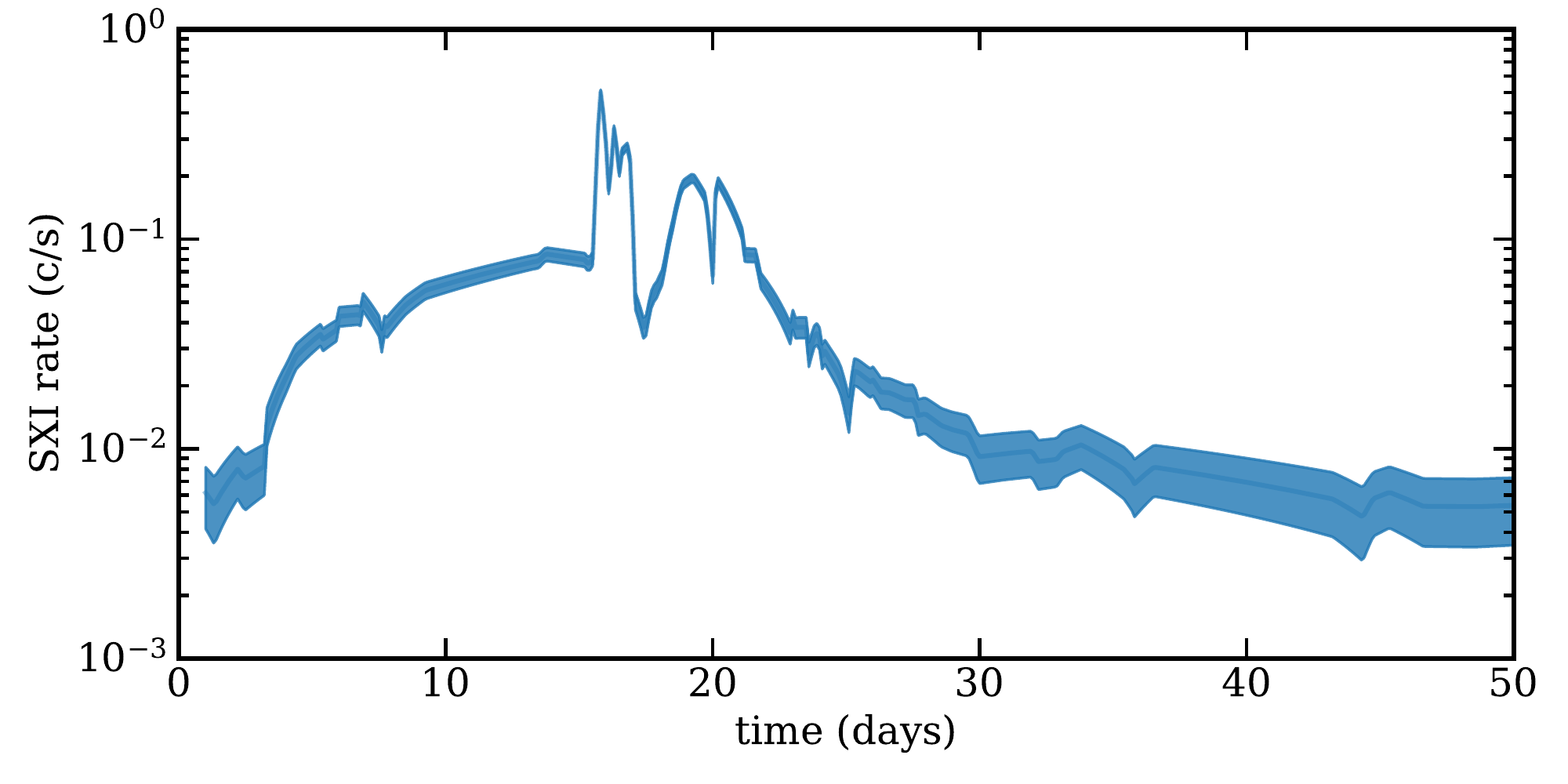}
    \includegraphics[width=10cm]{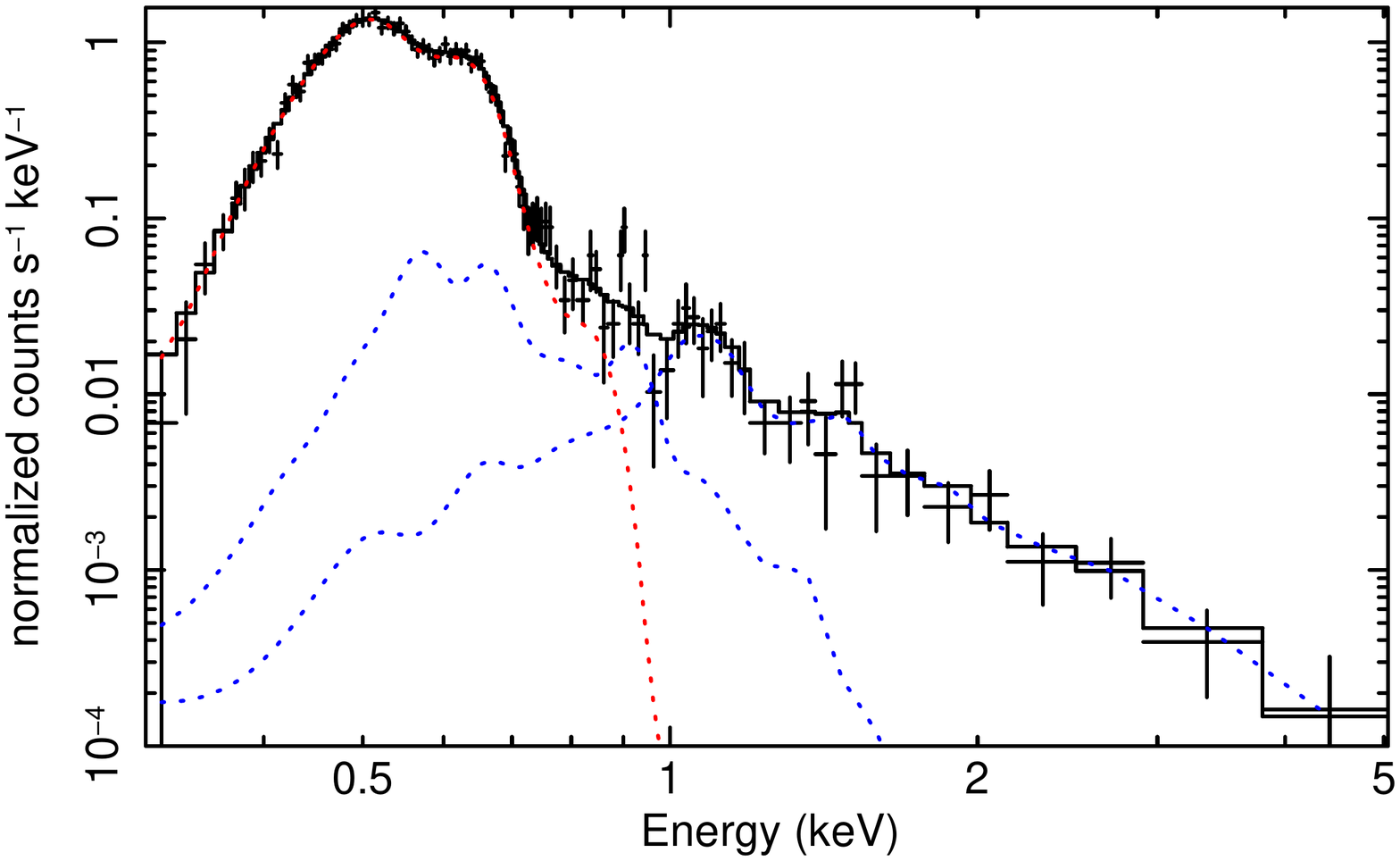}
    \caption{(a) The top figure shows the predicted SXI count rates that would have been seen from the recurrent nova V3890~Sgr, derived from  Swift/XRT observations \cite{KPage2020A}. The shaded region denotes the 1-$\sigma$ Poisson error range expected for an exposure time of 2.3 ks, typical for THESEUS visits. (b) The bottom figure shows the predicted SXI spectrum on day 20 of the outburst, as simulated for an exposure of 20 ks (i.e. multiple visits combined) to reveal the hard component. The supersoft source emission dominates the SXI count rate, with the shocked ejecta contributing only 10\% of the emission on day 20.}
    \label{fig:sxi-nova-v3890sgr}
\end{figure}

Figure~\ref{fig:sxi-nova-v3890sgr}(a) shows the SXI light curve
expected for a typical nova, in this case based on the Swift/XRT
observations of the 2019 outburst from the recurrent nova V3890~Sgr
\cite{KPage2020A}. In common with other recurrent novae containing a
red giant donor, the SXI emission comprises the supersoft component
from the nuclear burning WD and a harder component from the shocked
ejecta as it interacts with the red giant wind, as illustrated by the
simulated SXI spectrum in figure~\ref{fig:sxi-nova-v3890sgr}(b).  A
favourably positioned nova would receive multiple SXI pointings a day,
producing unique temporal coverage of the sometimes highly variable
supersoft source stage seen during some outbursts (e.g. 
\cite{KPage2020B} and ref. therein).

For both the supersoft phase and the shocked ejecta detections, we can make some clear predictions based on the Swift/XRT, which  detects 18\% of the novae it observes at a rate above 0.4 cts s$^{-1}$. 
These will be the novae detected with SXI in  2 ks detection leveI. Given the current uncertainties on the value of the galactic nova rate $\sim$20--100 yr$^{-1}$ \cite{2020A&ARv..28....3D,2021arXiv210104045D},  we expect to observe between 4 and 18 objects per year. Therefore, THESEUS observations have the capability to significantly decrease the uncertainty on the value of the frequency of occurrence of Galactic Novae and   provide the empirical basis to quantify the contribution of Novae to the Galactic nucleosynthesis. 
More detections will be due to the discovery of the expected fireball phase in some novae.

\section{Neutron Star X-ray Binaries}
\label{sec-NS}

Neutron star X-ray binaries (XRBs) are binary systems where the bulk of the energy is powered by   accretion of matter supplied by a donor component to a neutron star (NS).
NSs host the strongest magnets ever produced by nature, and have surface magnetic fields with strengths spanning from modest $\sim 10^{8}$\,G to truly extreme $\sim 10^{15}$~G. A field of this strength is capable to change the overall structure of the accretion flow up to  distances of  thousands of NS radii. A so-called ``magnetosphere'', where the motion of the accreting matter is governed by the magnetic field of the star, is therefore formed around the compact object. The accreting matter is channeled by the field lines onto two small regions close to the magnetic poles, which makes the emission strongly anisotropic, and, since the NS is spinning, gives rise to the phenomenon of X-ray pulsars (XRPs) \cite{1971ApJ...167L..67G,1972ApJ...172L..79S}. 

In weakly magnetized objects, when the radius of the magnetosphere is smaller than that of the NS, the accretion disk interacts directly with the surface of the  compact object. Therefore, no (or only weak) pulsations are observed, and the accretion physics is governed instead by processes which are intrinsic to the accretion disk and its interaction with the star surface (or with a very compact magnetosphere which extends for just a few NS radii). On the other hand, the observational properties of more stongly magnetized NSs are determined by their magnetic field and by the details of the (still poorly known) interaction of plasma and radiation in presence of  strong magnetic fields. 

Among  XRBs, strongly magnetized NSs (B$\gtrsim10^{11}$ G) are typically found in young massive systems, i.e. high mass X-ray binaries (HMXBs), while 
weakly magnetized NSs (B$\lesssim10^{11}$ G) are typically hosted in older, low mass systems and thus referred to as low mass X-ray binaries (LMXBs). A large fraction of both LMXBs and HMXBs are transient sources, and thus are interesting targets for a transient-focused mission like THESEUS. 
 
 \subsection{High mass X-ray binaries}
\label{sub-HMXB}

Among transient HMXBs, two classes of sources are of particular interest for THESEUS, i.e. Be X-ray binaries (BeXRB) and Supergiant Fast X-ray Transients (SFXTs). In both cases, extreme luminosity variations  up to a factor of $\sim10^6$ are observed  \cite{2017mbhe.confE..52S,2020MNRAS.491.1857D}, although on different timescales. 

In BeXRBs,  luminous outbursts lasting up to several months are observed when the NSs cross the circumbinary disk around  their fast-rotating    companions of Be spectral type, which possess  equatorial excretion disks due to their fast rotation.   BeXRBs may reach luminosities up to $\sim10^{38-39}$\,erg\,s$^{-1}$ and, when in outburst, are often among the brightest sources in the X-ray sky. 
Thus, bright BeXRB outbursts attract a strong interest and often lead to  extensive observational campaigns in X-rays and other bands. 
For less luminous BeXRBs, high cadence observations are still scarce, and thus spectral and timing variability on short timescales largely remains unexplored. Of particular interest in this context is the possibility to monitor with THESEUS  the evolution of the broadband continuum and of the cyclotron features observed in some bright BeXRBs during outbursts.
 
The  SFXTs  are high-mass binaries, with normal supergiant companions, which exhibit variability with a large dynamic range on short timescales, up to a factor of $10^4$ within minutes. It is unclear if SFXTs behave differently from persistent HMXBs because of the properties of their compact objects or because of the properties of the accreting wind  matter  (or both). 
The origin of their extreme variability is not well understood and models based on a particularly clumpy nature of the mass donor wind and/or  on the interaction of the accretion flow with the NS magnetosphere have been considered (see \cite{2017mbhe.confE..52S} for a review).

Both source classes are fairly large and include tens of objects, but their understanding, despite decades of studies, still presents several unresolved issues.  THESEUS can provide crucial contributions regarding the properties of the population at the lower luminosities, where the sensitivity of past and current all-sky monitors is insufficient to detect any X-ray emission, and, in general, to the science questions in which sensitive broadband high cadence observations are required. Most such observations, for instance those aimed at studying the evolution of spectral and timing properties of BeXRBs as a function of their luminosity, currently   require expensive dedicated observational campaigns with pointed instruments, and thus are not always feasible. Below we discuss how THESEUS can change that.

\subsubsection{Accretion physics and XRP magnetic fields}

\begin{figure}
    \centering
    \includegraphics[width=\columnwidth]{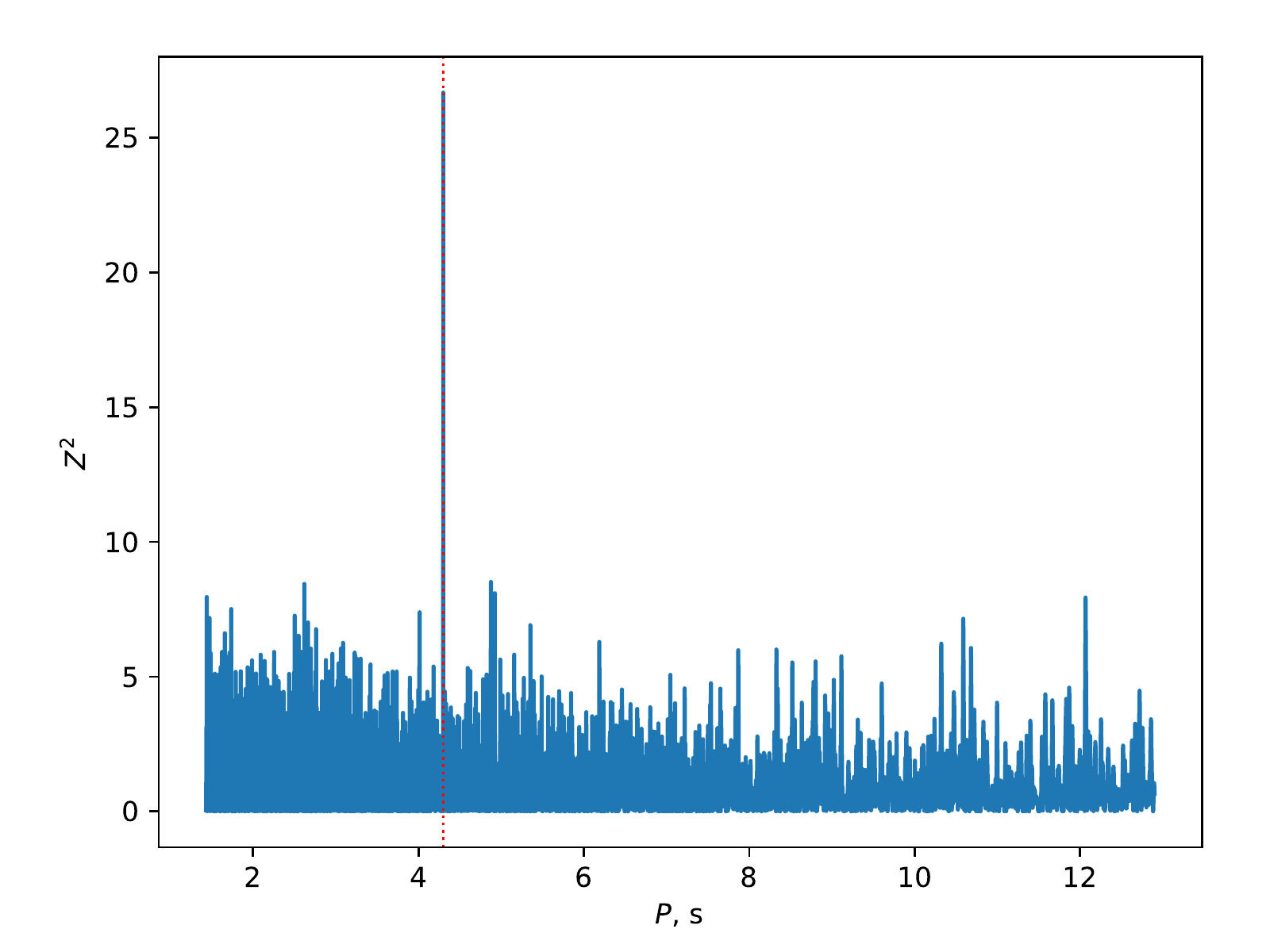}
    \caption{Simulated periodogram for 10\,ks THESEUS/SXI observation (typical daily exposure) for a representative transient X-ray pulsar with bolometric flux of $10^{-9}$\,erg\,cm$^{-2}$s$^{-1}$, spin period of $4.3$\,s and pulsed fraction of 20\% (excluding background). }
    \label{fig:periodogram}
\end{figure}

The interaction of the accreting matter with the NS magnetosphere   affects its spin evolution and variability properties (see, e.g., \cite{1975A&A....39..185I,1979ApJ...234..296G}).
Continuous observations of the transient XRPs with THESEUS during their visibility seasons, similar to the very successful Fermi/GBM pulsar program  \cite{2020ApJ...896...90M}, will allow us to study the effects of the plasma ionisation state on the NS spin-up/down process. In newly discovered, or poorly studied sources, orbital parameters will be determined as well. To illustrate the feasibility of such studies, we simulated source and background lightcurves for a representative XRP and performed a search for periodicity ignoring the contribution of background (worse-case scenario). As illustrated in Fig.~\ref{fig:periodogram}, the pulsations are clearly detected within a  single observation (10 ks), so the monitoring of the spin evolution can be used to determine the spin history of a pulsar during an outburst.
For real observations, an accurate modeling of in-orbit background will allow us to study fainter sources and objects with lower pulsed fraction, using both SXI and XGIS data.

The interaction between the accretion flow and the NS magnetosphere is also reflected by the variability properties of the source emission. It is now known that,  depending on the plasma ionization state  at the magnetosphere – accretion disk interface, two very different kinds of  phenomena can be expected at low-mass accretion rates. In the case of rapidly rotating NSs and highly ionized plasma, the accretion will be halted by a centrifugal barrier created  where the field lines rotate faster than the local Keplerian velocity. This is known as the ``propeller effect''  \cite{1975A&A....39..185I}, resulting in a sharp drop of the observed luminosity down to $\sim10^{33}$ erg s$^{-1}$,  produced  by the NS cooling \cite{2017MNRAS.470..126T}. 
Instead,  in slowly rotating pulsars the  quiescent luminosity right after the outburst was found to be about three orders of magnitude higher, i.e. $\sim10^{35}$ erg s$^{-1}$, followed by a slow decline. A model of stable accretion from a cold (nearly neutral) disk, consisting mainly of atomic hydrogen, with low viscosity and thus low and stable accretion rate, was proposed to explain these observations  \cite{2017A&A...608A..17T}. The existence of two distinct states with different viscosities is known to drive global instabilities in accretion disks of 
CVs and LMXBs, but this has not been studied in detail in the context of highly magnetized NSs,  due to the lack of suitable instruments.
A confirmation of the whole paradigm of the low-level accretion onto strongly magnetized NSs requires  long-term and  sensitive observations of transient XRPs during the fading parts of their outbursts. 
The THESEUS unique combination of large field of view and high sensitivity  will allow us to follow the oubursts down to fluxes of $\sim10^{-11}$ \flux on a daily timescale. This is about one order of magnitude better than that of current all-sky monitors, and will allow us to determine the transitional luminosity for all pulsars  within 5-7 kpc  (where the vast majority of known HMXBs reside),  as a function of spin period and magnetic field strength. It will thus be possible to determine the critical spin period (dividing two end-points of outbursts in transient XRPs theoretically expected at $P_{\rm crit}\sim$ 10-20 s), to explain the observed distribution spin periods,  to independently estimate magnetic field strength in XRPs and, ultimately, understand their rotational evolution under the influence of accretion torques.

\subsubsection{Cyclotron line studies}
\label{sub-cycl}

The magnetic field of accreting NS can be determined by several indirect methods based for instance on the study of their disk-magnetospheric interaction and the resulting variability properties,  or, more directly, through the observation of so-called cyclotron lines \cite{2015SSRv..191..293R}. 
These features originate due to resonant scattering of charged particles with the high-energy photons in the quantized Landau levels. The energy of the fundamental line corresponds to the separation between equispaced   levels. For electrons, it is given by  E$_{cyc}$=11.6B$_{12}\times(1+z){^{-1}}$\,keV, where  $B_{12}$ is the magnetic field strength in units of $10^{12}$~G and $z$ is the gravitational redshift. 
Cyclotron features have been observed to date in about 35 sources \cite{2017symm.conf..153J,2019A&A...622A..61S}, most of which are transients. 
THESEUS will expand this sample  through observations of cyclotron lines in newly discovered BeXRBs and other bright transients, thus providing better constraints on the distribution of the magnetic fields in NSs.

\begin{figure}
    \centering
 \includegraphics[width=\columnwidth]{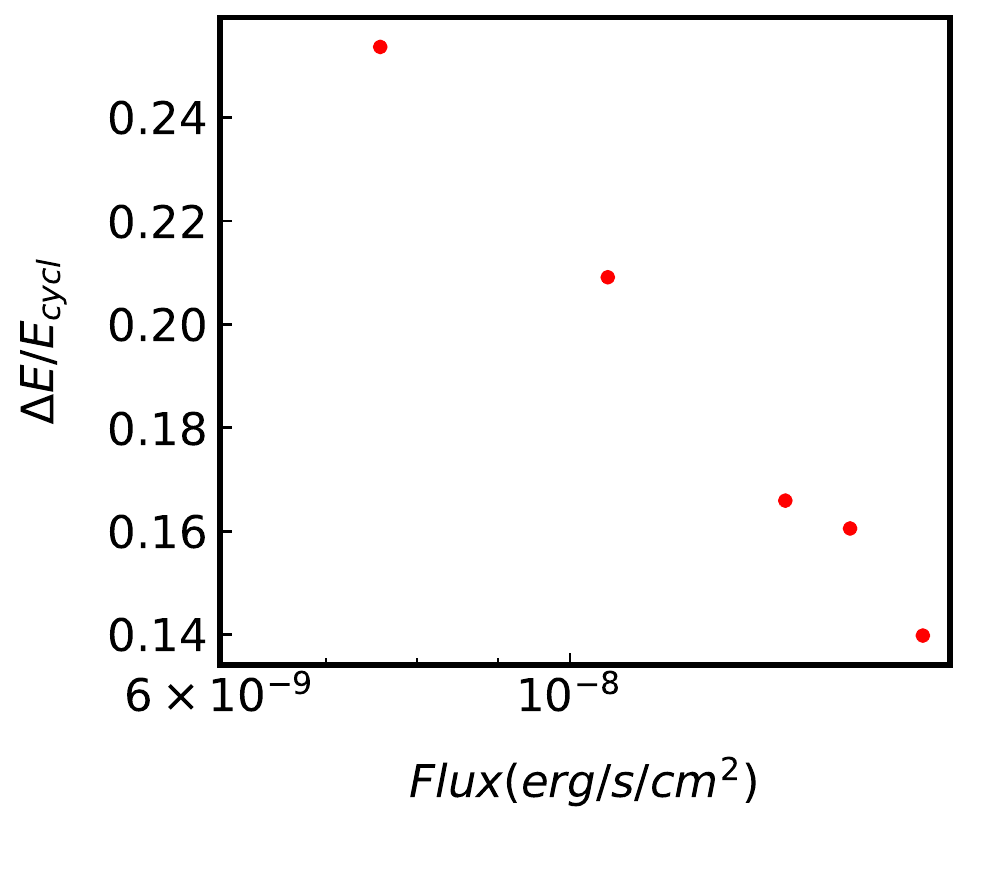}
   \caption{Relative error (90\% c.l.) on the determination of the cyclotron line energy as a function of source observed  flux (2-200 keV) for simulated XGIS  observations of 70 ks exposure time. We used a spectrum as observed in  HMXB Cen X$-$3 \cite{2008ApJ...675.1487S}, consisting of exponentially cut-off power law with an absorption line at E$_{c}$=30.7 keV.      
 }
    \label{fig:E_cyl_err}
\end{figure}

We explored the feasibility of this kind of studies taking  Cen X$-$3 as a representative of an intermediate-luminosity XRP with moderate field.  In Fig.~\ref{fig:E_cyl_err}  we show the line energy  accuracy that can be achieved as a function of the source flux for an exposure time of 70 ks. Note that the chosen   line energy of   E$_{c}$=30.7 keV, which falls between XGIS-X/S energy ranges  represents a worst-case scenario.

Some  BeXRBs  exhibit   a luminosity-dependence of the cyclotron line energy, which is believed to be associated with changes of the emission region geometry  \cite{2019A&A...622A..61S}. 
A positive  \cite{2007A&A...465L..25S}, negative  \cite{2006MNRAS.371...19T}, or no-correlation has been observed from a few sources. 
It was demonstrated \cite{2017MNRAS.466.2143D} that the analysis of  these correlations can be used to determine whether accretion occurs in sub- or super-critical regime  \cite{1976MNRAS.175..395B}. 
Theoretical studies suggest a number of hypotheses, such as variation in the line-forming region within the accretion column  \cite{2012A&A...544A.123B,2014ApJ...781...30N,2015MNRAS.447.1847M} or formation of the line by the reflected NS emission in polar and equatorial regions  \cite{2013ApJ...777..115P}. 
Considering all the current theoretical scenarios, an agreement on   where the lines are generated and in what detail they depend on the accretion column geometry has not been reached yet.
Such studies require extensive monitoring of outbursts, which are not always feasible with dedicated facilities. On the other hand, the broadband energy range covered by THESEUS will help to measure cyclotron line energy as a function of flux both for new as well as known BeXRBs. We emphasize that although THESEUS will not be able to directly compete with dedicated pointed observations with narrow field X-ray instruments, it will be able to provide almost continuous monitoring of the spectral parameters throughout the outburst, which sometimes leads to unexpected results \cite{2016MNRAS.460L..99C}.

\begin{figure}[!ht]
\centering
\includegraphics[width=7cm,angle=-90]{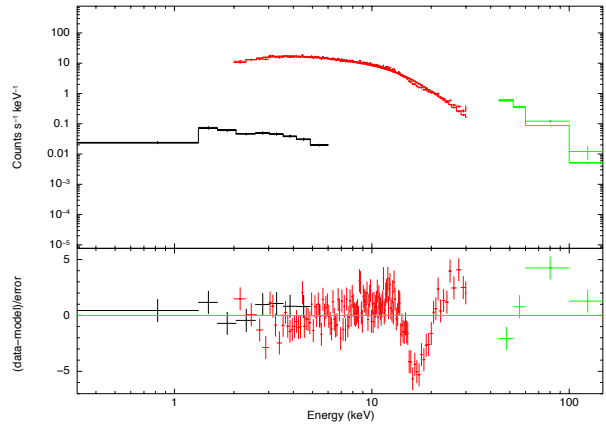}
\caption{Simulation of a SFXT spectrum in outburst (exposure time of 10 ks, SXI in black, XGIS-X in red, XGIS-S in green). The assumed model is an absorbed cutoff power law continuum modified by a cyclotron line at 17 keV, fitted by an absorbed cutoff power law with no line. }\label{fig:sfxt_cycl}
\label{fig:crsf}
\end{figure}

Cyclotron line studies are also relevant to better understand the nature of SFXTs. 
Up to now, the NS magnetic field is known only in one source of this class, thanks to the detection of a line at $\sim$17  keV    \cite{2015MNRAS.447.2274B}. 
THESEUS will permit to uncover cyclotron lines during very bright SFXT flares (F$>$10$^{-9}$ \flux , 0.3-6 keV) serendipitously caught during the survey. These are rather rare events, underlying the importance of large field of view instruments such as those of THESEUS.
In Fig.~\ref{fig:sfxt_cycl} we show the simulation of a 10 ks  spectrum  with an observed flux of 10$^{-9}$ erg cm$^{-2}$  s$^{-1}$ (0.3-6 keV).  We assumed a  cutoff power law continuum (photon index of 0.5, cutoff energy at 20 keV,  N$_{H}$=10$^{22}$ cm$^{-2}$), modified by a cyclotron line at 17 keV (line width of 2 keV, depth of 0.5, using the {\sc cyclabs} model in {\sc xspec}, similar to what reported by \cite{2015MNRAS.447.2274B}). 
The simulated spectrum has been fitted with the continuum model and no line, clearly showing negative residuals around 17 keV.   The presence of the cyclotron line is also evident when different absorbing column densities are assumed in the spectral simulations, from 10$^{21}$   to 10$^{23}$ cm$^{-2}$ (in the latter case, the observed flux in the 0.3-6 keV energy range reduces to 5.5$\times$10$^{-10}$ erg cm$^{-2}$  s$^{-1}$).

\begin{figure}
\center 
\includegraphics[width=8cm]{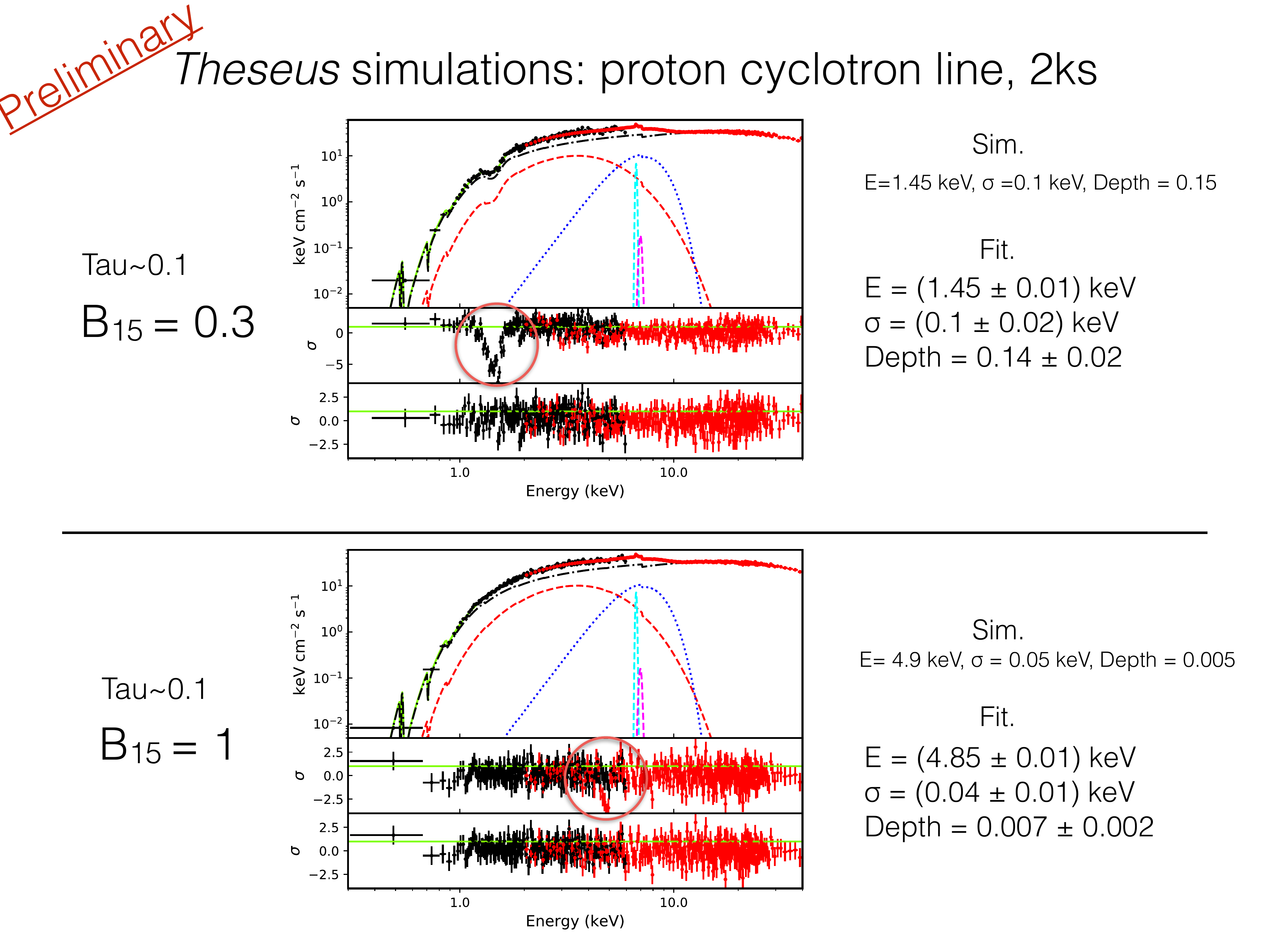}
  \caption{THESEUS simulation of a proton cyclotron absorption line in the continuum spectrum of the transient PULX Swift~J0243.6+6124 for an X-ray luminosity of $10^{39}$ \lum using  an absorbed black body and cut-off power-law plus some additional iron lines and its edge, based on the spectrum of the source during the 2017-2018 strong outburst \cite{2019ApJ...885...18J}. SXI in black, XGIS   in red.
   The simulated proton cyclotron line corresponds to a  magnetic field of $3\times10^{14}$~G. In the middle panel, the line at $\sim$ 1.45 keV is clearly identified. The residuals after a Gaussian absorption line is added are shown in the bottom panel.}
  \label{fig:ulx_p3_14}
\end{figure}

Of particular interest in the  context of cyclotron features are observations of the Galactic pulsating  ultra luminous X-ray sources (PULXs), for which a much better sensitivity for spectral studies can be achieved compared to extra-Galactic  PULXs (see Sect.~\ref{sec-ULX}). 
It has been claimed that PULXs may have magnetar-like fields \cite{2015MNRAS.448L..40E,2015MNRAS.449.2144D,2015MNRAS.454.2539M,2016MNRAS.457.1101T}. 
In this case, also  cyclotron features   due to  protons  (at an energy a factor 1836 lower than that of electrons) may be anticipated.
The recent discovery of the first Galactic PULX \cite{2018ApJ...863....9W,2020MNRAS.491.1857D},
which reached a peak flux of $\sim10^{-8}$ \flux , is very promising in this respect.
For this and similar sources  the SXI and XGIS, operating together in the 0.3 keV--20 MeV range, can detect electron cyclotron lines corresponding to fields in the $\sim 10^{11-15}$\,G range 
and proton cyclotron lines for fields above  $\sim 10^{14}$\,G. The  sensitivity to the detection of electron cyclotron  features shall be similar to that discussed above for normal XRPs. 
In Fig.~\ref{fig:ulx_p3_14} we present, as an example, the simulation for a proton cyclotron line corresponding to a $3\times10^{14}$ G field. The absorption feature is simulated in terms of a Gaussian absorption line for which the depth   and the width   of the line are related to each other through the line central optical depth (assumed to be $\tau \sim 0.1$) and other spectral parameters set to values reported by \cite{2019ApJ...885...18J}.

\subsubsection{Variability in super-giant fast X-ray transients}
\label{sub-SFXT}

Some of the models proposed to explain the SFXT behavior are based on gating mechanisms (centrifugal or magnetic barrier, or quasi-spherical settling accretion regime) able to halt accretion onto the NS most of the time. This ``gate" can be open or closed depending on the NS magnetic field and spin period.
Given the rarity, shortness and unpredictable nature of SFXTs outbursts, the properties of NS in SFXTs are elusive, and it is unclear if SFXTs behave otherwise from persistent HMXBs because of the properties of their compact objects or because of the different properties of the wind matter accreting onto them (or both). 

The SFXT  variability is likely associated with the interaction of the accretion flow with the NS magnetosphere,  but the details remain largely unclear due to the observational challenges:
the limited sensitivity of current all-sky X-ray monitors is insufficient to detect short low intensity flares and the low duty cycles and  unpredictable outbursts of these objects make dedicated observation challenging or even unfeasible. 
THESEUS, besides expanding the sample of known SFXTs,   will allow us to study in detail  their statistical properties  (e.g., duty cycles, duration and intensity distribution of the flares, and their dependence on orbital phase) at flux levels unfeasible up to now. 

The sky position of known SFXTs will be covered with total exposure times ranging from $\sim$1~Ms to more than 10~Ms (depending on the specific source, in 4 years of observations), with pointings with a typical duration of 2.3~ks, every $\sim$5.8~ks, during the seasonal observing windows. 
Since the typical duration of  the bright flares is about 2~ks \cite{2014MNRAS.439.3439P,2018MNRAS.481.2779S}, this observing strategy is particularly well suited to catch SFXT flares. A series of short flares usually compose an outburst, whose duration ($\sim$a few days, at most) has been constrained only for the very few best studied SFXTs, to date. The outburst duration is an important timescale: it has been suggested that it is linked to the properties of the magnetized wind clumps out-flowing from the  supergiant donor and accreting onto the NS \cite{2012MNRAS.420..216S,2014MNRAS.442.2325S,2016MNRAS.457.3693S}.
The search for orbital variability manifested by regular patterns in flaring behavior will also be possible. The orbital period is indeed a crucial quantity, unknown for several SFXTs.

An estimate of the number of SFXT flares which will be detected by SXI can be made considering the INTEGRAL results: for an exposure of 2 ks, a flare is detected at a flux of 2$\times$10$^{-10}$ erg cm$^{-2}$  s$^{-1}$ by INTEGRAL (18-50 keV; 5$\sigma$ detection). 
This high energy flux translates into an observed (not corrected for the absorption) flux of 6$\times$10$^{-11}$ erg  cm$^{-2}$ s$^{-1}$ (0.3-6 keV), assuming a typical SFXT spectrum with a cutoff power law model with a photon index of 0.5, a cutoff at 15 keV and an absorbing column density of 10$^{22}$ cm$^{-2}$. This represents a 5 $\sigma$ detection with SXI, as well. This allows us to calculate a rough number of flares detected by SXI, assuming similar duty cycles as observed by INTEGRAL in the 18-50 keV range \cite{2014MNRAS.439.3439P,2018MNRAS.481.2779S}. 
Therefore, assuming an average duty cycle of 1\% for bright flares,  and a total exposure time on known SFXTs of about 50 Ms with SXI, we  estimate 5$\times$10$^5$ s of exposure time during flares. This implies approximately  200-250 flares detected by SXI from the known SFXTs in 4 years of THESEUS observations. 

In case of very bright flares (0.3-6 keV flux $>10^{-9}$ \flux) it will also be possible to investigate spectral and column density evolution through the flare.  If both SXI and XGIS observe it, an exposure time as short as 1 ks is enough to obtain an uncertainty of 5\% (90\% c.l.) on the power law photon index (and cutoff energy), and $\sim$30\% error (90\% c.l.) on the absorbing column density, possibly revealing variations between the flare peak and the decay.

\subsection{Low mass X-ray binaries}
\label{sub-LMXB}

Neutron Stars LMXBs form a diverse group of XRBs, composed of weakly (with a few exception \cite{Dai2014}) magnetized NS accreting matter from solar or sub-solar companion stars in rather compact systems (with orbital periods ranging from few tens of minutes to a few days). They can be divided in persistent and transient sources, with the former having a higher time-averaged accretion rate despite the latter often being brighter for a short period of time. A further subdivision exists in both classes between Z-sources and atoll sources, named after the peculiar tracks they draw in X-ray color-color diagrams \cite{Vanderklis1989}. Both classes show spectral and temporal variability over a wide range of time-scales. Z-sources are found at higher X-ray luminosities and are characterized by soft spectra, well fit by multi-color blackbody model from an accretion disk plus a blackbody, possibly originating from the NS surface or boundary layer. On the other hand, atoll sources switch between two main spectral states, likely originating from different geometries of the accretion flow: the soft state, where emission is dominated as well by the contribution from blackbody components and the hard state, where the disk is likely truncated far from the NS and spectra are dominated by thermal Comptonization from a hot corona. Systems evolve in cycles through these spectral states \cite{MunozDarias2014}, going from hard-to-soft and then, after a decrease in luminosity, from soft-to-hard. While an analogous phenomenology is also observed in black hole LMXBs, atolls show remarkable and yet unexplained differences, e.g. they undergo  faster transitions and thus are rarely found in intermediate states \cite{Marino2019b}. THESEUS will be able to monitor the spectral evolution of almost the entire population of known persistent LMXBs during these loops, increasing enormously the amount of available data on the spectral properties of these and on how they rapidly evolve. 

The piling up of accreted material onto the surface of NSs (mostly in atoll sources) can trigger thermonuclear explosions, during which the luminosity of these systems might increase by several orders of magnitude. Such transient phenomena are called type-I X-ray bursts. Their spectral study  can be a particularly useful diagnostics of, e.g., compactness of the NS, nature of the companion star and (being standard candles) the distance of the system (see, e.g., \cite{2021ASSL..461..209G} for a recent review). The high sensitivity of SXI  in the range 0.3-5 keV combined with its large FOV make THESEUS particularly fit for the detection and the study of type-I X-ray bursts, particularly longer events where time resolved spectroscopy will be possible using THESEUS data alone.

Systems with lower time-averaged mass-accretion rates are typically transients, i.e. they show episodes of X-ray activity called outbursts (with L$_X \sim 10^{34}-10^{38}$ erg s$^{-1}$) while spending most of the time in a quiescent state (L$_X \sim 10^{31}-10^{34}$ erg s$^{-1}$). THESEUS will be able to catch the onset of outbursts from known and unknown sources, triggering   dedicated pointings from other multi-wavelength observatories. Most transient NS LMXBs in outburst behave as atolls and switch between hard and soft spectral states. An exception is represented by Accreting Millisecond X-ray Pulsars (AMXPs), NS binaries hosting rapidly spinning X-ray pulsars, which are very rarely found in soft state during their outbursts \cite{2020arXiv201009005D}. In these extremely intriguing systems the magnetic pressure of the NS   magnetosphere overcomes the pressure of the accreting material, so that the accretion flow is channeled along the magnetic field lines onto the poles of the star, giving rise to  pulsed X-ray emission. Finally a small number of transient NS LMXBs, defined as Very Faint X-ray Transients (VFXTs), have been observed to exhibit (typically long) outbursts with peak luminosities intermediate between outburst and quiescence \cite{Degenaar2017}. These sources thus remain faint even in outbursts, so catching them in this state is extremely challenging and indeed the amount of available data on these sources is rather limited. THESEUS will enable new observations of VFXTs, possibly shedding new light on their elusive nature.

\subsubsection{Accreting and Transitional Millisecond Pulsars}

Millisecond pulsars (MSPs) are fast rotating ($\lesssim$ 30\,ms), relatively weakly magnetized 
(B$\sim 10^8 - 10^9$\,G) NS found in binary
systems with  low-mass (M$\lesssim 1\,M_{\odot}$) companion stars.
Their low magnetic field, rapid rotation, and location in old binary systems (with many in 
globular clusters), suggest that MSPs are recycled
pulsars that were spun-up during a previous Gyr-long accretion LMXB phase. 
They are believed   to appear as rotation-powered radio MSPs (RMSPs) when the accretion is eventually stopped
\cite{1982Natur.300..728A,1982Natur.300..615B,1991PhR...203....1B}.
The discovery of the first accreting millisecond pulsar (AMXP) in outburst, 
SAX\,J1808.4-3658 \cite{1998Natur.394..344W}, confirmed the role of mass accretion in spinning up a NS. 
To date, about twenty AMXPs are known (see \cite{2020arXiv201009005D} for  a review).
These undergo occasional outbursts reaching X-ray luminosities up to  $\rm \sim 
10^{36}-10^{38}\,erg\,s^{-1}$, interleaved by long periods of  quiescence 
($\rm L_X \sim 10^{31}-10^{34}\,erg\,s^{-1}$) \cite{Marino2019a}. About one third of
them also displays thermonuclear type-I bursts, during which quasi-coherent oscillations are often observed.

The link between X-ray and radio MSPs was demonstrated by the discovery of a radio MSP binary that was previously found to possess an accretion disk \cite{2009Sci...324.1411A}. However only with the recent discovery of the transient 
AMXP, IGR\,J1824-2453 in the globular cluster M28, swinging from accretion to rotation-powered states, 
it has been possible to confirm the tight link between LMXBs and RMSPs \cite{2013Natur.501..517P}. 
Unexpectedly, this system also displayed an intermediate ``sub-luminous" state at   
$\rm L_X \sim 10^{33}-10^{34}\,erg\,s^{-1}$. Two other systems in the Galactic field were also 
found to undergo transitions  from  a disk (LMXB) state, characterised by the same sub-luminous 
X-ray level, to a radio MSP (RMSP) state and vice versa in the past few years, although never 
entering  into outburst. 
These MSP binaries, dubbed ``transitional'' MSPs belong to the class of MSP redbacks 
with non-degenerate companions ablated by the powerful pulsar wind.
They display a variety of modes   (flares, low and high modes), now posing more questions
than answers on the complex interplay between mass inflow and ejection of matter in the presence of moderate magnetic
fields (see \cite{2020arXiv201009060P} for a review). An additional unexpected finding is the detection 
of simultaneous X-ray and optical millisecond pulses during the sub-luminous  disk state 
only in the high mode
\cite{2019ApJ...882..104P}, exacerbating the question on how accretion and magnetic field rotation power loss compete. 
As of today there are a handful of candidate systems behaving in a similar manner in the sub-luminous X-ray state, though not yet  caught performing  a transition.

It is, therefore, very important to monitor AMXPs entering into outburst and uncover their evolution including reflares, found to occur just after the outburst decay \cite{2016ApJ...817..100P}. The timescale on which the disk instability occurs (days-weeks), the spectral changes
along the outburst, and the conditions for jet launching are crucial to understand the physical processes at the magnetospheric radius of these fast spinning NSs. In this respect the IRT instrument will be of great importance to obtain simultaneous NIR coverage of the outburst evolution.

In addition, it is essential to  catch as early as possible a transition from/to a RMSP to/from a 
LMXB state (over a few weeks), to track the disk formation/eva\-po\-ration.
Only wide field X-ray instruments such as THESEUS  will be able to monitor outburst evolution and to 
catch state transitions (see i.e. below), allowing follow-up observations with the high sensitive 
instruments on-board  Athena, as well as with ground-based optical and radio telescopes.

\begin{figure}[ht!]
\begin{center}
{\includegraphics[width=6cm,angle=-90]{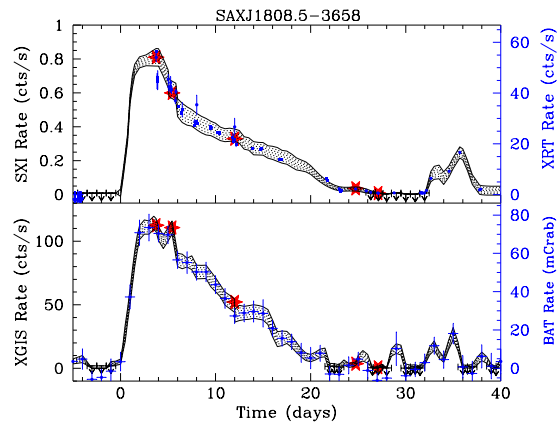}}
\caption{Simulated XGIS (2-30 keV) (bottom) and SXI (0.3-6 keV) (top)
light curves (black lines) of the AMXP SAX\,J1808.4-3658 as observed during
the outburst in 2015. The light curves were derived from interpolation among
5 epochs spectral fits (red stars) as derived from  Swift.
XRT and BAT light curves are also overimposed. The grey areas
cover 3$\sigma$ count rate  errors. SXI and XGIS upper limits are reported in black. THESEUS coverage, when long up to 20-25 days, will allow full or part of the outburst, including reflares at the end of the outburst, to be observed.}
\label{fig:saxj1808_outburst_L}
\end{center}
\end{figure}

\begin{figure}[ht!]
\begin{center}
{\includegraphics[width=8cm,angle=0]{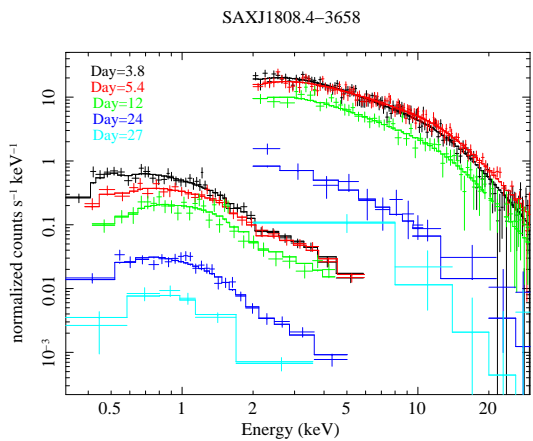}}
\caption{ Simulated
combined SXI and XGIS spectra of SAX\,J1808.4-3658 for the 5 epochs shown in Fig.~\ref{fig:saxj1808_outburst_L} adopting an absorbed power law
with parameters as derived from Swift/XRT spectra. Exposure times were 2 ks for the
first three epochs for both SXI and XGIS, while for the 4$\rm^{th}$ and 5$\rm^{th}$ epoch exposure times of 30\,ks and 100\,ks
were assumed, respectively}
\label{fig:saxj1808_outburst_R}
\end{center}
\end{figure}

Among hundreds of known MSP binaries, both in the Galactic field and globular 
clusters, about a hundred reside in close orbits (P$_{\rm orb} <$ 1 d). Of these about 65 are 
detected in  the X-rays consisting of the aforementioned  AMXPs, 23 redbacks and  21 systems with degenerate donors, dubbed ``black widows". Using the recent Swift/XRT point source catalogue \cite{2020ApJS..247...54E}, we have checked the observability of these objects with THESEUS.
We have found that given the large range (more than 5 order of magnitudes) of X-ray luminosities, THESEUS will be able to detect at least 13 AMXPs entering into outburst with both SXI and XGIS. 
An example of the evolution of the outburst observed in 2015 from the AMXP SAX\,J1808.4-3658, and simulated for both THESEUS instruments is shown in Fig.~\ref{fig:saxj1808_outburst_L}, based on Swift  XRT and BAT observations.
As most AMXPs are located in the Galactic bulge, THESEUS   will provide   observing windows lasting up to  tens of days and thus enable full or partial outburst evolution coverage with integration times ranging from 2 ks (peak of outburst) to 30 ks (decay) and 100 ks (return to quiescence). 
Simulated combined SXI and XGIS spectra at five different epochs are shown in  Fig.~\ref{fig:saxj1808_outburst_R}, adopting an absorbed power law with parameters as derived from  Swift. The recovery of power law index
and column density is at 2-4$\%$ level in the first three epochs and at 7$\%$ and 22$\%$ in the two later epochs, respectively.
The unique X-ray monitoring will be essential to trigger dedicated follow-up observations with  Athena and other facilities to study in details
the X-ray spin \cite{2020ApJ...898...38B} and spectral \cite{2019MNRAS.483..767D} evolution.

\begin{figure}[h]
\begin{center}
\includegraphics[width=\columnwidth,angle=0]{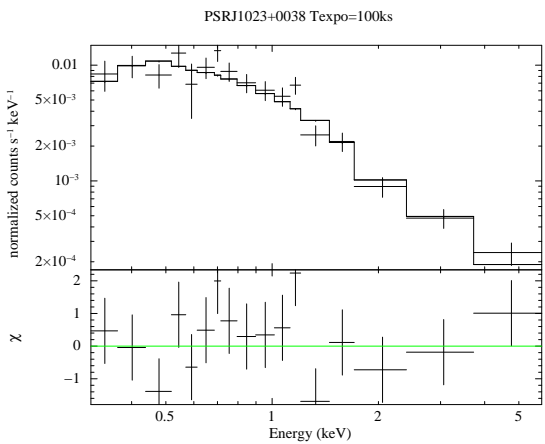}
\caption{Simulated SXI  spectrum for PSR\,J1023+0038 in its
current subluminous LMXB state adopting a 100 ks exposure and an absorbed
power law with $\rm N_{H}=3\times10^{20}\,cm^{-2}$ and $\Gamma$=1.7.}
\label{fig:psrj1023_sxi}
\end{center}
\end{figure}

While  the AMXPs  in outbursts will be at easy reach of both SXI and XGIS, about 6-8 of the
known MSP redback  binaries will be detected in 100 ks by the SXI, when either  entering/leaving or lingering in the LMXB sub-luminous state. 
On the other hand, no RMSP system, either redback or black widow, will be detected in their faint rotation-powered state. 
Fig. \ref{fig:psrj1023_sxi}  shows a simulated spectrum for the transitional MSP binary PSR\,J1023+0038 during its current sub-luminous LMXB state, adopting a 100 ks exposure and an absorbed power law, whose index can be recovered at 15\% level.
Therefore THESEUS will allow us to uncover state changes over a few weeks as well as keep track of the sub-luminous LMXB state of MSPs, known to last up to several years.

\subsubsection{X-ray bursts and super-bursts}

Prolonged accretion of matter on the surface of a NS leads to a steady accumulation of nuclear fuel. Due to  the steep dependence of the typical nuclear reaction rates on temperature and density, this can lead to unstable nuclear burning that for few tens of seconds outshines the steady X-ray emission powered by  accretion. These type-I bursts provide a unique access to the study of  nuclear burning, which is otherwise deeply hidden in the cores of normal stars.
Moreover, analysis of the spectral evolution during the bursts provides an opportunity to probe fundamental parameters of the NS, and ultimately the equation of state of dense matter \cite{1990A&A...237..103D,2011A&A...527A.139S,2017MNRAS.466..906S}.

Depending on the metallicity of the burning layers, their physical conditions (temperature, density, pressure) and the mass accretion rate, bursts show up with different properties (see \cite{2017arXiv171206227G} for a review). Given that at the moment we know about 115   LMXB bursting sources, the continuous sky survey of THESEUS will catch hundreds of bursting events, allowing for an unprecedented census of the bursting population of NS LMXBs.

To estimate the burst detection rate in the THESEUS survey, we considered first the bursting rate of each known LMXB using the estimates from the MINBAR archive \cite{2020ApJS..249...32G} and multiplied it for the respective expected daily-averaged source exposure times with the SXI, XGIS (fully-coded) and XGIS (half-coded). This computation resulted in a (possible) burst detection rate, on average, of 5.7, 0.5, and 12.1  bursts/day, respectively. Secondly, we took into account the sensitivity of the THESEUS instruments and the peak luminosity derived from the whole sample of detected bursts (Fig.~23 in \cite{2020ApJS..249...32G}). For the SXI (sensitivity of $\sim$\,1 Crab in one second), with about 50\% of burst peaks above 1~Crab, the detection rate decreases to about 2.9~bursts/day (although binning in a longer time window might allow to increase the signal-to-noise ratio and raise this estimate). Given the higher XGIS sensitivity, more than 90\% of the possible burst peaks can be detected at more than 5$\sigma$ (fully coded). The decrease in the effective area for the half-coded is well compensated by the larger FOV and the detection rate (conservatively) is foreseen $\geq$\,6 bursts/day.   

\begin{figure}[!ht]
\centering
\includegraphics[width=\linewidth]{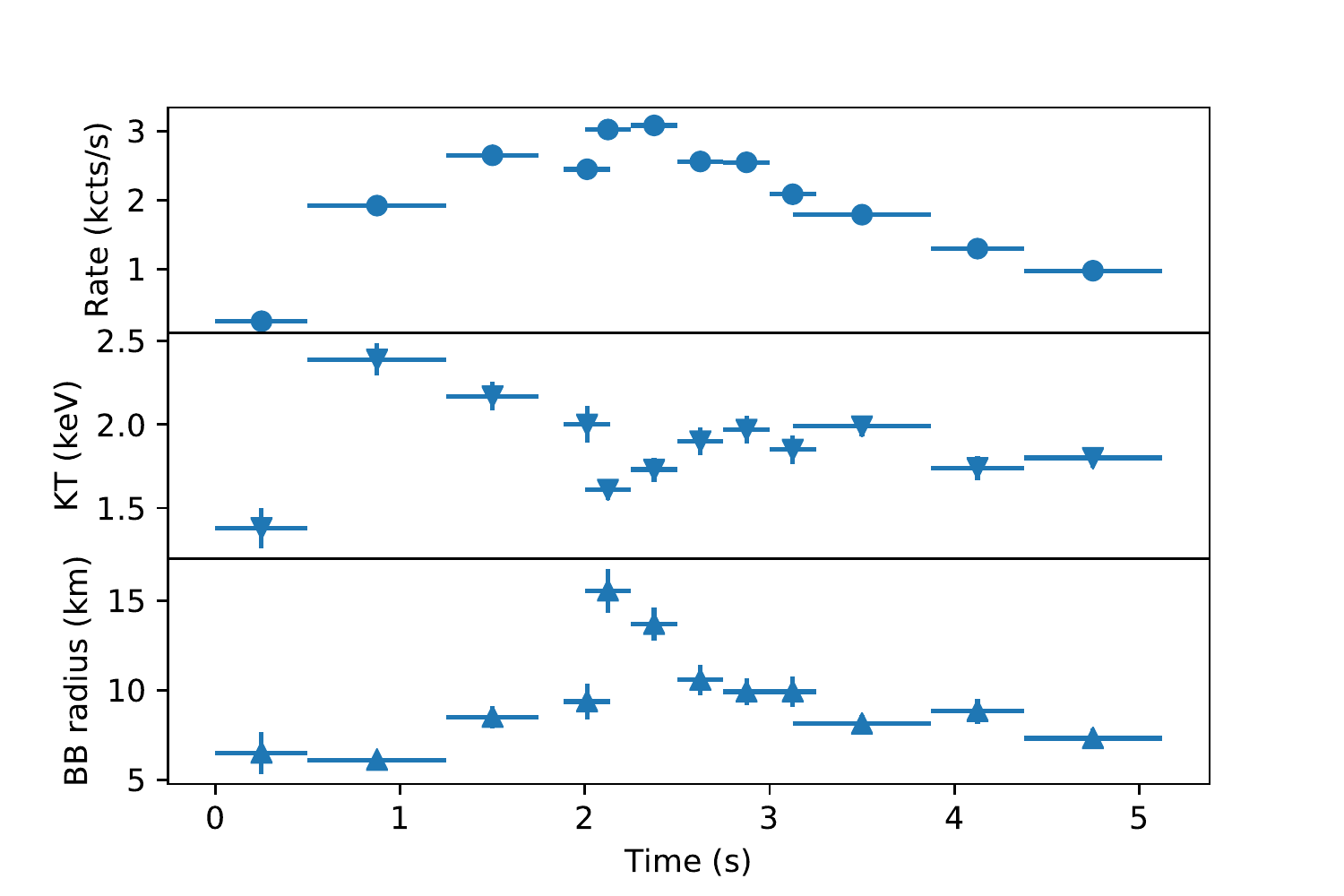}
\caption{Simulated on-axis XGIS light curve (upper panel) and 
spectral parameter evolution for a PRE burst (blackbody temperature in keV and black-body 
radius for an assumed 5 kpc distance in the middle and lower panels, respectively). Reference for this partcular PRE burst 
template in \cite{2000A&A...356L..45K}.} \label{fig:preburst}
\end{figure}

As a result, we estimate that more than 730 bursts will be detected in the first four years of the survey with XGIS (fully-coded) and about 150 of these will show photospheric-radius expansion (PRE) episodes. PRE bursts are of special interest, as they are Eddington limited, and so are considered as possible standard candles to determine source distances \cite{2003A&A...399..663K}. To demonstrate the feasibility of this kind of studies, we took as template a PRE burst from GX 3+1, a typical LMXB in the Galactic bulge, observed with RXTE \cite{2000A&A...356L..45K}. The time-resolved burst spectra evolve as a cooling black-body, but $\sim$\,2 s after the burst onset, a drop in the black-body temperature, together with a  significant increase in the apparent black-body radius clearly indicates a PRE. We simulated the same time-resolved spectra using the  XGIS on-axis response. The typical touch-down at time 2 s can also be easily detected in our simulations, as shown in Fig. \ref{fig:preburst}.

In addition to the typical short duration bursts, much rarer events called super-burst  \cite{2017symm.conf..121I} deserve special attention. They are ignited by unstable carbon burning and can last for hours; their structure and energetics strongly depend on the structure of the outer layers of the NS, the quantity of steady nuclear burning that precede the superburst event, and the long-term accretion history of the source, so their study can constrain different aspects of the accretion process at once.  
Up to now, only 25 superburst events from 15 different LMXBs have been detected, with only three sources showing a superburst more than once. Although the source list is still short, it contains a more significant fraction of persistent sources (typically accreting above 0.1 $L_{\rm Edd}$) with respect to transients (4 vs. 11). Furthermore, among the seven sources with known orbital period, two are super-compact systems. Some superburst sources are also copious emitters of short X-ray bursts, but a superburst event can dramatically quench the burst rate \cite{2002A&A...382..503K}. Given the unpredictable rate and origin of the superbursts, they could only be spotted through all-sky monitors (e.g. RXTE/ASM, BeppoSAX/WFC or MAXI), which, however, lacked timing and spectral resolution that THESEUS will provide.  Based on the number of detected superburst events with past all-sky X-ray monitors and the location of the sources where at least one superburst has been observed \cite{2017symm.conf..121I}, we estimated the chance occurrence of observing one such event with THESEUS in the first four years of operations. Assuming a superburst rate of one per year and the visibility of the sky containing superbursting sources, we derived a chance probability of $\sim$\,30\% of observing such an event.

\begin{figure}[!ht]
\centering
\includegraphics[width=\linewidth]{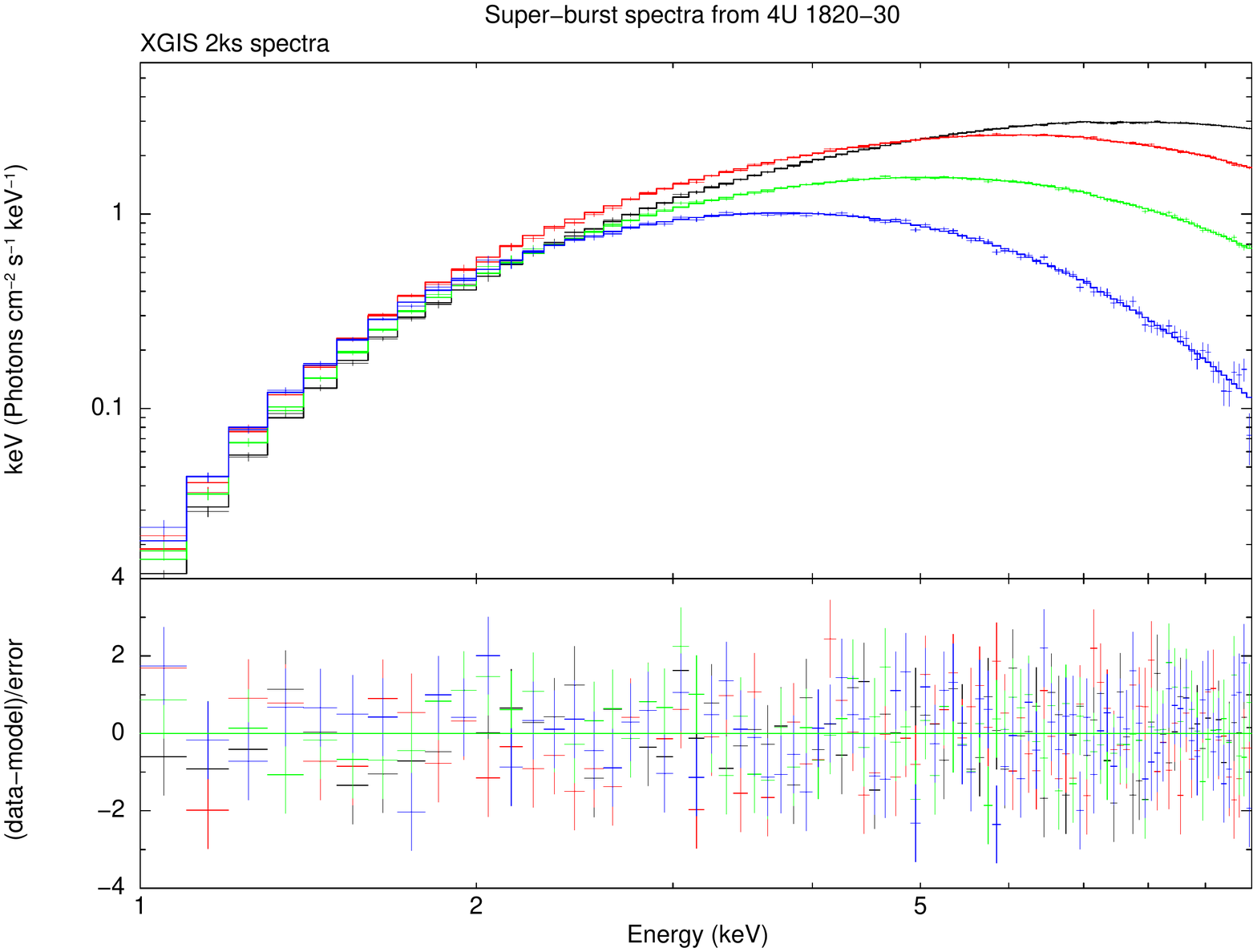}
\caption{Simulated XGIS time-resolved spectra of the cooling evolution of a superburst event. 
Each spectrum has an assumed exposure of 2 ks.}\label{fig:1820_superburst}
\end{figure}

To assess the capabilities of THESEUS to conduct spectral analysis for such events, we considered the properties of one of the superbursts as observed by the RXTE \cite{2002ApJ...566.1045S}. Taking average values for temperature and flux along steps of 2 ks each, we evaluated that the XGIS would be able to tightly constrain     evolution of temperature and 2--10 keV flux with (statistical) uncertainties of less than 0.4\%.  Simulated spectra with over-imposed best-fitting models are shown in Fig. \ref{fig:1820_superburst}.

\subsubsection{Monitoring of the light curve of LMXBs at high inclination angles}

The LMXBs at high inclination angles ($>60^\circ$) show  dips and/or eclipses in their light curves. The dip (or eclipse) arrival times allow us to infer  accurate orbital parameters \cite{2014RAA....14.1367C} and to estimate the orbital period decay rate. This permits to better understand the origin and details of mass-transfer, which is important for our understanding of evolution of these binaries. For instance, the study of the eclipse arrival times of the accretion disk corona source X1822$-$371   showed that its mass transfer is likely super-Eddington and non-conservative, with only a 14-22\% of the transferred mass from the companion accreted onto the NS \cite{2010A&A...515A..44B,2015A&A...577A..63I}. The eclipse arrival times of the transient eclipsing source MXB~1659$-$298 showed that the large orbital period derivative   implies a highly non-conservative mass transfer scenario, where more than 98\% of the mass provided by the companion star leaves the binary system  \cite{2018MNRAS.473.3490I,2017MNRAS.468L.118J}. The dip arrival times of  XB\,1916$-$053 also strongly support non-conservative mass transfer \cite{2015A&A...577A..63I,2001ApJ...549.1135C}. The large FOV of SXI and XGIS will allow to monitor the light curves of the LMXBs with high inclination, thereby providing an opportunity to determine arrival times for new dip events and eclipses, and to improve constrains on orbital period decay rate for other dipping sources such as 4U\,1323$-$619  \cite{2016A&A...589A..34G} to understand their mass-transfer scenario. Finally, residuals appearing in the orbital solutions for  XB\,1916$-$053 and MXB\,1659$-$298, suggest that these systems are hierarchical triple systems with a third body with a mass similarto Hot Jupiter extra-solar planets (see \cite{2018ApJ...859...40I} and ref. therein). This is a new and exciting area in the field of LMXB studies where THESEUS can also contribute by providing improved orbital solutions for more systems.

\subsubsection{Very Faint X-ray Transients}

The nature of the Very faint X-ray transients (VFXTs  \cite{Wijnands2006}), events with peak luminosities $<10^{36}$ erg s$^{-1}$  is still unknown. Although some have been associated to NS or BH low-mass X-ray binaries or symbiotic systems, the nature of most VFXTs is still uncertain \cite{2021MNRAS.501.2790B}. Current data suggest that they have very low mass-transfer rates, pointing towards the later (or unusual) stages in X-ray binary evolution \cite{2009A&A...495..547D,2015MNRAS.447.3034H}.  Furthermore, the family of VFXTs is not even homogeneous: some of them are quasi-persistent, i.e. showing a long-term outburst activity with low luminosities, while other sources have spent most of the time since their discovery in quiescence, showing up only during type-I X-ray bursts which are a signature of ongoing   accretion of matter (the so-called ``burst-only” systems \cite{Cornelisse2002}).

Thus, VFXTs allow to gather information that is not generally obtained with ``normal'', brighter, systems. THESEUS will detect many VFXTs and enable follow-up observations with Athena and  ground-based observatories at different wavelengths (e.g. \cite{2020MNRAS.492.4344S}). It will help to investigate the repeating behavior observed in several VFXTs, which will ultimately allow us to obtain a better understanding of their mass-transfer rates. The unique capacity of obtaining simultaneous X-ray and IR observations will also shed light on the donor stars.


\section{Black Hole   X-ray Binaries}
\label{sec-BH}

Black hole X-ray binaries (BHBs) are binary systems where a black hole accretes material from a  companion star.
While a dozen BHBs are persistent, most of them are transient sources (called black hole transients, BHTs), i.e. they alternate long periods in a low-luminosity quiescent state, with  luminosity as faint as $\sim$10$^{30}$ \lum, with episodic outbursts reaching peak luminosities in the $\sim$10$^{36-39}$ erg s$^{-1}$. 
BHTs show  X-ray spectral states characterized by  different spectral shapes (hard or soft and everything in between, \cite{zdziarski04,remillard06}) and timing properties \cite{belloni16} along their outbursts.
The different spectral states are explained in terms of changes in the geometry of the accretion flow around the central object \cite{zdziarski00,done07}. 
During the  hard state, BHBs show a characteristic energy spectrum dominated by the hard X-ray component physically interpreted as the result of the Compton up-scattering of soft,
thermal disk photons by a hot electron plasma ($\sim$100\,keV), called ``corona", located close to the BH \cite{zdziarski04}.
A weak soft component with an inner disk blackbody temperature k$T_{\rm in} \sim$0.1--0.3\,keV, emitted by an optically thick and geometrically thin accretion disk \cite{shakura73}, possibly truncated at a large radius from 
the innermost stable circular orbit, is also observed. During the soft state, the spectra show a strong soft disk component with inner disk temperature k$T_{\rm in}\sim$1\,keV. 
In addition, a steep power law tail ($\Gamma >$2.5), which often extends to  high energies ($\sim$500\,keV), is observed. Intermediate spectral states with parameters of the accretion flow in between the two main states are also observed.   
Disentangling the main components that contribute to the overall X-ray energy budget and follow the spectral evolution as a function of the accretion flow will be carried out by THESEUS thanks to its wide field of view combined with the broad band energy coverage.

While in Survey Mode, THESEUS will be able to discover new  BHTs in the Galaxy. Up to now, 66 BHTs have been catalogued\footnote{http://www.astro.puc.cl/BlackCAT/index.php} \cite{corral16}.
Taking advantage of the available X-ray data archives, Tetarenko et al. \cite{tetarenko16}   detected 132 outbursts from 47 BHTs in the period 1996-2015
and estimated an outburst rate as 7 and 8 per year for Swift/BAT and INTEGRAL/IBIS, respectively.
In the last twenty years, we have estimated that a mean value of 2 new objects per year have been discovered by  wide FOV X-ray telescopes.
Moreover, the majority of the BHT outbursts last more than 100 days, so that we are quite confident that XGIS will discover at least two new sources per year.
This will allow us to trigger simultaneous observations with the high spectral resolution X-ray telescopes (e.g., Athena), as well as radio and gamma-ray high sensitivity
telescopes (SKA and CTA) which will be complementary to THESEUS for a comprehensive accretion/ejection study of these sources.  

THESEUS offers the unique capacity to perform strictly simultaneous   X-ray and NIR observations. 
During outburst, IR fluxes are known to trace the X-ray emission \cite{coriat09}, and are thought to be an indicator of the strength of jet activity in BHTs \cite{casella10,gandhi11,corbel13,baglio18}.
Thus, simultaneous observations in these bands will lead to advancements in our understanding of the jet-disk coupling in BHBs.
Simultaneous X-ray and NIR observations with THESEUS will allow us  to monitor the evolution of X-ray properties in conjunction with low  ionisation (cold) disk winds that can be observed in the   NIR during 
large parts of an outburst \cite{2020A&A...640L...3S}.
The IRT will provide also important complementary constraints on the identification and nature of BHTs. Known BHTs are located at distances of several kpc, often at low Galactic latitudes suffering from high levels of dust extinction \cite{gandhi19}. IRT will allow BHT counterpart identification through higher levels of extinction than possible in the optical. Rapid identification, simultaneous with X-rays, will allow quick source characterisation. During quiescence, IR observations could help to constrain the spectral type of the donor \cite{khargharia13}. Only about 50\,\% of BHTs have measured quiescent fluxes \cite{corral16}, mostly in the optical with a median $G$ (Vega) mag of approximately 19.5 \cite{gandhi19}. In the IR, these sources are expected to be $\sim$\,1--2\,mag brighter, well within reach of the IRT.
Thus, IRT can enable complete IR sampling of the quiescent properties of the brighter end of the BHT population.

\section{Magnetars}
\label{sec-magnetars}

Magnetars are (isolated) neutron stars with extremely high magnetic fields, up to 10$^{15}$ G, and possibly higher in their interior \cite{2015SSRv..191..315M,2017ARA&A..55..261K}. They  are characterized by extreme variability phenomena on all timescales,  from short bursts of hard X-rays of few milliseconds to outburst lasting months or years. Although it is now well established that their persistent and bursting emission is powered by magnetic energy, many aspects of their rich multiwavelength phenomenology are still to be explained, both concerning the physical processes responsible for the bursts, giant flares, and outbursts, as well as for their evolutionary connection with other classes of neutron stars.  
 Currently, less than 30 confirmed magnetars are known in our Galaxy and in the Magellanic Clouds \cite{2014yCat..22120006O}, but this number will at least double before the launch of THESEUS. In fact, most magnetars are transient sources, difficult to identify during their long quiescent periods    at low luminosity, but easily detected when they become active. When X- and $\gamma$-ray instruments with good sky coverage are available,  about one new magnetar per year is discovered. 
 
Magnetars are also interesting because they have   been invoked in models for a variety of other sources, such as, e.g., GRBs (giant flares in nearby galaxies mimic short GRBs; extended energy injection from newly born magnetars), superluminous supernovae, and fast radio burst.  Finally, magnetars are relevant as sources of neutrinos, high-energy cosmic rays and gravitational waves.

\begin{figure}
\center
\includegraphics[width=8cm]{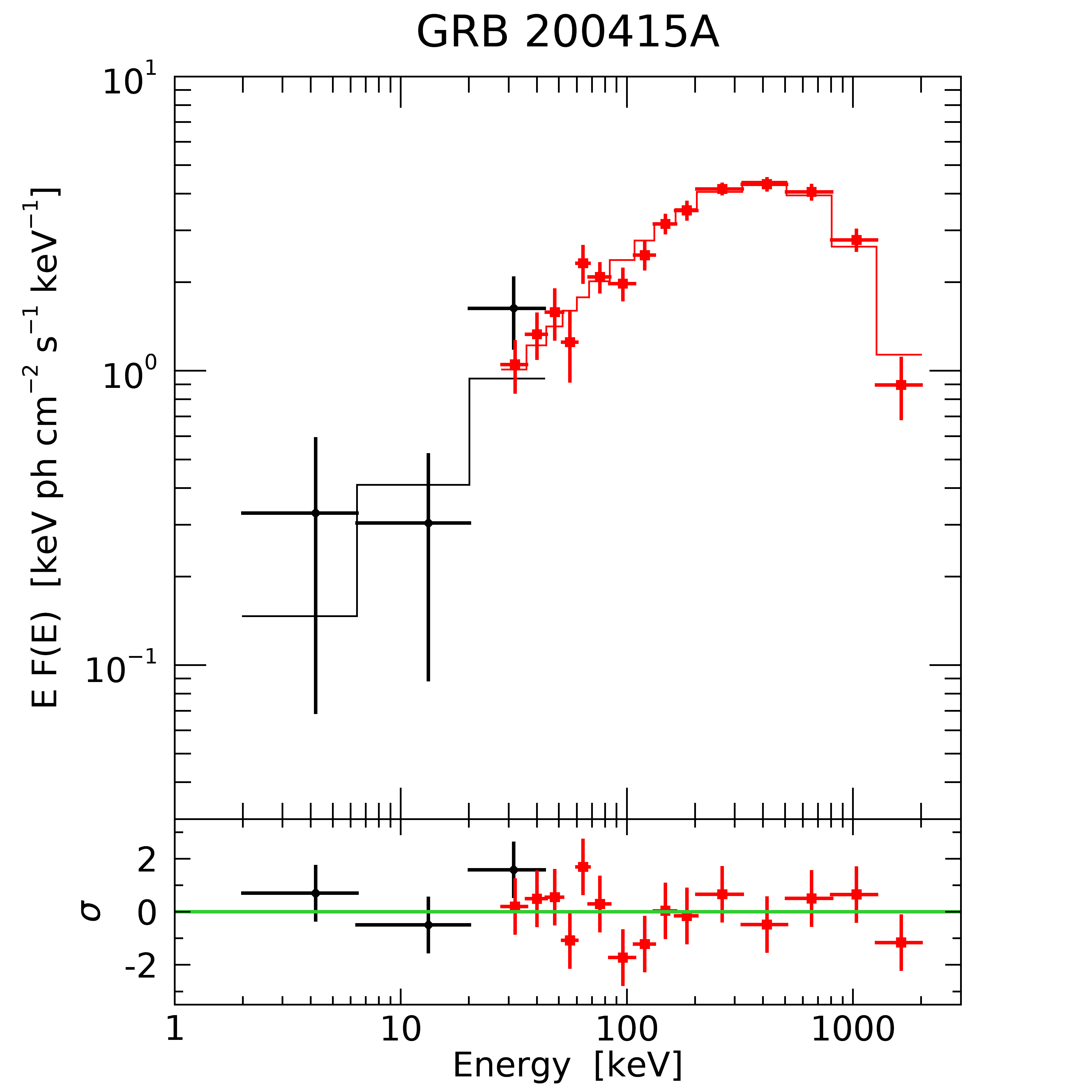}
  \caption{XGIS-X (black) and XGIS-S (red) simulated spectra of the short GRB 200415A, possibly due to a magnetar located in the Sculptor galaxy (NGC 253).}  
  \label{fig-magnetar1}
\end{figure}

With its frequent and sensitive coverage, THESEUS will provide robust constraints on the duty cycle of activity of magnetars, for example allowing the investigation of possible periodicities (relevant to their connection with FRBs, see Sect. \ref{sec-FRB}). THESEUS will give immediate alerts of outbursts from both known and newly discovered magnetars, through the XGIS detection of bursts and/or the SXI detection of the enhanced thermal emission that occurs when magnetars activate. Quick repointing with IRT is a unique feature useful to identify and study the counterparts, taking advantage of the small absorption in the NIR. Broad energy band coverage, rarely available up to now, is crucial to clarify the spectral shape of bursts and flares at low energies.

Magnetars are also able to emit Giant Flares,  reaching peak luminosities up to 10$^{47}$ erg s$^{-1}$. THESEUS will have the sensitivity to detect Giant Flares from magnetars in external galaxies, up to distances of  several tens of  Mpc. This is relevant because the extremely bright short ($<$0.2 s), and spectrally hard initial spikes of Giant Flares appear similar to short GRBs.  A few short GRBs have been proposed to be actually magnetar giant flares (e.g. GRB051103 in the M81 group at 3.6 Mpc \cite{2007AstL...33...19F}, GRB070201 in M31 at 770 kpc \cite{2008ApJ...681.1464O,2008ApJ...680..545M}, GRB200415A  in NGC 253 at 3.5 Mpc \cite{2021Natur.589..211S,2021Natur.589..207R} and GRB 070222 in M83 at 4.5 Mpc \cite{2021ApJ...907L..28B}). 
The simulation shown in Fig.~\ref{fig-magnetar1} shows how THESEUS would have detected the short   GRB200415A, possibly located in the Sculptor galaxy (NGC 253) at a distance of 3.5 Mpc.  More than 10,000 counts are detected above 30 keV with the XGIS-S, and the burst is visible also in the XGIS-X ($\sim$450 counts). Extrapolating this spectrum to the SXI energy range, we find that the initial spike is too faint to be detected ($\sim$0.1 cts s$^{-1}$, taking into account the Galactic absorption of $3\times10^{20}$ cm$^{-2}$ in this direction).  However, if this GRB were actually a Giant Flare, it should also have a long lasting tail of softer emission, that will be detectable in the SXI. In fact, rescaling the properties of the December 2004 giant flare of SGR 1806-20, we predict that a few tens of counts should be detected by the SXI.  A similar giant flare from a magnetar in M31 (as suggested for GRB070201) would yield a few hundreds of SXI counts, allowing to confirm the magnetar hypothesis and possibly  discover  spin period, as it happened for the famous giant flare of March 1979 in the LMC \cite{1979Natur.282..587M}.

\section{Ultra-luminous X-ray sources}
\label{sec-ULX}

Ultra-luminous X-ray sources (ULXs) are off-nuclear point sources that exhibit X-ray luminosities, $L_{\rm X}\gtrsim 10^{39}$ \lum\, exceeding the Eddington limit for a $10\,\msun$ black hole. For a long time, ULXs have been supposed to contain accreting black holes (BHs) of intermediate mass ($10^2$--$10^5\,\msun$). These systems are now widely believed to be stellar mass compact objects accreting above the Eddington limit, as initially proposed by \cite{2006MNRAS.368..397S,2009MNRAS.397.1836G,2012MNRAS.420.1107P} and further confirmed by the first results with NuSTAR \cite{2013ApJ...778..163B,2013ApJ...779..148W}. More recently, six extragalactic ULXs \cite{2014Natur.514..202B,2017Sci...355..817I,2016ApJ...831L..14F,2018MNRAS.476L..45C,2019MNRAS.488L..35S,2020ApJ...895...60R} and one galactic ULX \cite{2018ApJ...863....9W} have been shown to contain accreting neutron stars (NSs), thanks to the detection of X-ray pulsations at the NS spin period, thus confirming that these objects do undergo super-Eddington accretion.

Early spectral analyses of ULX data were performed using single component models, which led to the interpretation of the accreting compact object in terms of an intermediate mass black hole (IMBH) of $\sim 10^2$--$10^3\,\msun$. Subsequent observations of better spectral quality, showed that either complex single component models or two-component models, such as disk blackbody and cut-off power law or a hot thermal component, were required to adequately fit the 0.3--10 keV ULX spectra. Depending on the spectral state of the source, the spectrum can be dominated by   the flux of either  the soft   or       the hard component, leading to the identification of the spectrum with soft or hard ultraluminous states, respectively (see, e.g., \cite{2017ARA&A..55..303K} for a recent review).

\begin{figure}
\center
\includegraphics[width=8cm]{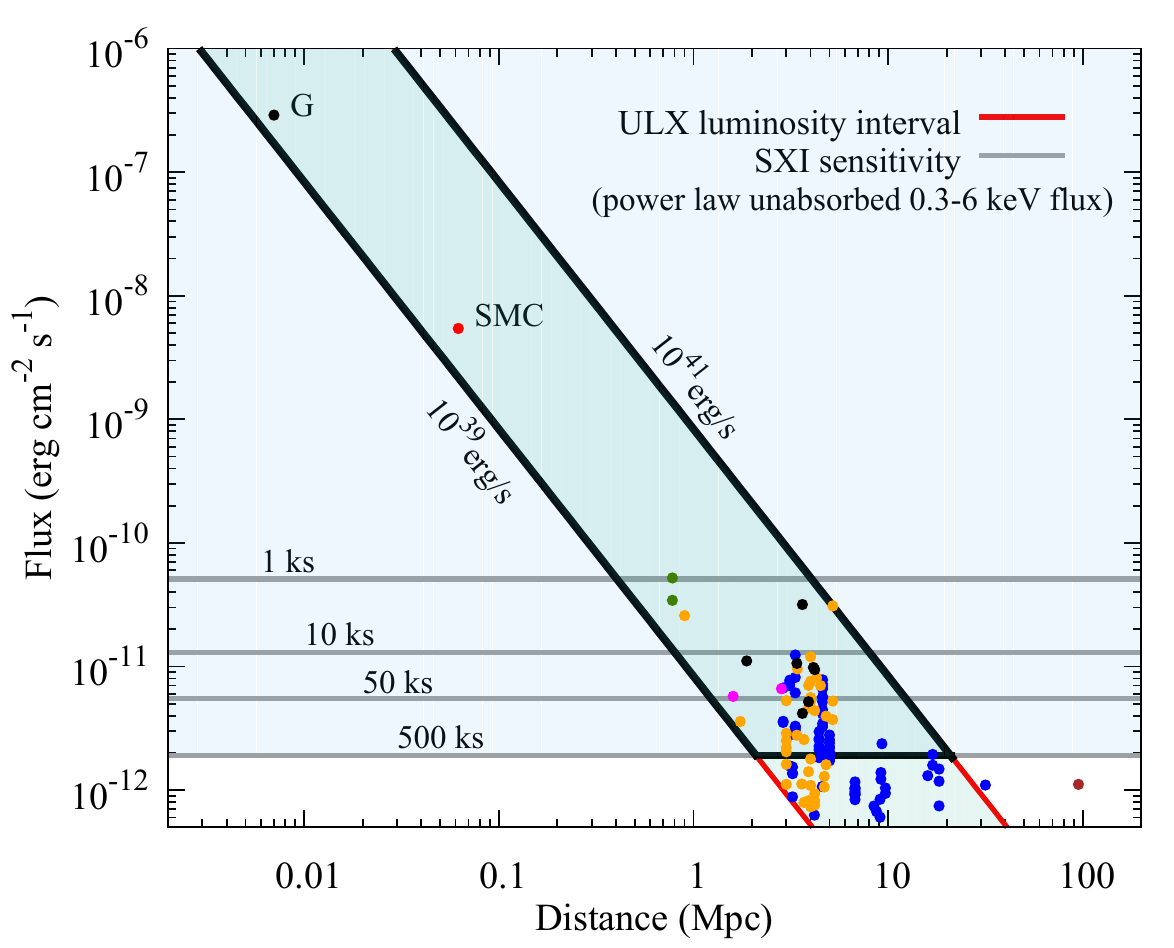}
  \caption{Estimate of the region in the flux vs. distance plane where SXI can detect the potential sources within the standard ULX luminosity interval ($10^{39}$--$10^{41}$ \lum). The SXI sensitivity (unabsorbed 0.3--6 keV flux) delineated by the horizontal gray lines labeled with minimum exposure times intersect with the ULX luminosity interval determined by the oblique red lines to define the region for the potential ULX targets THESEUS can monitor. The shaded green area bounded by the solid black curve displays such a sample region where ULXs with standard luminosities can be detected and monitored by SXI for an integration time of $\leq 500$ ks. Filled circles with different colors represent the observed ULXs from different catalogs/references:   \cite{2019MNRAS.483.5554E} blue; \cite{2013ApJS..206...14G} orange;  \cite{2018ApJ...863....9W,2014Natur.514..202B,2016ApJ...831L..14F,2018MNRAS.476L..45C,2019MNRAS.488L..35S} black; \cite{2012ATel.3921....1H,2012A&A...538A..49K} green; \cite{2017A&A...605A..39T} red; \cite{2017A&A...608A..47K} magenta; \cite{2009Natur.460...73F} brown. As the spectral energy range for each source group does not perfectly match the energy range of SXI and the spectral model for each source may differ from a pure power law, the source distribution is an approximation to the actual population of ULXs SXI can detect.
The  two  nearest ULXs are marked by G (Galactic) and SMC (Small Magellanic Cloud) labels. 
}
  \label{flxdist}
\end{figure}

Even though the ULX luminosities in the $\sim 10^{39}$--$10^{41}$ \lum$\,$ range can be explained by the combination of geometrical beaming and super-Eddington accretion onto NSs and stellar mass BHs in high-mass X-ray binaries, the high-luminosity tail of the population ($L_{\rm X}>10^{41}$ \lum) indicates the possible existence of IMBHs, which were originally suggested to account for the standard ULX luminosity range ($10^{39}<L_{\rm X}<10^{41}$ \lum). The ULXs with $L_{\rm X}>10^{41}$ \lum$\,$ are also known as hyperluminous X-ray sources (HLXs). The firmly established member of this family is ESO 243--49 HLX-1, which is thought to be an accreting IMBH candidate. Its observed peak unabsorbed 0.2--10 keV  luminosity   is $1.2\times 10^{42}$ \lum$\,$ \cite{2009Natur.460...73F}. Given the distance of 95 Mpc, the maximum X-ray flux of the source is $\sim 10^{-12}$ \flux. The source luminosity was observed to change from $2\times 10^{40}$ \lum$\,$ (low state) to $\sim 10^{42}$ \lum$\,$ (high state). Another HLX candidate, M82 X-1, with a luminosity level close to the lower limit of HLX luminosity range is only $\sim 3.6$ Mpc away.

\subsection{Monitoring nearby extragalactic ULXs with SXI}
\label{mon_ulx}

Detecting and monitoring the known ULXs with standard luminosities in the $10^{39}$--$10^{41}$ \lum $\,$ range with  SXI is crucial for the follow-up observations of the long-term flux variations in these sources. In Fig.~\ref{flxdist}, we illustrate a sample region in the unabsorbed flux versus distance plane where all the potential targets with standard ULX luminosities can be detected and monitored by SXI above a certain sensitivity threshold,  based on a power-law spectrum of photon index 2 and $N_{\rm H}=2\times 10^{21}\,{\rm cm^{-2}}$.

\begin{figure}
\center
\includegraphics[width=8cm]{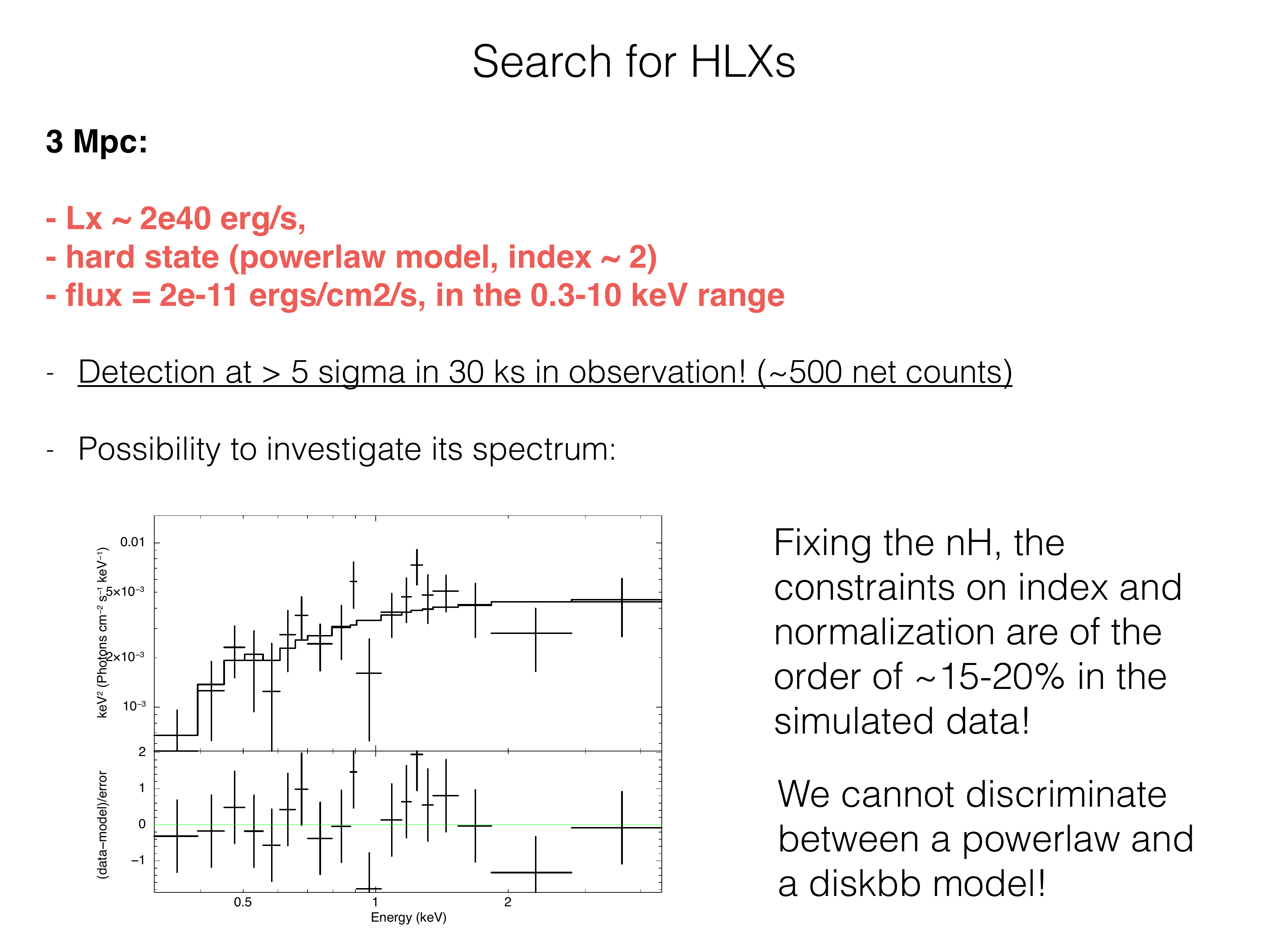}
  \caption{THESEUS-SXI simulation of a nearby HLX in the low/hard state for an exposure time of 30 ks based on the observational properties of ESO 243--49 HLX--1  from \cite{2012ApJ...752...34G}.}
  \label{hlx_lh}
\end{figure}

We also display in Fig.~\ref{flxdist} the distribution of observed ULXs in the local universe (within d $\sim100$ Mpc). 
Most ULXs except the nearest ones, i.e. the Galactic transient source Swift J0243.6+6124 (see Sect.~\ref{sub-cycl})  and the transient SMC X-3 in the Small Magellanic Cloud, are clustered in a distance range of 1--10 Mpc. According to the luminosity intervals shown in Fig.~\ref{flxdist}, the standard ULXs, in particular with distances $\lesssim 5$ Mpc, are the best SXI targets for exposure times $\lesssim 500$ ks. As a standard ULX, we consider NGC 1313 X-1 with an average spectrum based on an absorbed {\sc diskbb+diskpbb+cutoffpl} (in Xspec), which consists of two thermal blackbody components and an exponentially cutoff power-law component \cite{2020MNRAS.494.6012W}. The average flux of $\sim 4\times 10^{-12}$ \flux $\,$ implies a luminosity of $\sim 8\times 10^{39}$ \lum$\,$ for a source distance of $\sim 4$ Mpc. In 30 ks, we can detect the source at $5\sigma$ significance with $\sim $ 100--200 net source counts. The monitoring with SXI is therefore viable for the known standard ULXs that lie within a distance of $\sim 5$ Mpc. THESEUS would indeed provide us with the opportunity of following the source flux evolution of the nearby ULXs with daily exposures in an unprecedented way.

\begin{figure}
\center
\includegraphics[width=8cm]{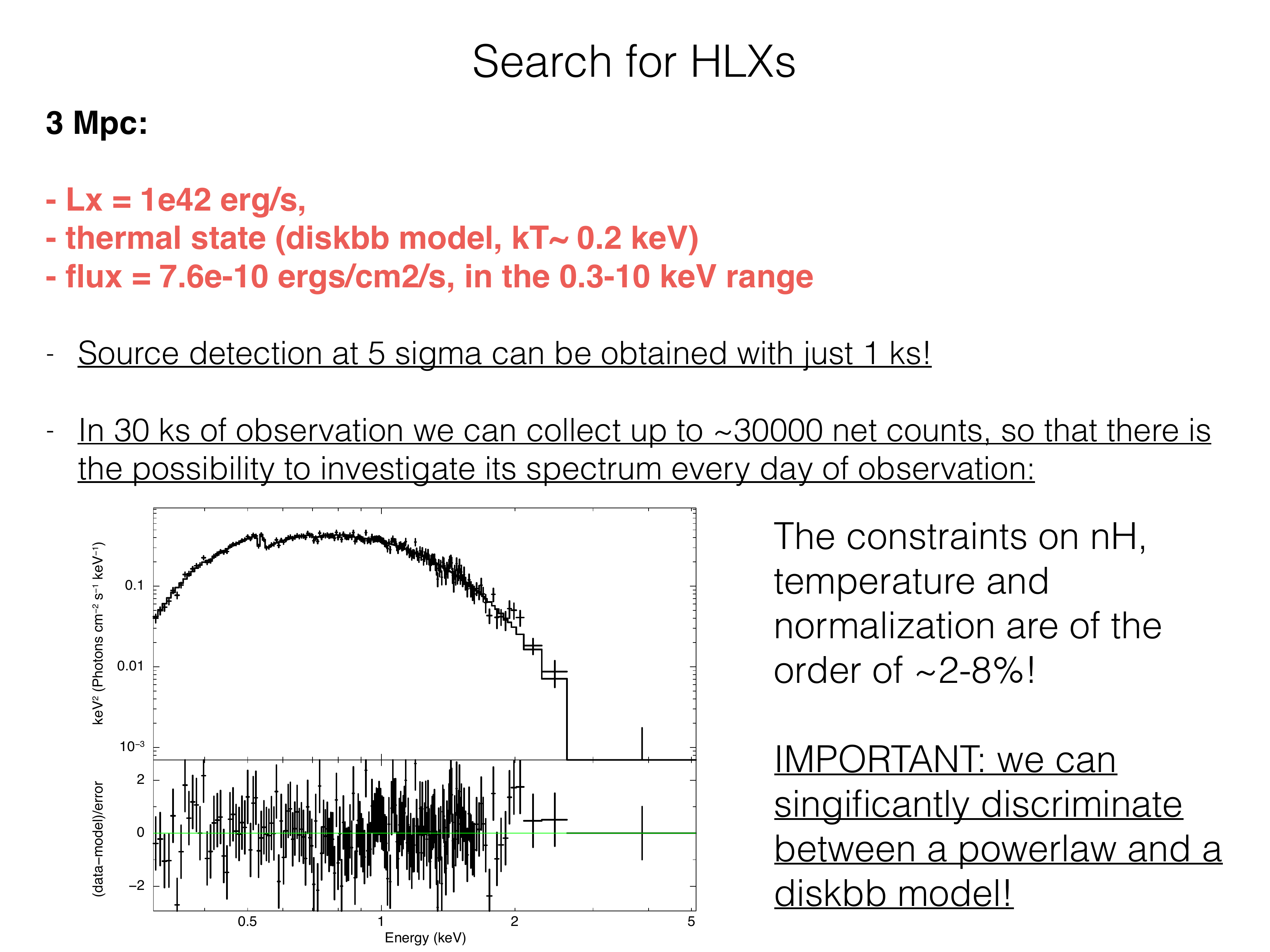}
  \caption{THESEUS-SXI simulation of a nearby HLX in the high/soft state for an exposure time of 30 ks based on the observational properties of ESO 243--49 HLX--1 from \cite{2012ApJ...752...34G}.}
  \label{hlx_hs}
\end{figure}

There are two or even three (if M82 X-1 is included) HLX candidates, among which ESO 243--49 HLX--1 is the best known and the well established source, at a distance of $\sim 100$ Mpc with a flux of $\sim 10^{-12}$ \flux, as also shown in Fig.~\ref{flxdist}. We highlight that the search for relatively nearby hyperluminous members of ULXs is also possible with SXI. Using ESO 243--49 HLX--1 as a benchmark for hunting out such HLXs, we simulate sources at two luminosity states, namely low/hard and high/soft states. ESO 243--49 HLX--1 was observed at both $2\times 10^{40}$ \lum$\,$ with a power-law spectrum of index $\sim 2$ and $10^{42}$ \lum$\,$ with a disk blackbody of temperature $\sim 0.2$ keV \cite{2012ApJ...752...34G}. In Figures~\ref{hlx_lh} and \ref{hlx_hs}, we display the simulated spectra of a nearby HLX at a reference distance of 3 Mpc when the source is in the low/hard and high/soft states, respectively. Even though we cannot discriminate between a power-law and a disk blackbody model in the low/hard state, we can detect the source at $>5\sigma$ significance in 30 ks with $\sim 500$ net counts. In the high/soft state, an exposure time of only 1 ks is sufficient for the detection of the same source at $5\sigma$ significance. In the high state, it is also possible to investigate the source spectrum on a daily basis as one can collect up to $\sim 3\times 10^4$ net counts in an observation of 30 ks. As shown in Fig.~\ref{hlx_hs}, we can significantly discriminate between disk blackbody and power-law models for an exposure time of 30 ks. As might also be deduced from Fig.~\ref{flxdist}, the highly luminous ULXs or HLXs with luminosities close to $10^{40}$ \lum, which are as far away as 20 Mpc, can be detected at $5\sigma$ significance provided we stack several days of observations ($\sim 500$ ks). The detection of an HLX at a distance of 20 Mpc when it transitions to the high state with a luminosity of $\sim 10^{42}$ \lum$\,$ is compatible with that of an HLX at a distance of 3 Mpc when it is in the low state.

\begin{table}
\begin{tabular}{llcc}

\hline 
Host Galaxy & ULX & $d$ (Mpc) & H mag  \\ 
\hline

NGC 4258 & J121844+471730 & 7.2 & 17.79  \\ 
M 101 &  J1402+5440 & 6.5 & 19.3 \\ 
 & J140248+541350 & 6.5 & 17.72 \\ 
 & J140314+541807 & 6.5 & 18.25 \\ 
NGC 4136 & J120922+295559 & 9.7 & 19.2  \\ 
NGC 925 & J022721+333500 & 9.3 & 18.7  \\ 
&  J120922+295559 & 9.3 &19.2 \\
\hline 

\end{tabular} \caption{Sample of ULXs detectable by THESEUS with H band magnitudes brighter than 20.6 mag. Distance values are taken from the  NED Extragalactic Database, magnitudes  from \cite{2014MNRAS.442.1054H} and \cite{2016MNRAS.459..771H}.}
\label{ulx_IRT}
\end{table}

\subsection{Near Infrared Observations of ULXs with IRT}
\label{IR}

One of the open problems is to determine the nature and mass of each binary component in a ULX. Except in cases where it is possible to detect coherent pulsations, and clearly identify the compact object as a NS,  the mass determination for binary components can be an arduous task for ULXs, most of which are extragalactic objects located in crowded fields.

The IR and optical emissions from ULXs  arise  mainly from the donor star and the accretion disk \cite{2008MNRAS.386..543P}. The fraction of emission from either component strongly depends on the evolutionary phase of the donor and that of the binary itself \cite{2018MNRAS.480.4918A}. Detecting and analyzing the IR counterpart would help to shed light on the origin of such emission and also on the accretion state of the system.

Tao et al. \cite{2011ApJ...737...81T} studied the optical emission of nearby ULXs and found that some of them do not point to a given stellar type but can instead be dominated by the reprocessed emission from an extended accretion disk (e.g. the red excess in the SED of Holberg IX-X1).  This was also claimed by the subsequent studies of optical and near IR counterparts \cite{2013ApJS..206...14G,2020MNRAS.497..917L}, which occasionally revealed the presence of red super giants in nearby ($d<10$ Mpc) ULXs. 
L{\'o}pez et al. \cite{2017MNRAS.469..671L} found 15 ULXs with H band magnitudes brighter than 20.6,  which are consistent with the presence of   red super giants (we list some of them in Table~\ref{ulx_IRT}). The ULX counterparts are therefore bright enough to be studied with the IRT.

Among  the galaxies where the ULX optical/IR counterparts are bright enough, NGC 5204, NGC 3034 and NGC 300 are all close to the baseline Survey mode pointing directions of THESEUS. NGC 253 is slightly further from the baseline Survey mode pointing directions, however, making pointed observations will be feasible (see Fig.~14 in \cite{2018AdSpR..62..191A}). ULXs are not bright enough in IR for doing spectroscopy with THESEUS (with the exception of NGC 300 ULX1 if the source is in the bright state and possibly NGC 3034 which has an I band magnitude of 18.5), but the photometry can be done to search for the periodicities (orbital or superorbital). The Visible/IR photometry can be performed in all of the THESEUS bands and will allow us to constrain the nature of the star from which the compact object accretes. As very few of the donors have been identified so far, the visible/IR photometry with THESEUS would be of great value. Once we know the nature of the donor, using the mass can help to constrain the mass of the compact object (black hole or neutron star), especially if we can determine the orbital period of the system through monitoring of the optical counterpart.

\section{Active Galactic Nuclei}
\label{sec-AGN}

Active Galactic Nuclei (AGN) are the signatures of material feeding
supermassive black holes (SMBHs; $10^6 - 10^9 M_{\rm \odot}$) at the
centers of large galaxies.  AGN are the most persistent and
highly-luminous sources ($10^{40}-10^{47}$ erg s$^{-1}$) in the
Universe and their activity is both a tracer for the luminous growth
of SMBHs and for the co-evolution of SMBHs and their host galaxies.
Their copious radiative energy as well as mechanical energy in the
form of outflowing, radiatively- or magnetically-driven winds and collimated,
relativistically-moving jets likely provide a form of ``feedback''
that impacts host galaxy star formation and infalling gas, thus likely
self-regulating both BH  growth and host galaxy evolution (e.g.,
\cite{2007ARA&A..45..117M}). 
Roughly $40\%$
of all SMBHs are accreting at some level (e.g.,
\cite{1997ApJ...487..568H}). 

An accretion disk, with thermal emission peaking in the
optical/UV/EUV, feeds the BH \cite{1989ApJ...346...68S}; 
disk sizes in a few nearby Seyferts have recently been traced via
multi-wavelength reverberation-mapping campaigns
\cite{2019ApJ...870..123E}.   
At scales of a few to $\sim$10 gravitational radii, a high-temperature ($\sim10^9$ K) 
corona emits rapidly variable hard X-rays, though its exact structure is still 
debated, e.g., the base of a jet \cite{2005ApJ...635.1203M,2017MNRAS.471.4436W}.
  %
X-ray reverberation measurements obtained with XMM-Newton and NuSTAR track
the response between the coronal X-ray emission and
that reflected off nearby structures such as the 
accretion disk, providing constraints on corona 
height and structure \cite{2016MNRAS.462..511K}. 
Doppler-broadened optical emission lines indicate clouds of gas near the inner accretion disk; this ``Broad Line Region'' (BLR) may be an
outflow from the disk (e.g., \cite{2011A&A...525L...8C}). 
At scales of parsecs and farther, a dusty, clumpy ``torus'' of gas may act as a reservoir for feeding the disk and BH; recent years
have seen progress in exploring its morphology and extent (e.g., \cite{2012AdAst2012E..17B,2017NatAs...1..679R}). 

Intriguingly, similarities in broadband spectral and timing properties between Seyfert AGN and BH XRBs suggest that both classes of systems host similar accretion processes, scaled by mass, luminosity, and timescale (e.g.,\cite{2003MNRAS.345.1057M,2006Natur.444..730M}). 
For example, the broadband SEDs of low-luminosity AGN are similar to those of BH XRBs in the low/hard state, and those of Seyferts are similar to those of BH XRBs in the high/soft states in terms of the dominance of the disk's thermal emission and the presence/lack of radio jets \cite{2005A&A...435..521N}. 
The conventional wisdom is that an optically-thick, radiatively-efficient disk dominates in the high/soft state, while a
radiatively-inefficient flow dominates in the low-hard state. This notion is supported by, e.g., changes in the correlation between photon
index and accretion rate relative to Eddington $L_{\rm  Bol}/L_{\rm Edd}$ above and below a critical value of accretion rate
for individual BH XRBs in outburst/decay and across samples of AGN \cite{2008ApJ...682..212W,2015MNRAS.447.1692Y}. 

However, the analogies in spectral states are difficult to solidify: while we have the luxury of observing BH XRBs undergoing outbursts and
decay accretion events on timescales of days--weeks, AGN duty cycles are considerably longer.  Accretion in AGN likely occurs sporadically
\cite{2015MNRAS.451.2517S}, 
though observational constraints on duty cycles are sorely lacking.  We thus do not know if persistently-accreting AGN are ``stable'' or ``unstable'' enough for an accurate comparison to BH XRBs -- are accretion flows in AGN capable of evolving as rapidly as those in BH XRBs?

Recent years have seen initial insights into these questions, courtesy of new \textit{transient} AGN phenomena which hint at complexity and
diversity in accretion flows.
X-ray monitoring on timescales of days--months, such as that to be provided by THESEUS, can give insight into open questions about
accretion in such {\it rapidly-evolving} AGN: do AGN exhibit structural changes in accretion flow as a function of system
parameters such as accretion rate? How rapidly can those structures --- disk, X-ray corona, BLR --- evolve in response to changes in the
global accretion supply? How long is the typical AGN duty cycle?

Our first clues into AGN duty cycles and rapid evolution in AGN come from a small number (so far) of Seyfert AGN and
quasars, in which radical changes in optical continuum luminosity, X-ray luminosity, and optical spectroscopic classification, usually over timescales of years have been observed (e.g., \cite{2014ApJ...788...48S,2015ApJ...800..144L,2018ApJ...862..109Y,2019MNRAS.485.4790S}).
 %
Optical and/or X-ray fluxes generally change by factors of 10--30, while optical spectral types shift between types 1--1.5 (strong broad
lines) to 1.8--2 (broad lines weak or absent).  Changes in
line-of-sight obscuration can be excluded; instead, the global
accretion supply was either cut-off (an accretion shutdown event) or
re-established (accretion ignition event).  We call these
changing-state AGN (CSAGN;  
changes between optical Seyfert classifications are frequently also called 
Changing-Look AGN in the literature). Over 60 have been identified so far, mostly through
optical photometric and/or spectroscopic monitoring over timescales of
years
\cite{2018ApJ...862..109Y,2019ApJ...874....8M,2020MNRAS.497..192H}.

CSAGN are our first observational window into AGN duty cycles, how quickly AGN accretion flows can undergo major structural changes, and whether AGN undergo state changes like those observed in BH XRBs.
Typically, X-ray and optical/UV continua track each other roughly in tandem over timescales of years in both accretion ignition and
shut-down events (e.g., \cite{2014ApJ...796..134D}). 
The BLR is generally expected to respond to changes in the optical/UV ionizing thermal
continuum of timescales of tens to $\sim$a hundred days \cite{2018MNRAS.480.3898N}.
However, the X-ray behavior can sometimes strongly deviate from that observed in the optical/UV (e.g., \cite{2019ApJ...883...94T}), 
suggesting the X-ray corona may be intermittent and/or highly beamed during such events.                
To make further progress in understanding this phenomenon, we must accumulate a statistically meaningful sample of CSAGN
detections, and build a database of their multi-wavelength variability properties, including X-ray luminosity evolution.


Recently,  a new timing phenomenon has been discovered in AGN: so-called Quasi-Periodic Eruptions (QPE, \cite{2019Natur.573..381M,2020A&A...636L...2G}). 
The X-ray flux in these sources was seen to increase by factors of 50--100 in discrete bursts,   occurring roughly regularly (every $\sim$9 hr in GSN69 and every $\sim$4--6 hr in RX~J1301.9+2747), and lasting for $\lesssim$ one hour, separated by times of quiescent emission.
Their X-ray spectra are dominated by thermal emission, with temperature of order 30--50 eV in the quiescient state, and corresponding to the  $L  \propto T^4$ relationship expected in standard accretion disk theory.
During eruptions, the temperature  increases to $\sim$ 100--150 eV, similar to how the ``soft excess'' of persistently-accreting AGN
is sometimes modeled (e.g., \cite{2011A&A...534A..39M,2018A&A...609A..42P}).  
An additional hard X-ray 
non-thermal component, which may indicate the presence of a  corona was also detected \cite{2020A&A...636L...2G}.
These QPEs have continuum variability patterns similar to those observed in the ``heartbeat'' states of the Galactic micro-quasar
GRS 1915+105 and BH candidate IGR~17091$-$3624 \cite{1997ApJ...488L.109B,2011ApJ...742L..17A}. 
Potential relevant explanations include a radiation-pressure disk 
instability \cite{1974ApJ...187L...1L,2011MNRAS.414.2186J}, 
a magnetically-elevated accretion flow \cite{2019MNRAS.483L..17D}, 
or close SMBH binaries. However, given that only two
QPEs have been confirmed so far, basic questions regarding disk
structure and properties remain. To understand the nature of this
phenomenon, a statistically meaningful sample of QPE detections is
needed, allowing their distribution of periods,
luminosities, waveform profiles, etc.  to be measured.

In addition to transient AGN, there remain questions on the properties of circumnuclear structures in persistently-accreting Seyferts.
 Variations in the line-of-sight column density on timescales from hours to years, have been observed in roughly
two dozens of objects across both optical classes of Seyferts (e.g.,\cite{2005ApJ...623L..93R,2007ApJ...659L.111R,2009ApJ...696..160R,2014MNRAS.439.1403M,2016ApJ...820....5R,2020MNRAS.492.3872Z}).
 %
These events are due to discrete clouds of gas transiting in the line of sight to the X-ray-emitting corona, and the AGN in which they occur are denoted changing-obscuration AGN (also sometimes   called Changing-Look AGN in the literature).              
These occulting clouds can be Compton-thin or Compton-thick, and they indicate that the structure of the circumnuclear material is
clumpy or filamentary, as opposed to being a smooth continuous medium, at distance scales of light-days to light-years \cite{2006ApJ...648L.101E,2008ApJ...685..160N}. 
Strong support for clumpy tori comes from high-spatial resolution IR SED fits \cite{2011ApJ...731...92R,2014MNRAS.439.3847R}. 
However, the origin of these clouds is unclear; for instance, they might be part of an outflow from the disk  \cite{2010ApJ...715..636F} 
or IR radiation-driven winds \cite{2012ApJ...747....8D}. 
It is also possible that the BLR and dusty torus comprise one continuous distribution of clouds spanning inside and outside the dust
sublimation radius, respectively \cite{1993ApJ...404L..51N,2008MmSAI..79.1090G}. 

To address these questions,  additional cloud events must be observed to accumulate statistics on both the properties of
individual clouds and on how frequently such eclipse events occur --- and on what timescales --- across Seyferts to
better constrain the parameters of clumpy-torus models (cloud distribution parameters and distance scales).  
Using RXTE, it was shown that sustained long-term X-ray monitoring of a large sample of Seyferts is essential for obtaining such X-ray-based statistical constraints \cite{2014MNRAS.439.1403M}.

The major challenge, particularly with CSAGN,  is that on a per-object basis, such events can occur very rarely (for our lifetimes)!  The solution is to monitor large fractions of the sky in order to cover as large a starting
sample as possible; this amplifies small-number events to reasonable numbers, and catches transient AGN ``in the act".  THESEUS will trigger additional
observations (e.g., monitoring in optical and/or radio; optical spectroscopic monitoring) on these ``highlighted" transient events
in-progress and catch state changes in action, enabling us to improve our understanding about the structure of both the accretion disk and the corona, and how they interact and evolve.  


During the 2030s, THESEUS will be the premiere instrument to detect new X-ray-transient AGN in the local Universe, monitor them in X-rays, and trigger multi-wavelength follow-up observations.  The currently-ongoing eROSITA mission is expected to boost initial insight
into transient AGN and ``open the door,'' but its field of view is small: during its all-sky surveys, objects get visited six times in
one day, and that occurs only once every six months for four years.
In contrast, THESEUS will provide \textit{sustained} X-ray monitoring covering a wider range of variability timescales:
from satellite orbital timescale to weeks--months within a given observing window, as well as on timescales of months--years between
observing windows. As a consequence, THESEUS is sensitive to variability mechanisms spanning a much wider range of BH
masses.

\subsection{Detecting new CSAGNs and studying their coronal evolution}

By comparing accumulated X-ray fluxes within SXI observation windows,
or from one observation window to the next, we can detect new
accretion ignition or shut-down events as they are occurring,
including quiescent galaxy to Seyfert transitions (or vice versa) or
major variations in X-ray luminosity associated with strong
accretion changes in Seyferts.
After the end of the  eROSITA mission, THESEUS will take over transient AGN detection 
with the major advantage  of flux monitoring on timescales of days--weeks, enabling us to track the behavior of the
corona and establish time constraints for replenishing or depleting it when major structural changes to the flow occur
(e.g. \cite{2013MNRAS.433.1764M,2020ApJ...898L...1R}). 

For the best-profiled ignition events, THESEUS can trigger follow-up optical photometric monitoring (e.g., if not already available from LSST) and spectroscopic monitoring from multi-meter class facilities to discern how the accretion disk and BLR, respectively, evolve.           
For example, it will be possible to track BLR structural changes as disk luminosity or structure evolves (e.g., \cite{2019MNRAS.485.4790S}) 
or to determine if a previously-present reservoir of BLR gas is illuminated/de-illuminated by a changing continuum (e.g., \cite{2019ApJ...883...94T}). 
We also expect to use THESEUS detections to trigger ATHENA for deep follow-up X-ray observations
on selected events for further spectral and timing constraints on the structure and location of the corona.
In this regard, the combination of X-ray and multi-wavelength monitoring, with unprecedented time coverage, will improve our view on
how the various accretion components all interact and structurally evolve during major accretion changes.
We also aim to compare CSAGN accretion flow evolution to evolving flows in BH XRBs that undergo outburst/decay events. For instance, we can test for correlations or anti-correlations between coronal X-ray photon index  and accretion rate   to probe accretion mode and constrain state-transition models in AGN (e.g., \cite{2019ApJ...883...76R}).  
In addition, with THESEUS and SKA concurrent monitoring to probe radio jet--X-ray coronal coupling in CSAGN, we can explore the extent to which state changes in CSAGN track those in outbursting and decaying BH XRBs (e.g., the ``fundamental plane of accretion", \cite{2003MNRAS.345.1057M}). 
Finally, the accumulated statistics on how frequently ignition/depletion events occur will yield sorely-needed  
observational constraints on the duty cycles of supermassive BH feeding.

X-ray identification of a CSAGN candidate requires a minimum flux change of $\sim$ 10--20.  A 90 ks SXI
exposure (3 days) can detect 0.3--5 keV fluxes above $4\times10^{-13}$ erg cm$^{-2}$ s$^{-1}$.  If flux changes by a factor
of at least 20 (10) between a low-state flux $\leq4\times10^{-13}$ erg cm$^{-2}$ s$^{-1}$ and a high state flux of at least $7\times10^{-12}$ ($4\times10^{-12}$) erg cm$^{-2}$ s$^{-1}$, the flux change will be registered at 5$\sigma$ (2.5$\sigma$) confidence.  
After a CSAGN ignition event is detected, the  SXI can track its   flux variability down to timescales of a few days as long as its 0.3--5 keV flux remains above $\sim 2-4 \times10^{-12}$ erg cm$^{-2}$ s$^{-1}$.
Long-term monitoring with SXI will enable us to distinguish between CSAGN and Tidal Disruption Events (see Sec.~\ref{sec-TDE}), as the latter typically show characteristic luminosity decays as a function of time as well as extremely soft   spectra.

Expected detection rates for new CSAGN events are highly uncertain, as this is still new discovery space.  Assuming there are of order $10^6$ currently-quiescent galaxies across the whole sky capable of ignition events in which they brighten to $F_{0.3-5} \geq 4 \times10^{-12}$ erg cm$^{-2}$ s$^{-1}$, and if the average AGN duty cycle is $10^5$ years,
we may expect of order six events/year visible by THESEUS.

 
\subsection{Detecting and monitoring QPEs}

XMM-Newton and eROSITA  can explore relatively narrow temporal frequency
ranges, and can monitor only short-period (hours) recurring bursts,
which may skew detection towards relatively low-mass BHs.  For
example, GSN69 and RX~J1301.9+2747, with masses in the range
$\sim(1-6)\times 10^6 M_{\rm \odot}$, have recurrence times of 4--9
hours.  THESEUS represents a great opportunity to explore a region of
frequency parameter space not currently covered.  If recurrence times scale linearly with BH mass,
THESEUS can potentially detect longer and slower QPEs than those
observed so far --- recurrence times of few to tens of days, and
possibly associated with more massive (and more typical) SMBHs.
That is, the temporal frequency space covered by THESEUS complements that
of XMM-Newton and eROSITA, and provides the missing parts of the
picture of the (currently wholly unknown) period distribution of QPEs.

\begin{figure}
\centering
\includegraphics[width=8cm]{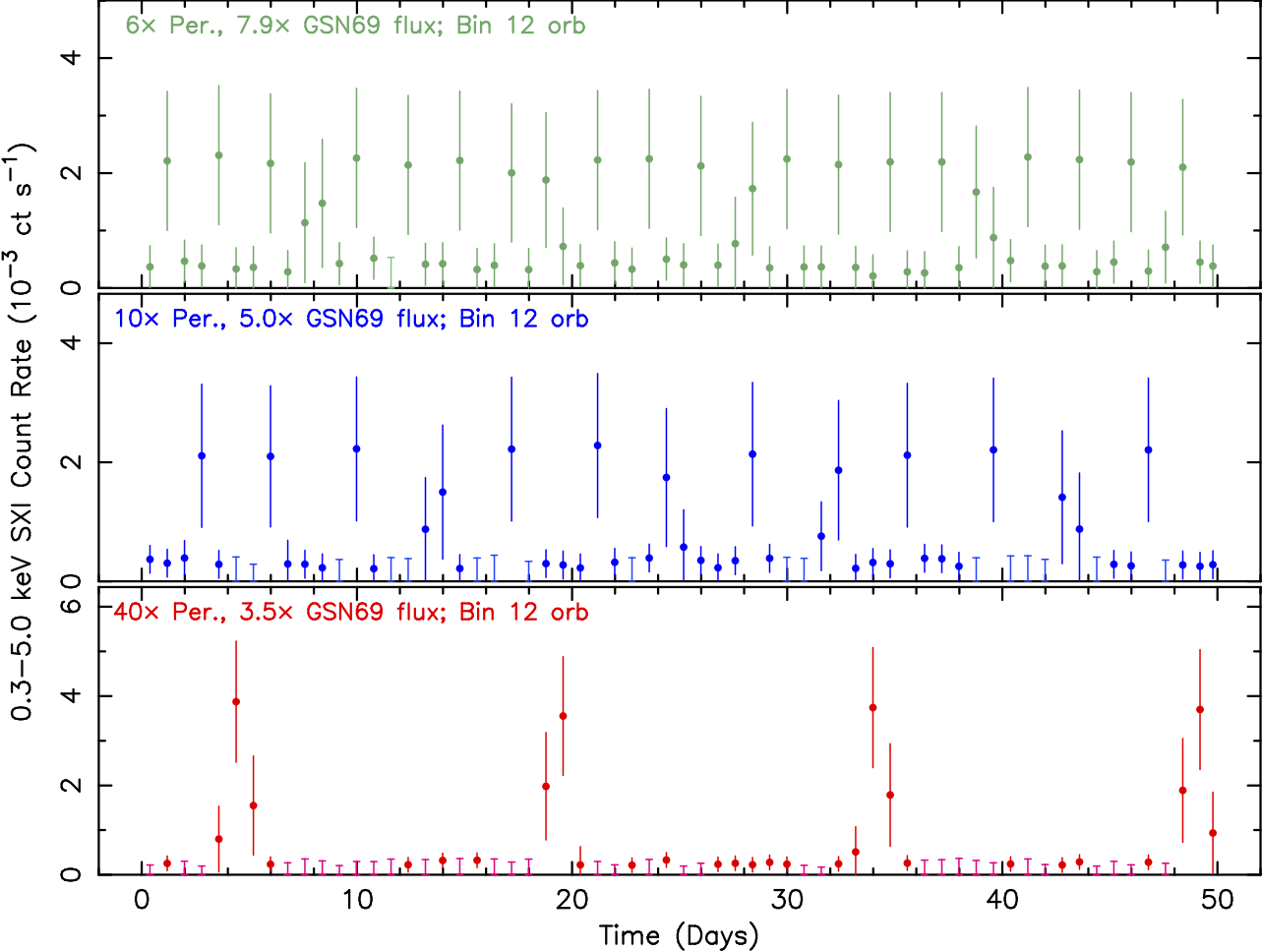}
\caption{A sample of simulated QPE light curves, based on the waveform observed in GSN~69 \cite{2019Natur.573..381M}, and
with binning every 12 satellite orbits.  These light curves were produced
by multiplying both the waveform and recurrence time observed in GSN~69 by factors of 6, 10, and 40 (top, middle, and bottom panels,
respectively), and simulating peak 0.3--5 keV fluxes that are factors of 7.9, 5.0, and 3.5 times that observed in GSN~69.  THESEUS SXI will
be most sensitive to QPEs with periods of order tens of hours or longer, and fluxes a factor of at least a few times higher than that
of GSN~69.}
\label{fig:qpelightcurves}
\end{figure}

For each newly-detected source,  the SXI can be used  to constrain waveform, peak luminosity, and burst recurrence times (see
Fig.~\ref{fig:qpelightcurves}); first-look X-ray spectra with SXI can confirm if X-ray emission during flares is thermal-dominated.
Follow-up multi-wavelength observations (and/or concurrent monitoring from LSST, SKA, etc.) can be done to
probe the behavior of accretion disk thermal emission and jets, and test if BLRs are routinely present and stable or absent in such objects.
THESEUS detections will also trigger observations with ATHENA for additional, detailed X-ray spectral constraints and to track the
behavior of individual X-ray emission components.
Near-continuous monitoring such as that provided by THESEUS is key for
accumulating statistics on flare profiles and amplitudes,
flare--recurrence time correlations, X-ray spectral behavior, and
comparing burst behavior to that found in microquasars' heart-beat
states.  Various scenarios for the outbursts, such as
radiation-pressure instabilities, Lense-Thirring precession of the
inner disk, or interaction with a secondary body can then be tested.  In summary, THESEUS
will likely be the community's \textit{only} opportunity to accumulate
statistics on long-period X-ray QPEs to ascertain how frequently they
occur in accretion disks across the local Universe.

A sample of simulated  QPE light curves 
based on the observed peak flux and waveform of GSN~69, is displayed in Fig.~\ref{fig:qpelightcurves}.  
Binning SXI data every 12 satellite orbits (about one day), the peak flux can be constrained well, and quiescent-state fluxes of $<5\times10^{-13}$ erg cm$^{-2}$ s$^{-1}$ rejected at $2.5\sigma$ confidence, if 0.3--5 keV peak flux
is at least $\sim 6\times10^{-12}$ erg cm$^{-2}$ s$^{-1}$, $\sim$3 times the peak flux of GSN 69.  
THESEUS will be most sensitive to QPEs if each peak lasts 10--tens of hours, with successive peaks occurring
at least a couple days apart, $\gtrsim$10 times slower than in GSN 69 and RX~J1301.9+2747.

Estimates of the expected QPE detection rates are highly uncertain: not only is this new discovery space, but the distribution of pulse
periods is completely unknown. We can only assume that the two QPEs detected with XMM-Newton in XMM-DR9 (1152 deg$^2$) are representative of the class, and we can assume a uniform pulse period distribution for simplicity.  
We consider a peak luminosity threshold $\sim$three times higher than the peak luminosity of GSN 69 to be firmly detected
with the SXI.  For simplicity, we can also assume that the QPE luminosity distribution in the low-redshift Universe mirrors that of
soft X-ray-emitting AGN; soft X-ray AGN luminosity density functions for relatively low-luminosity AGN \cite{2001A&A...369...49M} 
suggests that increasing relative luminosity by $\sim$3 yields a corresponding decrease in number of the same factor $\sim$3.  THESEUS SXI thus may detect of order 10 new QPEs.
Note that even non-detections of QPEs with THESEUS SXI will be valuable for deriving constraints on the period and luminosity distributions of QPEs.


\subsection{Detecting new Changing-Obscuration AGN events in progress}

THESEUS   will allow us to monitor a sample of $\sim10^3$ sources to watch for sudden changes between unobscured, Compton-thin
obscured, or -thick obscured states.
We will likely need to accumulate SXI spectral data on timescales up to days--weeks; this provides sensitivity to obscuration events due to clouds residing in the outer BLR/inner dusty torus or farther out.
When a new cloud obscuration event is identified, deeper follow-up X-ray observations (e.g., with ATHENA) will be triggered
to determine the properties of individual occulting clouds: ionization, column density, and location from the BH.
Follow-up observations with ground-based optical facilities can assess if dust extinction is also present along the lines of sight to the accretion disk and BLR.

Over the lifetime of the THESEUS mission, we will accumulate statistics on how frequently obscuration events occur, including as a
function of BH mass and luminosity.  
These statistics will expand the X-ray  constraints accumulated  in the RXTE survey \cite{2014MNRAS.439.1403M}
to lower column densities.  That is, THESEUS can provide a more complete picture of the distribution of
clouds and their properties within the context of clumpy-torus models.

\begin{figure}
\centering
\includegraphics[width=8cm]{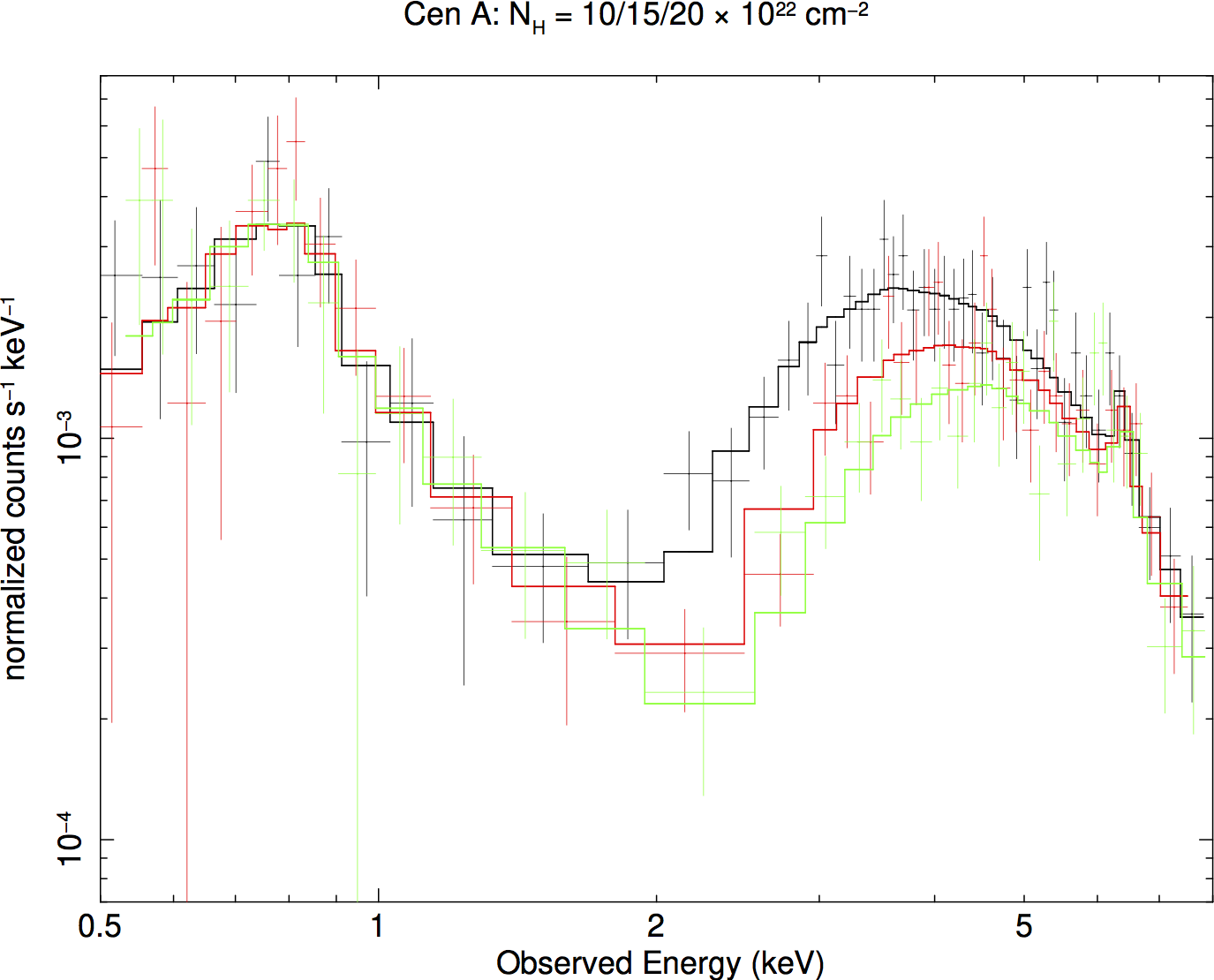}
\caption{
Simulated SXI spectra of Cen~A.  THESEUS SXI can track fluctuations in  line-of-sight column density $N_{\rm H}$ as density variations in its clumpy torus drift across the line of sight.  Shown here are spectra corresponding to columns of 10, 15, and 20$\times10^{22}$ cm$^{-2}$, (black, red, and green, respectively, based on variations observed by \cite{2011ApJ...742L..29R}),
each 70 ks accumulated exposure.}
\label{fig:cenanh}
\end{figure}

In 90 ks exposures (3 days) of an unobscured source,   SXI spectra can reveal major increases in obscuration   at $\geq$2.5$\sigma$ confidence if the unobscured continuum flux is brighter than roughly $F_{0.3-5} \sim 1.3\times             
10^{-11}$ erg cm$^{-2}$ s$^{-1}$ (for column densities $N_{\rm H}$ of roughly $10^{21.5-22}$ cm$^{-2}$) or $F_{0.3-5} \sim 1.9\times                
10^{-11}$ erg cm$^{-2}$ s$^{-1}$ (for $N_{\rm H} \sim 10^{22.5}$ cm$^{-2}$).
Spectral fits to   SXI will discern full-covering obscuration events, though not partial-covering events. However, simple power-law
model fits to SXI spectra of full- or partial-covering Compton-thin obscuration events in Seyferts as faint as $F_{0.3-5} \sim 2.5\times          
10^{-11}$ erg cm$^{-2}$ s$^{-1}$ can yield photon indices $\lesssim1.5$ or lower, which can alert suspicion.
For perpetually Compton-thin-obscured sources such as Cen~A (see Fig.~\ref{fig:cenanh}), increases in full-covering $N_{\rm H}$ of
order $10^{22.5-23}$ cm$^{-2}$ or more can be detected in sources brighter than $F_{0.3-5} \sim 8\times 10^{-12}$ erg cm$^{-2}$
s$^{-1}$.
The complex soft and hard X-ray spectrum  of Cen~A, which includes thermal and non-thermal contributions, was studied extensively  by many missions \cite{1998A&ARv...8..237I,2003ApJ...593..160G,2004ApJ...612..786E,2016ApJ...819..150F}, and provides a case study for spectrum formation and variability of more distant radiogalaxies, and an un-beamed reference template for blazars.

Changing-oscuration AGN  event detection rates can be based on the logN-logS relation of \cite{2012ApJ...749...21A}, 
and the flux thresholds listed above: we estimate starting samples of roughly 300--1000 unobscured Seyferts and 25 Compton-thin-obscured Seyferts that can be monitored by THESEUS.

\subsection{Blazars}
\label{sec-rlagn}

AGN are the most abundant extragalactic X-ray emitters, and those belonging to the radio-loud subgroup ($\sim$10\% of all AGN)  are the strongest ones.  In blazars, a small subset of radio-loud AGN, kpc-sized jets are nearly aligned with the line of sight (within 5-10$^{\circ}$). The ultra-relativistic conditions of the plasma propagating in these jets and the small viewing angles boost the broad-band luminosities up to $\sim 10^{49}$ erg~s$^{-1}$ and Doppler-foreshorten the time-scales (typical Doppler factors are $\delta \simeq 10-20$), making blazars the most extreme radio-loud, and most violently variable persistent extragalactic objects at all wavelengths. In outburst, blazars get typically 10 times, and occasionally up to 100 times, brighter than in quiescence, and their variability is generally chromatic, with TeV emission exhibiting the largest amplitudes and shortest timescales.

Blazar emission is dominated, at IR to soft X-ray energies, by
synchrotron radiation caused by electrons gyrating in a magnetic
field.  In a subgroup of sources, this emission component can extend up
to hard X-rays \cite{1998ApJ...492L..17P,2019MNRAS.486.1741F}.
At higher energies ($>$100 keV),   if the jet composition is purely
leptonic the radiation is due to inverse Compton scattering of particles
off synchrotron or ambient photons.
The presence of an additional hadronic component may engender proton
synchrotron emission, neutral pion decay photons, synchrotron and
Compton emission from secondary decay products of charged pions, and the output from pair cascades initiated by these high-energy photons intrinsically absorbed by photon-photon pair production (see \cite{2017SSRv..207....5R}, and ref. therein for a review). 
In a $\nu f_\nu$ representation, these radiation components are seen as two maxima, which give blazars their characteristic multi-wavelength ``double-humped" spectral appearance.

Depending on the frequency of the synchrotron  spectral peak, blazars are subdivided into classes that form actually a continuum (blazar sequence, \cite{1998MNRAS.299..433F,1998MNRAS.301..451G,2008MNRAS.387.1669G}; see however \cite{2012MNRAS.422L..48P}), with Flat Spectrum Radio Quasars (FSRQ) synchrotron spectra peaking at the lowest frequencies (IR), Low-frequency-peak BL Lacs (LBL) peaking at optical frequencies, intermediate-frequency-peak BL Lacs (IBL) peaking in UV, and High-frequency-peaked BL Lacs (HBL) peaking at soft X-rays or even  - during outbursts -  hard X-rays (``extreme" BL Lacs).  

The location of the humps is related to the magnetic field and energy of the particle distributions that are injected in the emitting region, so that  the  characteristic frequencies of the spectral maxima provide a first direct hint to the power injection and cooling in the jet.
In FSRQ and LBL, the X-ray spectrum represents approximately the rising part of the second, high energy spectral hump, coinciding with the inverse Compton component in a leptonic scenario, while in HBL soft X-ray spectra are produced by the synchrotron process and  therefore are generally steeper and brighter  than in FSRQ and LBL.  In objects with intermediate properties, the two spectral components contribute in the X-ray domain with comparable strength.  The X-ray band is thus crucial to explore the particle acceleration, the interplay of the radiation mechanisms, and, together with information over the rest of the electromagnetic spectrum, to discern leptonic and hadronic composition of the jet.  

More specifically, HBL X-ray spectra map the {\it high-energy tail} of the relativistic particles distribution, and thus encode critical information on the inner mechanisms and efficiency of the central engine.  Furthermore, these sources are either confirmed or prime candidates as TeV emitters, as the spectrum at GeV-TeV energies  traces the same electron distribution tail that is responsible for the X-rays.  Despite the severe suppression caused by the far-infrared extragalactic background, state-of-the-art Cherenkov telescopes could detect flaring blazars up to z$\sim$1 at energies larger than $\sim$300 GeV \cite{2015ApJ...815L..23A,2016A&A...595A..98A}.

As multi-wavelength observing campaigns on blazars become more complete and sophisticated, and the astroparticle angle of their  investigation becomes more effective, thanks to the improved sensitivity of present high-energy neutrino detectors, 
fundamentally new observational and theoretical issues are emerging, critically driven by X-ray data.  These include the finding that blazar variability  may be occasionally due to a multiplicative process originated in the accretion disk, rather than in the jet \cite{2020arXiv201201348M}; the evidence that multi-component models  describe the blazar spectral energy distributions and variations thereof more comfortably than homogeneous, single zone models \cite{2019MNRAS.490.2284M,2020A&A...637A..86M,2020A&A...640A.132M,2020ApJ...890...97A}, even down to sub-hour X-ray variability time-scales \cite{2020ApJS..248...29A}; the presence of a hadronic component in the jet, either suggested by a model-driven approach \cite{2020A&A...638A..14M}  or proven by the direct detection of high-energy ($>$ TeV) neutrinos, so far achieved with the IceCube array in a few  cases, with attendant multi-wavelength outbursts \cite{2018Sci...361.1378I,2018Sci...361..147I,2020ApJ...893..162F,2020GCN.26655....1L,2020A&A...640L...4G}.

Inhomogeneity is envisaged by current interpretive scenarios, either in the form of two geometrical components (spine + layer), or of the interplay between an acceleration region and a more extended emission region, or of conical zones where the physical quantities show a dependence on the distance from the jet apex \cite{1985A&A...146..204G,2008MNRAS.385L..98T,2018MNRAS.473.4107P,2019ApJ...887..133B}.   However, these models have a large number of free parameters, that could be well constrained by time-resolved broad-band spectral energy distributions obtained through intensive monitorings over long periods of time.  Temporal lags measured through accurate cross-correlation of the X-ray light curves with those at MeV-GeV energies from satellites and at TeV energies from Cherenkov telescopes hold the potential to map the emitting regions and thus probe the physical evolution  of the flares.
Similarly, the  X-ray data have been recognized to have a pivotal role in determining the relative importance of the leptonic and hadronic jet components in shaping up the blazar spectrum \cite{2019NatAs...3...88G,2018ApJ...864...84K,2019MNRAS.483L.127R,2019MNRAS.489.4347O,2020arXiv200904026R,2020ApJ...899..113P,2020ApJ...893L..20L}.

THESEUS, with its simultaneous soft- and hard-X-ray month-long looks, made possible by the large SXI and XGIS FOVs and regular sky survey, can measure time delays between various X-ray bands and between these and signals at lower or higher frequencies in a more accurate way than it is done with the combined operations of different instruments (e.g. Swift/XRT, or XMM-Newton, and NuSTAR) that often only afford quasi-simultaneity owing to scheduling contraints (see, e.g., \cite{2020MNRAS.499.2094G} for an example of the superiority of {\it strictly} simultaneous observations in constraining physical parameters).

With its prospect of enabling accurate time correlation analysis of blazar X-ray light curves and spectral slopes, THESEUS is well poised to tackle the problems posed by increased blazar modelling complexity.  
A case in point is the delay between light curves of different energies  measured for the BL Lac object Mkn421 with uninterrupted  INTEGRAL  monitoring over 6 days.  
By correlating IBIS and JEM-X light curves, a  progressively larger time-lag is found between signals of increasing energy separation, up to a maximum of 70 minutes, measured between the light curves at 3-5.5 keV and 40-100 keV (see Fig. 5  in \cite{2014A&A...570A..77P}).   This source is one of the X-ray brightest blazars and the brightest of the list of HBL we have selected as primary blazar targets of THESEUS (Table \ref{tab:xbl}).  

\begin{figure}
\centering
\includegraphics[width=8cm]{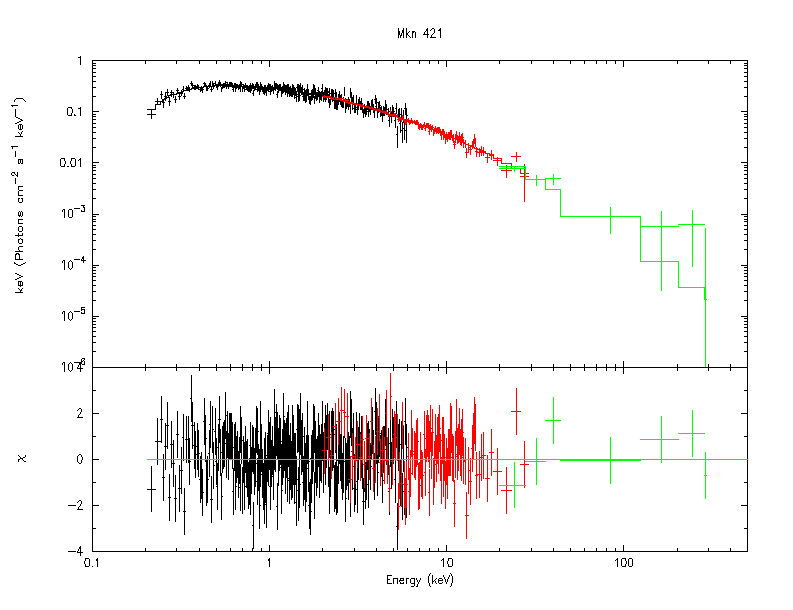}
\caption{Simulated SXI and XGIS spectra of Mkn421 (35 ks exposure, corresponding to 1 day of observation).  An emission state similar to the one observed in April 2013 by  INTEGRAL has been taken as a reference (i.e. the spectrum in the considered energy range is described by a  logarithmic parabola with an energy cutoff at 10 keV (see \cite{2014A&A...570A..77P}).}
\label{fig:mkn421simulspectrum}
\end{figure}

A simulation made for the three bright HBLs Mkn421, Mkn501 and 1ES1959+650 shows that in 500 s a good (at least 10 $\sigma$) and decent (3-4 $\sigma$) significance can be obtained for light curves of these sources in flaring state in the 0.3--2 keV and 2--6 keV ranges, respectively.   The same integration time will give a satisfactory signal over the whole XGIS energy range.  This will enable correlation of 3 independent signal trains over more than 2 decades in energy, which represents a good diagnostic of models in a spectral region dominated by the competition of particle acceleration and cooling.  By integrating the SXI and XGIS signals over a typical survey mode pointing (2.3 ks), one obtains a good estimate of the hardness ratio, whose time series can also be used for cross-correlation with flux.  In a net exposure time of 35 ks  (that can be acquired in one day), spectra of excellent S/N can be constructed during outbursts, to provide a reliable estimate of the shape and parameters (see example of Mkn421 in Fig. \ref{fig:mkn421simulspectrum}).

Thanks to the month-long observing periods allowed by the THESEUS pointing  strategy (see percentages of total observing time in Table \ref{tab:xbl}),  HBLs  will be followed over the entire excursion of their brightness (up to factor 100) to study their X-ray duty cycle.  The broad energy range simultaneously  covered by SXI and XGIS will make it  possible to investigate the change in spectral curvature of the synchrotron spectra of these sources, directly probing the mechanism of acceleration of highly energetic electrons.

The extended observing periods also guarantee coordination with  TeV coverage by atmospheric Cherenkov telescopes (eminently CTA) and at MeV-GeV energies with satellites, to study high-frequency variability down to timescales of hours or less.  Complementary long monitorings at optical (e.g. Rubin Observatory-LSST,  THESEUS IRT) and radio (Meerkat, ASKAP, SKA, ngVLA) wavelengths will enhance the results and probe the connection of multiple emitting regions.  

At redshifts higher than $\sim$1, where TeV detection is difficult owing to pair-creation opacity, the THESEUS instruments can be used in  coordination with future $\gamma$-ray satellites, to follow the blazar emission in the brightest part of their spectrum. Besides providing insight into the largest and most distant engines in the Universe, this simultaneous monitoring will build statistics for the assessment of blazar cosmological evolution.

THESEUS' primary asset in blazar research is the large FOV of its cameras, which sets it entirely apart from facilities with a small FOV.  The latter must be turned around  in a target-of-opportunity mode after outburst notification, which - given the typical $\sim$1 day duration of  a blazar flaring episode -  results in their ability to only  follow the decaying part of the flare. 
For blazars in visibility periods of the  XGIS and SXI cameras, these will follow the outburst since inception and will monitor it for its entire duration.

A list of optimal radio-loud AGN for SXI and XGIS monitoring is provided in Table \ref{tab:xbl}.  The well-known and frequently monitored source 3C273 is a bright multi-wavelength emitter that will be well detected. In the X-ray spectrum of this source, both the beamed non-thermal (jet) and unbeamed thermal (accretion flow) radiation were   detected (e.g. \cite{2004Sci...306..998G}). The combination of SXI and XGIS  will allow us to better characterize these two  components.  It is  a bright Fermi-LAT source, but it has never been detected at TeV energies.  The CTA sensitivity may make its TeV detection possible, so that correlated X-ray and MeV-to-TeV variability studies may become  finally affordable.   The other  sources are all bright HBL and TeV emitters.  Their X-ray brightness and long THESEUS visibility periods will open unprecedented windows into their multi-wavelength variability.

\begin{table*}[]
    \centering
    \begin{tabular}{lcrcrrc}
    \hline
Source    & redshift & N$_{H}^a$                     & Flux$^b$                                                  & SXI$^c $        & XGIS$^c$ & Visibility$^d$  \\
             &                 &                                   & 0.3-10 keV                                              &      0.3-5 keV    &  2-30  keV&    \\           
                 &        & 10$^{20}$ cm$^{-2}$ &   10$^{-10}$ erg cm$^{-2}$ s$^{-1}$  & counts s$^{-1}$  &  counts s$^{-1}$ &     \\
         \hline
Mkn 501            & 0.034   & 1.3      & 8    & 0.38  & 36.3 & 11\% \\
Mkn 421            & 0.031   & 1.5      & 52   & 3.41   & 156.1 &  8\% \\
1ES 0033+595  & 0.467   & 38.1    & 6     & 0.11 & 29.2 & 3\% \\
PG 1553+113  & $>$0.09 & 3.7    & 2     & 0.11 & 4.9 & 8\% \\
3C 273             & 0.158   & 1.6       & 2     & 0.07 & 7.3 & 7\% \\
PKS 2155-304 & 0.116   & 1.4       & 3    & 0.29  & 6.4 & 8\% \\
H1426+428    & 0.129   & 1.1       & 1    & 0.05 & 1.6& 9\% \\
1ES 1959+650  & 0.048  & 10.7     & 9    & 0.41  & 24.1 & 2\% \\
    \end{tabular}
    \caption{X-ray bright Radio-loud AGN. $^a$  column density  from \cite{2005A&A...440..775K}; $^b$ Unabsorbed flux;   $^c$ Expected   count rate;   $^d$ Fraction of total  time in which the source will be in the SXI FOV.
    }
    \label{tab:xbl}
\end{table*}

\section{Tidal Disruption Events}
\label{sec-TDE}

Tidal disruption events (TDEs) are associated with the catastrophic disruptions of stars passing too close to the supermassive BHs (SMBHs) in the centres of their host galaxies. At the tidal radius of the BH, the tidal forces exceed the binding energy of the star causing it to be torn apart \cite{1988Natur.333..523R}. Upon the encounter, about half of the stellar mass is expected to be captured and accreted by the BH on a timescale of about one  year \cite{1975Natur.254..295H}, producing powerful flares with peak luminosities of up to 10$^{45}$ erg s$^{-1}$ \cite{2009MNRAS.400.2070S}. However, the duration of TDEs is highly variable. Some of the TDEs detected in X-rays are observable for only half a year (e.g. Swift J2058.4+0516 \cite{2015ApJ...805...68P}), whereas others, such as 3XMM J150052.0+015452 \cite{2017NatAs...1E..33L} have stayed bright for more than a decade. Why this should be the case is unclear. The duration depends on the viscosity of the matter \cite{1988Natur.333..523R}, but such a long duration would require a very (too) low viscosity. It may be linked to the mass of the accreted star, or the circularisation of the debris stream may be very inefficient due to weak general relativity effects, so that there is a high mass fallback rate \cite{2018ApJ...867...20C}. Increasing the sample of TDEs observed in X-rays and modelling their X-ray lightcurves will help determine the physical mechanism behind TDEs.

The emission from   TDEs is expected to be thermal in nature \cite{1988Natur.333..523R}, with a temperature peaking in the soft X-ray/UV, depending on the mass of the BH. The effective temperature scales as T$_{eff} \propto r^{-3/4}$ \cite{1973A&A....24..337S}, where $r$ is the radius, so the more massive BH show their peak temperature in the UV. Indeed, the majority of TDEs do show very soft thermal emission, peaking at  $\lesssim$0.1 keV \cite{2011ApJ...738...52L,2018MNRAS.474.3000L,2019A&A...630A..98S,2019MNRAS.488.4816W}. However, a small percentage have shown much harder X-ray emission, notably Swift J164449.3+573451 \cite{2011Natur.476..421B} and Swift J2058.4+0516 \cite{2012ApJ...753...77C}, which have X-ray spectra that can be fitted with power laws and   emission extending to hundreds of keV.  Why some TDEs show these much harder spectra is unclear. It may be due to a jet pointing towards us, e.g. \cite{2017ApJ...838..149A}, but other mechanisms are also suggested \cite{2016A&A...586A...9H}. Determining why we see some TDEs with hard spectra is one area where THESEUS will be able to make headway, thanks to the hard X-ray/$\gamma$-ray telescope XGIS which will be able to both discover and study these sources and  enlarge the very small sample of only 4 or 5 that we know today. Furthermore, if the hard emission is due to a jet pointing towards us, determining the fraction of hard TDEs will give clues to the opening angle of the jet.

Besides the stark differences between some TDEs in   X-rays, we also note that some   are not detected in X-rays at all, but only at longer wavelengths (UV, optical, IR and/or radio). 
The origin of the UV/optical TDE emission is unclear. It may be from reprocessing of the X-ray emission from the accretion disk by optically thick material surrounding it   \cite{2013ApJ...767...25G,2016ApJ...827....3R}, or from shocks between the debris streams as they collide  \cite{2015ApJ...806..164P}, or a combination of both   \cite{2016ApJ...830..125J,2020MNRAS.492..686L}. Some TDEs are detected only in   X-rays and not in the optical, and others still are seen in both bands \cite{2017ApJ...838..149A}, as suggested by \cite{2018ApJ...859L..20D}. To further complicate things, a TDE event was also observed in the radio domain, but not in optical or X-rays \cite{2018Sci...361..482M}.  Why this should be the case is another open question regarding TDEs. It may simply be due to the viewing angle, or obscuration by dust, or related to an as yet unknown physical mechanism. Observing TDEs with THESEUS with the X-ray and NIR telescopes simultaneously will provide contemporary data in several wavebands and therefore help us addressing this question.

Observing TDEs from non-active galaxies is of particular interest, as this is perhaps the only way to directly reveal and study nuclear massive BH there. The high luminosities of TDEs allow us to observe them up to $z\sim$1 \cite{2012ApJ...753...77C}, which, together with their long duration, implies that TDEs are not rare events, despite a low probability for such an event to occur in any individual galaxy. Rates of tidal disruptions are estimated to be (1.7$^{+2.85}_{-1.27}$) $\times10^{-4}$ gal$^{-1}$ yr$^{-1}$ (90\% confidence limit) \cite{2018ApJS..238...15H}. To date, only about 90 TDEs have been detected\footnote{https://tde.space/}, about half of which in X-rays and most of the others in the optical/IR. The soft emission can be modelled to determine the mass of the BH, the accretion mechanism and,  if the event is close and therefore bright, to detect an iron line, allowing us to put limits on the BH spin \cite{2014bhns.work..129K}. As the tidal radius of the BH must be outside of the Schwarzschild radius for us to observe the  event, it is only for the lower mass BHs ($<$ 10$^8$ M$_\odot$) that we detect these events, which may allow us then to detect the extremely rare intermediate mass BHs (IMBH, 10$^{2-5}$ M$_\odot$). Finding these objects is important in understanding the origin and growth of SMBHs. SMBHs can not form from stellar mass BH, as even continuously accreting at the Eddington limit, they would never reach masses as high as $\sim$10$^9$  M$_\odot$ observed in a massive quasar at $z\sim$7.1 \cite{2011Natur.474..616M} or the 8$\times$10$^8$ M$_\odot$ BH found at $z$=7.54 (0.69 Gyr) \cite{2018Natur.553..473B}. Different theories propose that the IMBH  could either merge and/or accrete to create SMBH (see \cite{2012Sci...337..544V,2012NatCo...3.1304G,2017IJMPD..2630021M}  for reviews). However, few good IMBH candidates are known.  The large sample of TDEs discovered with THESEUS will help us to determine the demographics of massive BHs as well as to find IMBHs (there should be several detected per year, see below), which are thought to be the building blocks of supermassive BHs. Identifying the host galaxies and their masses will also help identify the formation mechanism of IMBHs which is still disputed.  Another possibility is that accretion could occur beyond the Eddington limit for a long period of time, but to date, the physical mechanism is not really understood. Again, observing TDEs in which a period of super-Eddington accretion is expected  \cite{2019ApJ...882L..25L} and modelling the emission could help us understand the physical mechanism behind super-Eddington accretion. Studying TDEs is therefore essential for understanding the origin and growth of  SMBHs.

\begin{figure}
\center
\includegraphics[width=8.5cm]{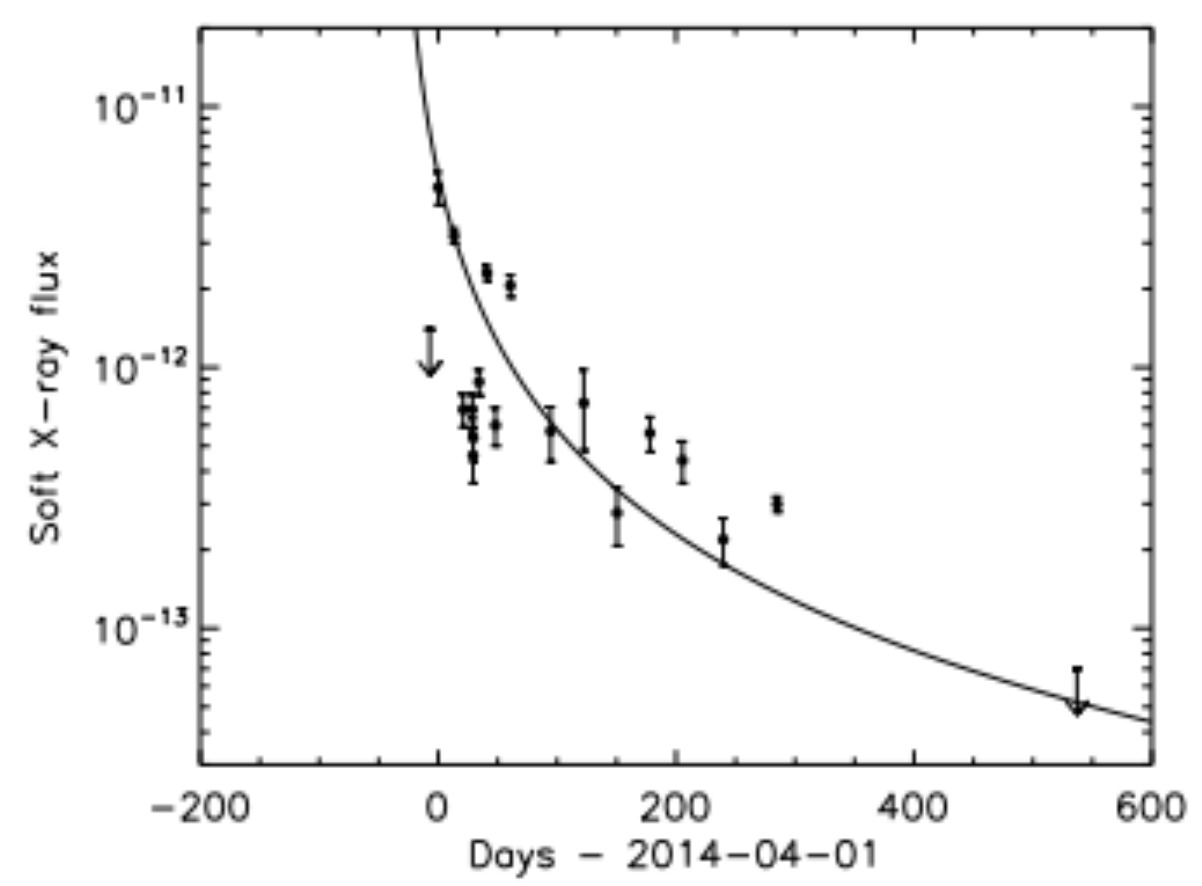}
\includegraphics[width=8.5cm]{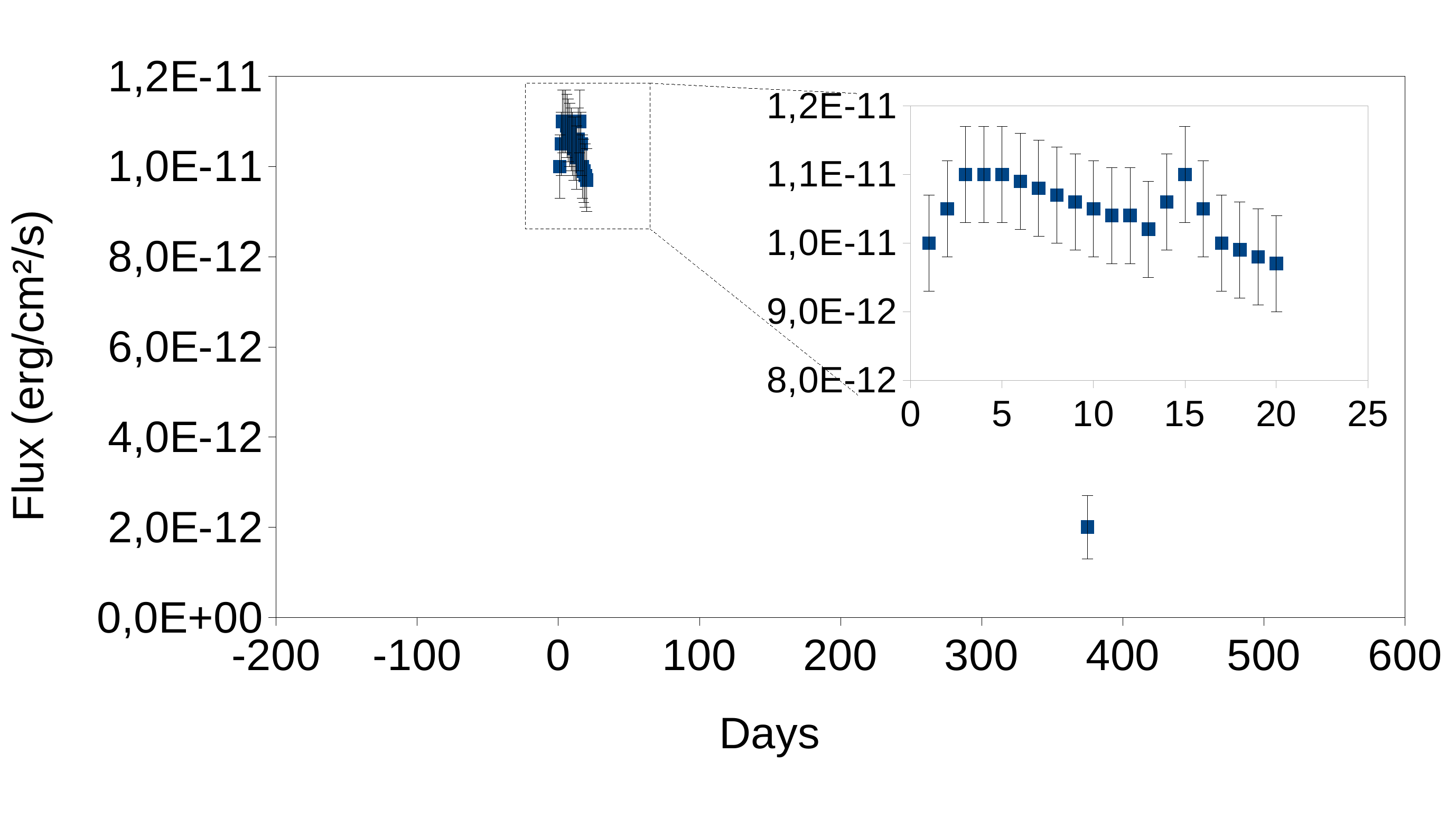}
  \caption{Top: The lightcurve of XMM SL1J0740-85 observed with the XMM-Newton slew survey in the 0.2-2 keV band (reproduced with permission from \cite{2017A&A...598A..29S}). Bottom : Simulation of a TDE lightcurve similar to XMM SL1J0740-85 as seen with the SXI. Details such as flares can  be discerned, and the decay rate can easily be determined. Further dedicated pointed observations could also be taken to refine the long term lightcurve.}
  \label{fig:TDElc_alaXMMSL1J0740}
\end{figure}

High-cadence observations of sources over a month every year with the SXI, see Fig.~\ref{fig:TDElc_alaXMMSL1J0740}, and over two months every six months with the XGIS (thanks to its larger FoV), will allow many TDEs to be discovered (see below) and a detailed lightcurve to be recorded. This is essential to determine the BH and the disrupted star's mass through modelling  \cite{2016ApJ...827..127K}. Through modelling the lightcurves, it will also be possible to identify if the SMBH is in fact in a binary system, an important information for the future LISA mission. 
The spectral evolution  will also be monitored  to study how the accretion evolves, either through standard accretion disk state changes  \cite{2017NatAs...1E..33L}, or in some other way. Furthermore, the evolution of any hard emission could also be followed to understand the origin of this emission. 

THESEUS will  fly at a similar time to Athena, which has a small field of view and will therefore not discover many TDEs. THESEUS will therefore be able to discover TDEs that can be followed up with Athena, and notably with the high resolution Integral Field Spectrometer, X-IFU. This will allow the iron line to be detected and modelled to determine the BH mass and spin, absorption and emission lines to determine the metallicity of the environment using the light from the TDE and to detect outflows, to study reverberation to probe the structure of the accretion flow.

The IRT will be very useful in following up NIR counterparts that will help us to understand if the viewing angle causes some TDEs to be seen in the optical domain and not in the X-ray whilst others in the X-ray and not the optical, and others still are seen in both. 87\% of the TDEs detected in the optical to date\footnote{https://tde.space/} could be detected with the THESEUS IRT. The NIR counterparts can be detected simultaneously with the IRT, but follow-up with other optical telescopes, such as Zwicky Transient Facility  or IFUs such as WEAVE on the William Herschel Telescope,  will also provide an optical spectrum.   In addition, TDEs offer a unique opportunity to study the  interstellar medium within a few parsecs around them. The UV and soft X-rays from TDEs can ionise the surrounding gas and the subsequent recombination produces spectral signatures, which can be absorbed and re-emitted by the local dust and provide ``echos'' observable in the IR, lasting a few years. Such phenomena will be readily detectable with THESEUS.

\begin{figure}
\center
\includegraphics[width=9cm]{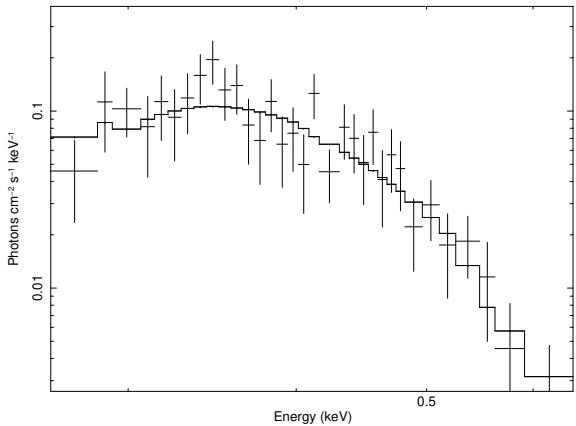}
  \caption{Simulation of a typical, soft TDE (kT=50 eV, N$_H$=1$\times$10$^{21}$ cm$^{-2}$) observed for 20 ks with the SXI with a flux of $\sim$1$\times$10$^{-11}$ erg cm$^{-2}$ s$^{-1}$.}
  \label{fig:TDEsoft}
\end{figure}

To estimate the number of TDEs observed by THESEUS per year, we will consider both the SXI and the XGIS. As described above the majority of tidal disruption events are soft, where the thermal emission peaks at $\lesssim$0.1 keV, see Fig.~\ref{fig:TDEsoft}. They will be detected mainly with the SXI. The majority (approximately 66\% \cite{2017ApJ...838..149A}) of the sources remain bright for about a month or so and then the flux drops off slowly, broadly following a decline in flux proportional to t$^{-5/3}$  \cite{1988Natur.333..523R}. The large, FoV of the SXI  and the  pointing strategy that covers approximately 19\% of the sky several times per day (average exposure per pointing, 2.3 ks) for approximately one month, before returning to the same point a year later, is ideally suited to detecting TDEs. In a single pointing, fluxes of the order of 5$\times$10$^{-11}$ erg cm$^{-2}$ s$^{-1}$ are reached. Over a day,  fluxes of the order of 1$\times$10$^{-11}$ erg cm$^{-2}$ s$^{-1}$ can be achieved. Cumulating the exposures over a month, the sensitivity is $\sim$ 1$\times$10$^{-12}$ erg cm$^{-2}$ s$^{-1}$. Taking the average maximum luminosity observed of the TDEs, $\sim$2.8$\times$10$^{43}$ erg s$^{-1}$ \cite{2017ApJ...838..149A}, TDEs can be detected out to a luminosity distance of $\sim$480 Mpc. In theory, TDEs can be detected out to higher distances as a BH of   10$^8$ M$_\odot$ accreting at the Eddington luminosity would have a luminosity of $\sim$ 10$^{46}$ erg s$^{-1}$, so this could be detected out to $\sim$1.3 Gpc. The average distance is therefore a conservative value. Taking the number density  of galaxies to be 0.02 galaxies Mpc$^{-3}$ \cite{2020SSRv..216...85S,2002AJ....124.1308D,2008A&A...489..543E}, there are about 2 $\times$10$^{6}$ galaxies in the volume out to 480 Mpc.  Using the rate of tidal disruption events given above, this gives about 340 TDEs per year. We will assume 300 in the following. Assuming a conservative 15\% of the sky covered in a month, whilst the TDE is bright, this translates to 30 (300 $\times$ 0.15 $\times$ 0.66 $\times$ 45) soft, ``short'' TDEs per year.

The other third of TDEs remain bright for a longer duration before fading, on average 3 months \cite{2017ApJ...838..149A}. Using the same values as above, but considering that a larger fraction of the sky will be observed during the three months ($\sim$40\%), then we expect 40 longer TDEs to be detected per year. As mentioned above, we now have examples of TDEs that remain bright for more than a decade, so again this is a conservative estimate.

We also expect the XGIS to detect TDEs. Given that this detector is less sensitive than the SXI, it is unlikely that it will detect many of the soft TDEs, but it is very likely that it will detect the harder TDEs that have spectra that can be described by hard power laws. After finding 9 candidate hard TDEs in the Swift/BAT data, \cite{2016A&A...586A...9H} propose a rate of 2 $\times$10$^{-5}$ galaxy$^{-1}$ yr$^{-1}$ for these hard events. The large, field of view of the XGIS   and the  pointing strategy that covers approximately 44\% of the sky per day for approximately two months, before returning to the same point six months later, is also well suited to detecting TDEs. In a single pointing, fluxes of the order of 1$\times$10$^{-10}$ erg cm$^{-2}$ s$^{-1}$ are reached.  Cumulating the exposures over two months, the sensitivity is $\sim$ 1$\times$10$^{-11}$ erg cm$^{-2}$ s$^{-1}$. Taking the peak luminosity observed for this handful of TDEs, $\sim$1$\times$10$^{44}$ erg s$^{-1}$ \cite{2016A&A...586A...9H}, TDEs can be detected out to a a luminosity distance of $\sim$290 Mpc, so 4.8 $\times$10$^{5}$ galaxies in this volume.  Using the estimated rate of tidal disruption events for hard TDEs, this gives about 10 hard TDEs per year. These objects are bright for approximately two months, in which time 88\% of the sky is covered so almost 10 hard TDEs will be detected per year. Some of these will also be observed with the SXI.

We therefore expect of the order 80 TDEs to be detected per year with THESEUS. Some of these will occur around IMBH, although their number is uncertain. Taking the method  proposed by \cite{2018ApJ...867...20C}, 4$\pi$d$_{lim}$/3 $\times$ n$_{GC}\times$ 10$^{-7}$, and supposing that there may be an IMBH in every globular cluster and then calculating the number of globular clusters per Mpc$^3$ (n$_{GC}$) to be 100 \cite{2018ApJ...867...20C}, we may expect 1000 per year all sky, as these TDEs have soft spectra and are detected with luminosities up to the Eddington limit for a BH  of mass 1 $\times$ 10$^5$ M$_\odot$, namely a luminosity of $\sim$1$\times$10$^{43}$ erg s$^{-1}$ (e.g. \cite{2020ApJ...892L..25L}). Therefore, they can be detected out to d$_{lim}\sim$288 Mpc.  These TDE have durations similar to the TDEs around SMBH and therefore we could expect 150 detectable with the SXI.
This seems very high, probably due to the assumption that there is an IMBH in every globular cluster. Taking the number of IMBH detected in the known population of TDEs, $\sim$5\%, then we could perhaps more realistically expect around 4 of the observed TDEs per year to be around IMBH. 

To validate this estimate of the number of TDEs per year discovered by THESEUS, we make the comparison with eROSITA  \cite{2014MNRAS.437..327K}, which has an energy response and  sensitivity similar to that of the THESEUS SXI.  \cite{2014MNRAS.437..327K} adopted a limiting flux of $\sim$10$^{-12}$ \flux\ for eROSITA, which is comparable to the expected SXI sensitivity although a significantly longer exposure is required in the latter case (i.e. 1 month compared to 1 ks), which is, however, not an issue for events lasting months. This implies that this estimate can also be directly applied to SXI which should detect a comparable number of TDE candidates, i.e. $\sim$100-1000 TDE/yr for the assumed limiting flux of  10$^{-12}$ \flux\ (confusion-limited sensitivity) and using Eq. 16 from \cite{2014MNRAS.437..327K}. Observations during the first year of eROSITA indicate a value closer to the lower boundary (P. Jonker, priv. comm.). This is in line with the numbers estimated above. A more recent study \cite{2019MNRAS.488.4042T}, indicates that $<$10 of these TDEs will be around IMBH.

\begin{figure}
\center
\includegraphics[width=9cm]{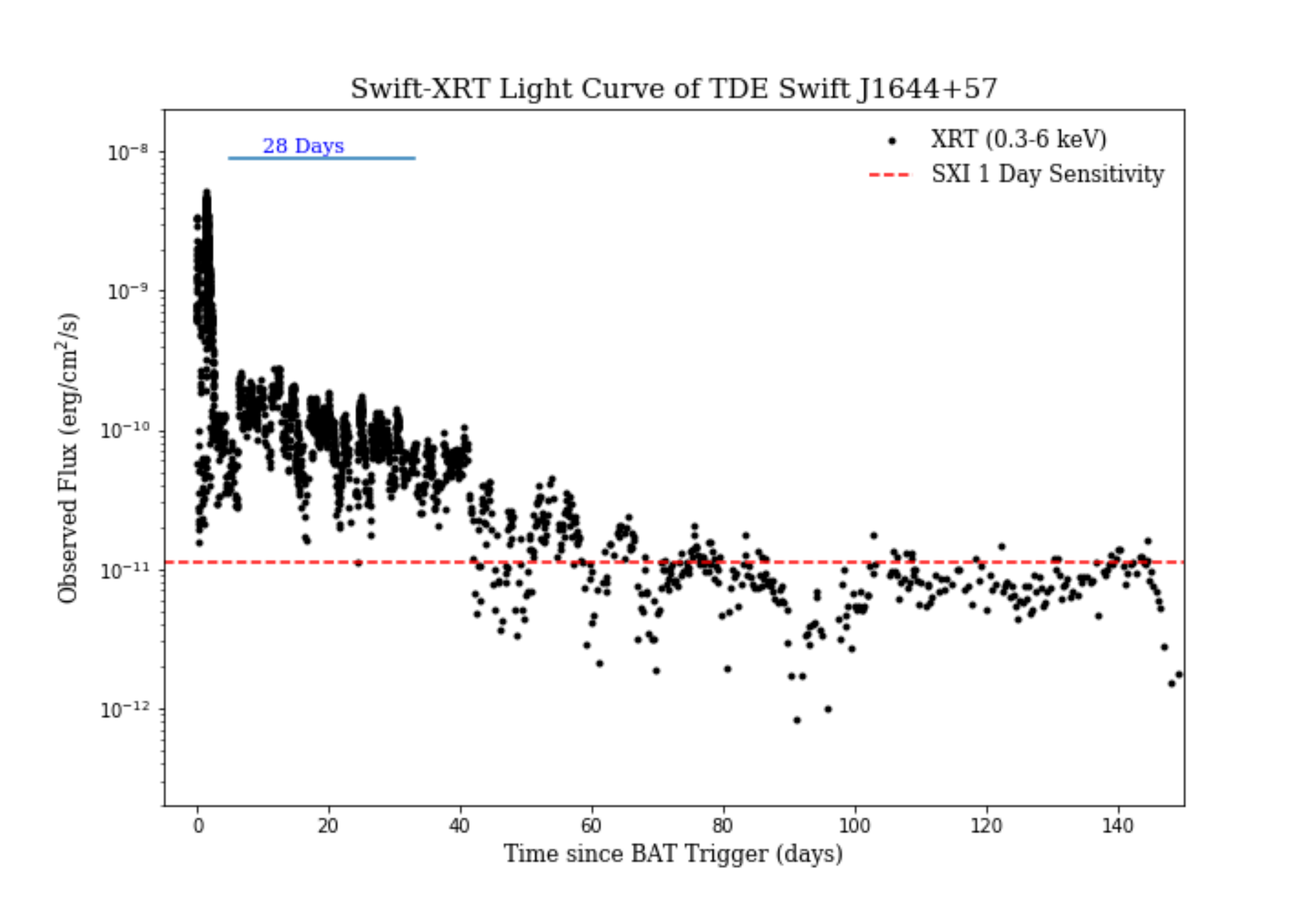}
  \caption{Observations of the hard TDE Swift J1644+57. The red  dashed line shows the detection limit for each pointing of approximately 2.3 ks several times a day with the SXI (0.2--6 keV).  It is obvious that a highly detailed lightcurve can be obtained of a similar event.}
  \label{fig:TDElcHard}
\end{figure}

Considering that the lifetimes planned for the THESEUS and eROSITA missions are also comparable, one may expect, therefore, that THESEUS may double the sample of TDEs detected by eROSITA. It may also be possible to follow-up any long duration TDEs detected with eROSITA towards the end of its lifetime. More importantly, THESEUS will also provide detailed lightcurves for TDE candidates, see for example Fig.~\ref{fig:TDElcHard}, which is not possible with the eROSITA all-sky survey which scans a given position only every six months. These lightcurves could be further improved using dedicated follow-up observations with THESEUS. Searching for TDEs in real time, will allow us to make good quality follow-up observations with other observatories, rapidly.  Finally, THESEUS will also provide X-ray spectral information (also providing a handle on SMBH mass) and distance to the host galaxy thorough follow-up with IRT, which will allow a deeper study of individual TDEs, and make THESEUS a very capable machine for study of these extraordinary events.

 
\section{Supernovae}
\label{sec-SNSBO}

\subsection{Thermonuclear SNe}

Type Ia SNe  can be understood in terms of explosions of white dwarfs (WD) in binary systems, disrupted by thermodynamical burning fronts. The apparent homogeneity in their optical light curves and spectra is a direct result of the nuclear physics  which governs the WD structure and the explosion. 
Potential progenitor systems include those composed of two WDs (double degenerate) and those composed  of a WD with a non-degenerate donor (single-degenerate). The triggering mechanisms of the explosion of the C/O WD may be compressional heat in the center, leading to the favored  Chandrasekhar mass  (M$_{\rm Ch}$) scenario,    the surface  He detonation triggering the C/O ignition in sub-M$_{\rm Ch}$ WD, or friction during the dynamical merging of two WDs. 

The connection between progenitors system and environment is largely unknown. A direct probe of the connecting layers requires advances in the time domain observations, with  thermonuclear burning temperatures and interactions with velocities in excess of 10,000 km s$^{-1}$, i.e. X-ray observations. The physical size sets the time-scales from $\sim$0.5 s to weeks.  
THESEUS will directly probe this mostly uncharted  territory. 
The yearly SNe rate is 0.75, 6  and 750 for distances of 10, 20 and 100 Mpc \cite{2015A&A...584A..62C}. 
Depending on the realization, the values below are ‘best’ estimates in the realm of unknowns.

The relevant phases for X-ray emission 
are:   \textit{a)} the shock-breakout, \textit{b)} the interaction with the material bound in the system,  and \textit{c)} the interaction with low density material in the environment produced during the progenitor evolution.

\subsubsection{
Shock breakout}

\begin{figure}
\includegraphics[width=8cm]{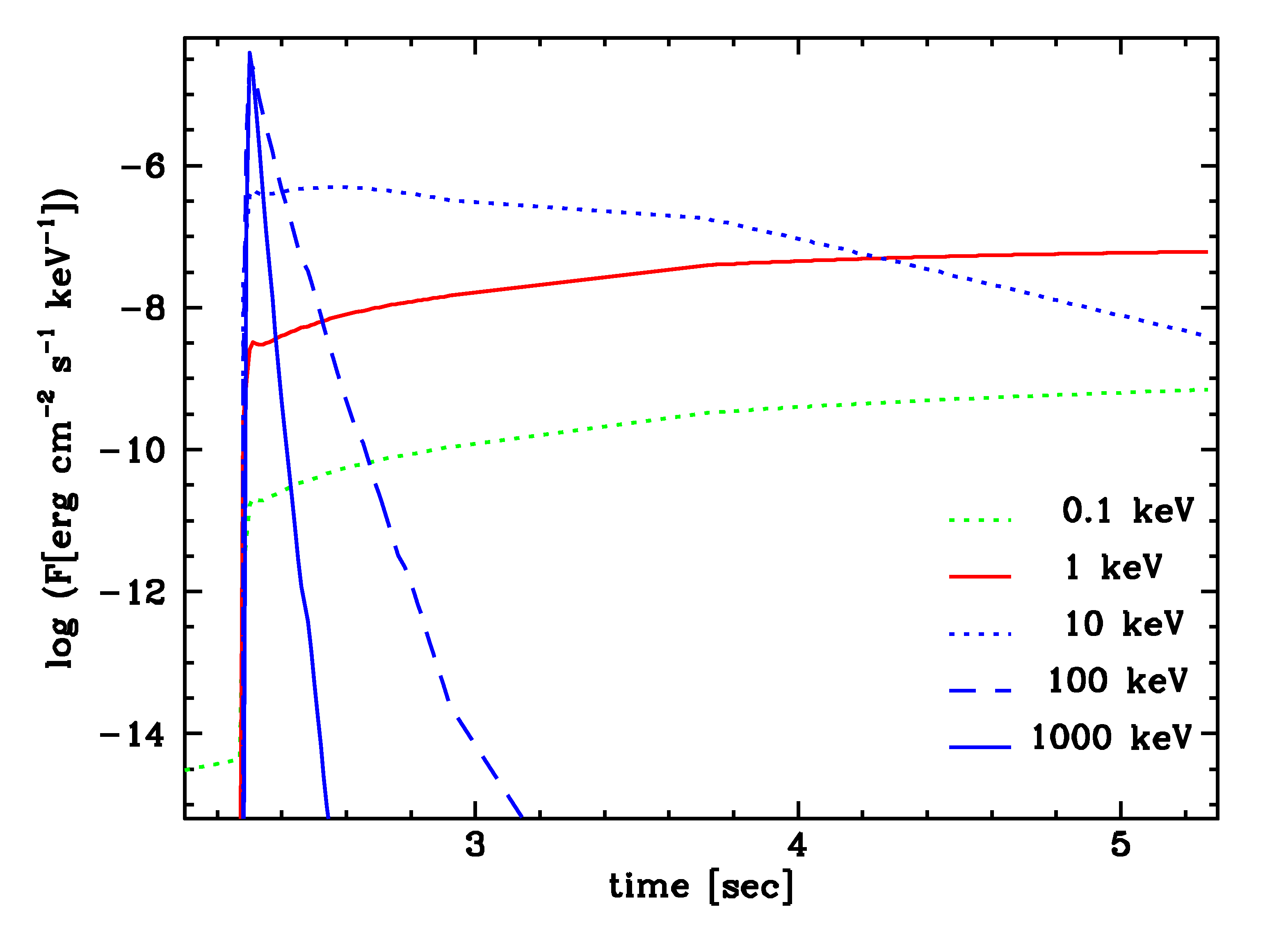} 
\caption{Expected flux as a function of time for a Type Ia SN at 10 Mpc  for the thermonuclear shock break out with a H/He/C mixed layer (peak luminosity = $5 \times 10^{48}$ \lum ), which can be detected with XGIS over 1 to 10 s
\cite{2009ApJ...705..483H,2004ApJ...607..391G,2020ApJ...900..140H}.
}
\label{fig-snial}
\end{figure}

The time-scale for the shock breakout is governed by cooling and nuclear burning \cite{2009ApJ...705..483H}. Similar to short GRBs, the spectral peak shifts in the XGIS band on time-scales of seconds, followed by a tail in the soft X-range covered by the SXI. Differences in the time scale may allow to distinguish the progenitors and explosion scenario. 
For pure C/O mixtures for C-accreting for single or dynamical double degenerate systems, nuclear burning time scales ($^{12}C(^{12}C,\gamma )^{24}$Mg) are longer than 10 s, preventing ongoing nuclear burning leading to an evolution shown in Fig.~\ref{fig-snial}. 
Since the volume increases by a factor of $\sim$1000 during the first second, adiabatic cooling leads to rapid flux decline, below the detection limit for XGIS, within several tenth of a second.  Possible detections with SXI are limited to 5-10 minutes, putting into reach only SNe~Ia within 5 Mpc. This corresponds to an expected rate of about 0.5 to 1 SNe~Ia  in the entire sky, during the nominal mission duration of $\sim$4 yrs.  However, in both currently favoured scenarios (i.e. He-triggered detonations of sub-$M_{\rm Ch}$ and $M_{\rm Ch}$ explosions), surface He-layers of $(0.5-5)\times10^{-2} M_\odot$  and $(0.5-5)\times 10^{-4} M_\odot$ respectively, are expected for most progenitors \cite{2014ApJ...797...46S,2017hsn..book.1955S,2019nuco.conf..187H}, producing mixed $C$/He layers. As a consequence, the nuclear burning  time scales of $\sim$0.1 s via  $^{12}C(\alpha,\gamma)^{16}O$ results in strong continuous heating, which lasts up to several seconds during the expansion phase and delays  the phase of rapidly adiabatic cooling. The boost  of the early luminosity and the longer duration lead to a factor $\sim$100 increase in the  X-ray flux. We may then expect about   3-4   detections with XGIS and SXI. Even a non-detection will provide valuable information about the progenitor because the position and timing of nearby SNe~Ia are known from optical surveys.
  
\subsubsection{
 Interaction with  bound material and the nearby environment}
  
\begin{figure}
\includegraphics[width=8cm]{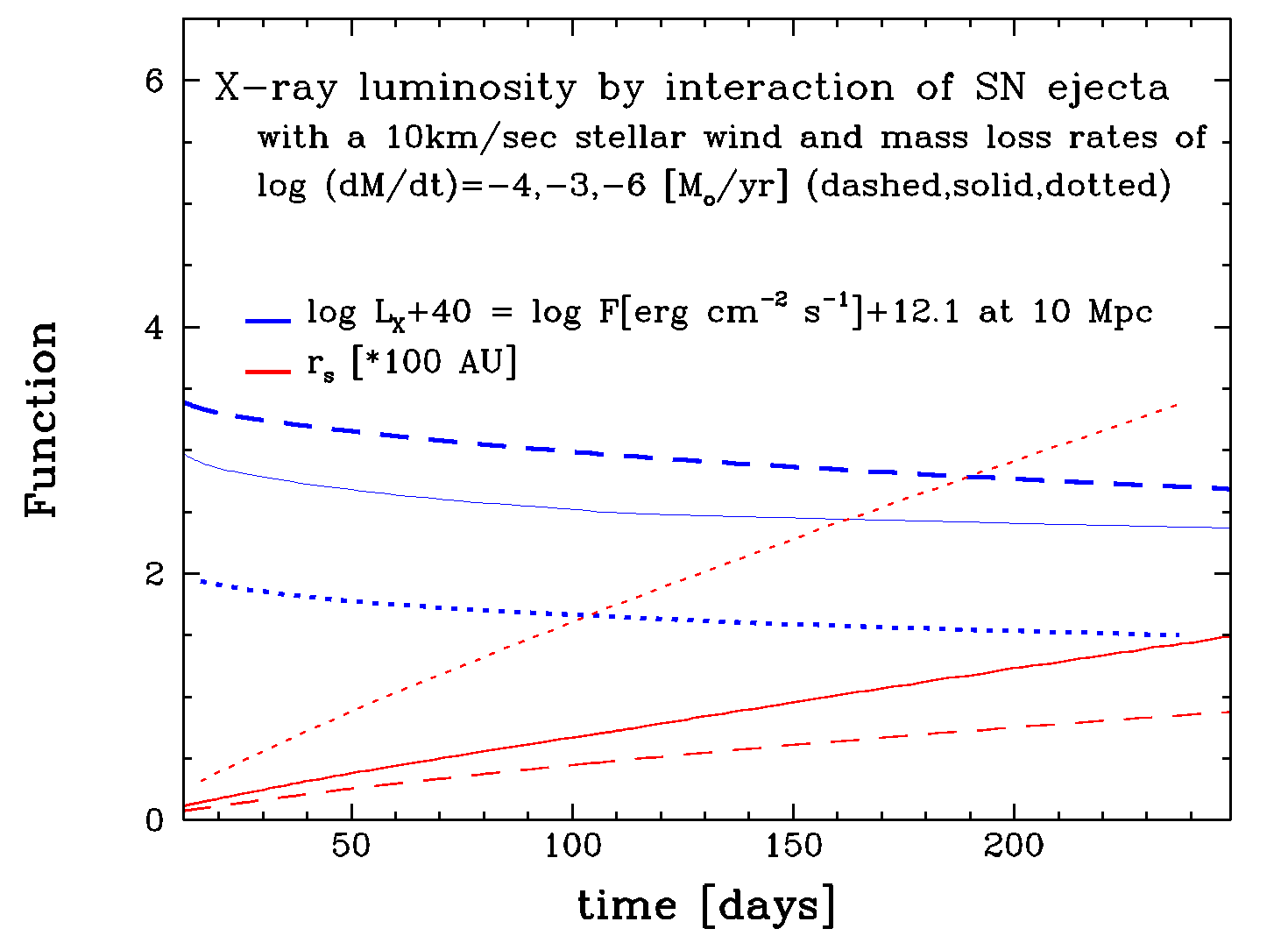}
\caption{Expected flux as a function of time for a Type Ia SN at 10 Mpc due to  the interaction within the  Roche-lobe with a mass-outflow
$10^{-3...-6} M_\odot$, which may be within reach of SXI for integration times of 1 to 10 ks  
\cite{2009ApJ...705..483H,2004ApJ...607..391G,2020ApJ...900..140H}.
The onset of the interaction provides a direct measure of the size of the cavity and wind-properties of the progenitor.}
\label{fig-sniar}
\end{figure}

The interaction between the SNIe~Ia ejecta and the matter bound in the system with the progenitor \cite{2000ApJS..128..615M,2010ApJ...708.1025K,2014ApJ...794...37M},
 and the interaction between the SN ejecta and the progenitor wind will produce X-rays \cite{2004ApJ...607..391G,2012ApJ...749...25G}. The expected size of the relevant regions are between $\sim10^{12}$ cm for the matter bound in the system to a few times $\sim10^{15} $~cm for the close environment which shows the
imprint of progenitor system and the acceleration region of the wind just prior to the explosion. This translates to a duration ranging from minutes to days. 
The flux shown in Fig. \ref{fig-sniar} 
translates to about $10^{-10}$ \flux for  SNe distances of 20 to 30 Mpc, which is well within the detection limits of SXI. Depending on the mass loss of the progenitor system and the material bound in the system, we may detect up to few dozens SNe~Ia over the lifetime of THESEUS. Note that even a non-detection would constrain the configuration and, thus, provide
important information about these objects.
 
\subsubsection{
Interaction with circumstellar material}
 
 At larger distances, the environment is dominated by the progenitor wind/interstellar medium interaction. The interaction with the ISM produces a forward shock region and an inner, low-density void \cite{1983ApJ...270..554C,2016ApJ...818...26D,2020ApJ...900..140H}.
We expect no significant interaction of SN-ejecta within the void. For example, in case of SN2014J, a nearby SN at 6.5 Mpc, X-rays have not been detected by XMM-Newton and   Chandra.
However, when the SN ejecta, some 10 to 300 years after the explosion, reach the  forward shock front at a distance between 0.1 to 30 pc \cite{2016ApJ...818...26D},  there is revival of X-ray emission similar to the wind interaction of  phase b) described above.

\subsection{Core-Collapse supernovae}

\subsubsection{Shock breakout}

Shock breakouts (SBOs) in CC SNe produced by Wolf–Rayet stars, as well as by red and blue supergiants, are expected to appear as bright X-ray bursts lasting 10–1000 s and with luminosity in the range 10$^{43–46}$ \lum .  Such stars are the likely progenitors of Type Ibc and most Type II SNe, which occur at rates of $2.6\times 10^{-5}$ and $4.5\times 10^{-5}$ Mpc$^{-3}$yr$^{-1}$, respectively \cite{2011MNRAS.412.1473L}.  SXI and XGIS can detect shock breakouts for SN  up to distances of 50 Mpc, leading to an estimated rate of at least one event per year \cite{2018AdSpR..62..191A}. Simulations of the only robust evidence of a shock breakout  (SBO 080109 at 25 Mpc, serendipitously detected with Swift/XRT \cite{2008Natur.453..469S}) show that this source could have been detected with the SXI  at $\sim$5$\sigma$ level with an exposure of 400 s (assuming a power law spectrum with photon index 2.3 and absorption column of $7 \times 10^{21}$ cm$^{-2}$ in excess to the Galactic one  \cite{2008Natur.453..469S}). The SN associated with SBO 080109 has been identified as a type Ib (i.e. showing conspicuous He lines) from a Wolf-Rayet progenitor, with ejected mass of $\sim$ 7 $M_\odot$ and a kinetic energy of $\sim$6$\times 10^{51}$ erg \cite{2008Sci...321.1185M}, while the radius was estimated of the order of 1-10 $\rsun$ depending on the assumed model \cite{2009ApJ...702..226M}.  In fact, since the estimated blackbody radius is smaller than that of a typical WR star, the X-ray emission could be related  to the thermal cocoon of a jet or outflow, rather than to a spherical SBO.  

Features moving at $\sim10^5$ km s$^{-1}$ have been recently detected in the early spectra of a SN associated with a GRB, which have been interpreted as signatures of a jet-cocoon \cite{2019Natur.565..324I}.
The diagnostic power of soft X-rays (and possibly also hard X-rays) detected with THESEUS may be crucial in discerning the nature of the emission.   The frequency of these events \cite{2020ApJ...898...37N} justifies some optimism about the probability that at least a handful may be seen by THESEUS.
Particularly, these observations have the potential to clarify whether or not the relativistic jets can provide an alternative mechanism, other than neutrino re-heating, to driving CC-SNe to the final explosion \cite{2019ApJ...871L..25P}. Also, the cocoon X-ray emissions associated with short GRBs might be one of the best promising electromagnetic counterpart for GWs merger signals \cite{2017ApJ...834...28N}. 

Wolf-Rayet stars are expected to produce harder SBO with respect to the more absorbed emission from red or blue supergiants and, in this regard, THESEUS is better suited to observe these events with respect to    X-ray all sky monitors limited to the soft band below a few keV. 
SBOs are temporally closer to the possible GW events than the optical CC SNe counterparts, thus their detection can mark with more precision the start time of the GW signals. This would prove extremely helpful, if not crucial, for the challenging signal search process \cite{2016A&A...587A.147A} and THESEUS in synergies with the 3G network will perfectly respond to this need.  

Early X-rays from the nearby  SN2006aj ($z = 0.033$), manifesting as the X-ray Flash 060218 \cite{2006Natur.442.1008C}, have a debated origin, as classical SBO emission would be too luminous \cite{2007MNRAS.382L..77G}.  Again, the strict simultaneity of soft and hard X-ray measurements with THESEUS for a similar case may remove the uncertainty by favouring a more accurate mapping of the circumstellar material (CSM) in which the shock is expanding, thus providing  a clearer identification of the emission mechanism.

\subsubsection{Late time emission}

Medium-late time soft and hard X-ray emission is produced by interaction of the shock with the CSM and ejecta and can last from days to years, before leaving room to a fully fledged SN remnant nebula \cite{2020ApJ...901..119R}.  The luminosities, typically around $\sim 10^{37} - 10^{38}$ erg s$^{-1}$, 
convert, however, to elusive flux levels at typical SN distances (100-200 Mpc).  
More rarely, luminosities up to a few $\sim10^{40}$  erg s$^{-1}$ can be reached (see,e.g., the case of  SN2014C \cite{2017ApJ...835..140M}). 
Thus, very nearby (10 Mpc) CC SNe may be detected by THESEUS and would, together with radio data, map the CSM radial distribution, which in turn is related to the progenitor's evolution.  
X-ray emission results also from the interaction of the SN shock front with the environment. In some cases (e.g., type IIn SNe, but also stripped-envelope SNe), the X-ray emission implies the presence of massive shells ejected from the progenitor star years to centuries before the stellar collapse. The structure and dynamics of these shells (and, from there, their origin) may be studied by detailed   X-ray  coverage of a sample of SNe.

Superluminous SNe are equally elusive in X-rays (barring the high energy emission from a GRB occasionally associated with them \cite{2015Natur.523..189G}), and represent an almost entirely uncharted territory \cite{2013ApJ...771..136L,2018ApJ...864...45M}.  The detection and study of (rare) nearby sources with THESEUS would represent a big breakthrough in the opportunity they provide to get insight into the Superluminous SNe engine.

\section{Fast Radio Bursts}
\label{sec-FRB}

Fast Radio Bursts (FRBs, see \cite{2019ARA&A..57..417C,2020Natur.587...45Z} for reviews) are currently one of the most mysterious class of astronomical sources. Despite having been recorded by radio telescopes since 2002 (although recognized only later \cite{2007Sci...318..777L}), their true nature is still unknown and numerous models involving very different explanations have been proposed \cite{2019PhR...821....1P}. This resembles the case of GRBs in the first two or three decades after their discovery, but with the important difference that it was soon realized that FRBs are extragalactic.  From the observational point of  view, FRBs are very intense and short radio pulses, characterized by a dispersion measure\footnote{The dispersion measure is the integral of the electron density along the line of sight, $DM = \int n_e \,dl$. Photons of frequency $\nu$ are delayed proportionally to  $DM/\nu^2$. } in excess of that attributable to the Milky Way in their direction. A subset of FRBs have been seen to repeat, and two of them with a periodic pattern of activity \cite{2020Natur.582..351C,2020MNRAS.495.3551R}. It is not clear yet if the remaining ones, detected only once, are also repeating sources with much longer duty cycles or really one-off events. 
Thanks to the FRB localisation improvement by the current radio interferometers, thirteen galaxies have been firmly identified as FRB hosts. These allowed us to precisely measure their redshift and to get information on host stellar population and local FRB environment, which may help in understanding the progenitor nature  (e.g. \cite{2020ApJ...903..152H,2021ApJ...907L..31B,2020arXiv201211617M}).
Knowledge of the distances also allowed to determine the energy requirement, which corresponds to isotropic radio luminosities in the  range 10$^{38}-10^{46}$ erg s$^{-1}$.

Several of the models proposed for FRBs are based on magnetars, strongly magnetized neutron stars powered by magnetic energy (see Sect. \ref{sec-magnetars}). These models have recently received strong support with the discovery, on 2020 April 28,  of an extremely bright radio pulse from the known magnetar SGR 1935+2154 \cite{2020Natur.587...54C,2020Natur.587...59B}. This radio pulse had many similarities with FRBs (the only difference being a DM consistent with the magnetar location in the Galaxy) and it occurred simultaneously with an X-ray burst which had a particularly hard spectrum \cite{2020ApJ...898L..29M}. 

The likely association with magnetars makes FRBs extremely interesting targets for THESEUS.  The new generations of radio telescopes will detect hundreds of FRBs each day, many of which will fall in the SXI/XGIS field of view.  Combining THESEUS with these wide-field,  sensitive radio monitoring observations will allow us to investigate the connection between magnetars and FRBs. Even if it is possible that the mystery of FRB’s nature will be solved  by that time, multi-wavelength observations of a large sample of FRBs will be fundamental to study these sources and to apply them to cosmology and fundamental physics problems, as it is now well demonstrated in the analogous field of GRB research.  

\begin{figure}
\center
\includegraphics[width=8.5cm,angle=0]{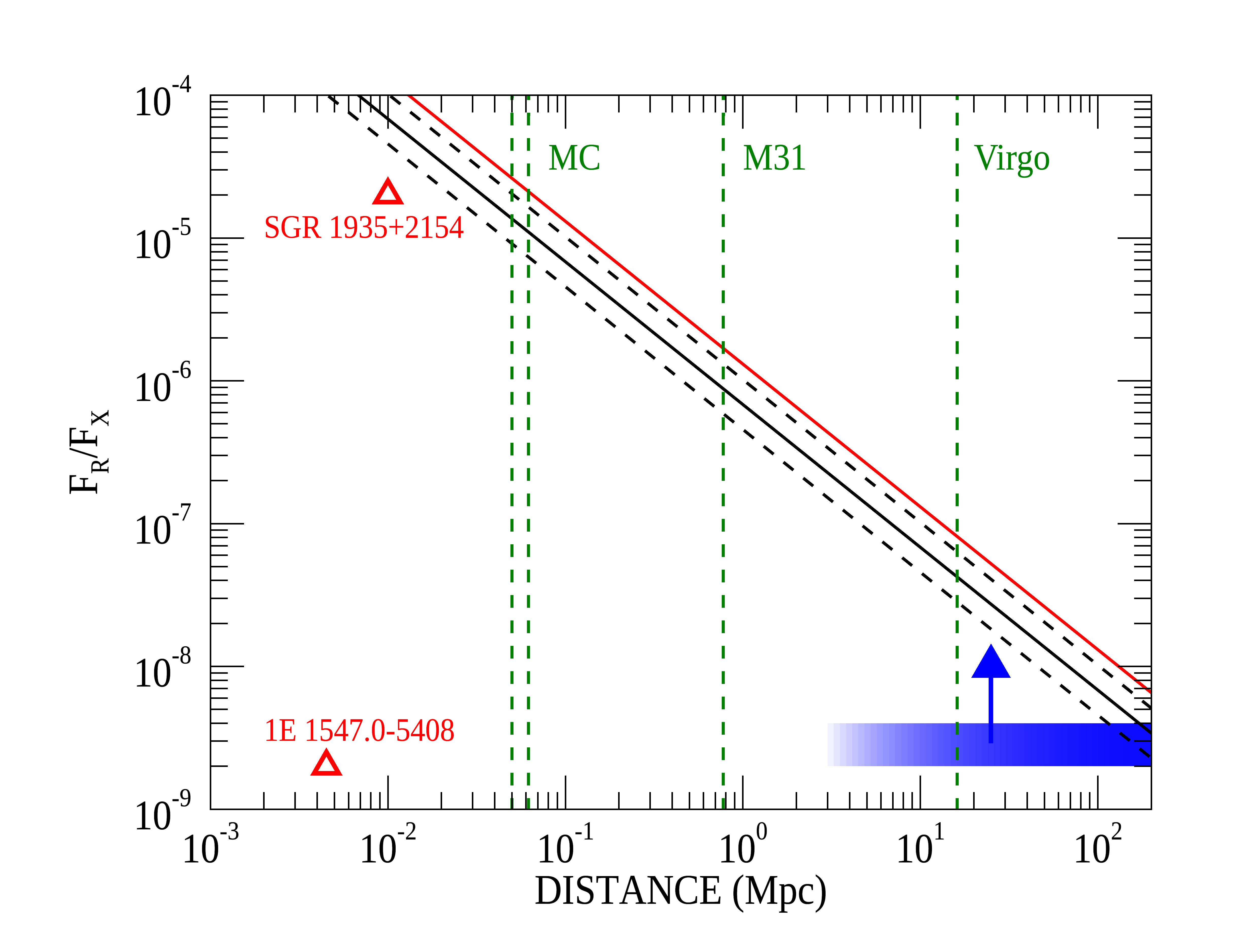}
  \caption{Maximum distance for   detection of a burst from a FRB for SXI 
  (black) and XGIS (red) as a function of the ratio between radio and X-ray fluence $F_R/F_X$ (see text for details). For these estimates we scaled the properties of the 2020 April 28 burst, using the Hard X-ray Modulation Telescope   X-ray spectrum \cite{2020arXiv200511071L} for the SXI estimates,  and  the INTEGRAL spectrum \cite{2020ApJ...898L..29M} for the XGIS ones. For the SXI we have required a minimum fluence of 5 counts, which constitutes a significant detection in the observed duration ($\sim$0.5 s) of these short events.  
  The dashed lines indicate the range of uncertainty due the unknown absorption (they refer to N$_H$=10$^{20}$ and 10$^{22}$ cm$^{-2}$).
  The red triangles indicate the only values of $F_R/F_X$ measured to date from two Galactic magnetars, while the blue region indicates typical lower limits inferred from the non-detection of high energy emission from extragalactic FRB. The green dashed lines indicate the distances of the Magellanic Couds, the Andromeda galaxy and the Virgo cluster. 
   }  
  \label{fig-frb}
\end{figure}

As discussed above, THESEUS will certainly be able to detect bursts from Galactic magnetars. On the other hand, the possibility to reveal high-energy emission from FRB at extragalactic distances is more difficult to assess. 
In Fig.~\ref{fig-frb} we show the estimated maximum distance at which the SXI (black lines) and the XGIS (red lines) can detect bursting emission from a FRB as a function of the ratio between radio and X-ray fluence $F_R/F_X$ (the data available up to now indicate that this ratio spans a very large range). 
The dashed lines indicate the range of uncertainty due to the unknown absorption. They refer to the cases of 10$^{20}$ and 10$^{22}$  cm$^{-2}$. 

\section{Other extragalactic transients}
\label{sec-FELT}

\subsection{Fast Evolving Luminous Transients}
\label{sec-FELTS}

The Fast Evolving luminous transients (FELTs) are a new and rare class of extragalactic transients that have peak bolometric luminosities $\gtrsim10^{43}$ \lum and reach the peak in a timescale of days. Their origin is debated, but scenarios involve an engine driven explosion, a supernova in a dense environment or a TDE. In 2018, the Cow (short name for AT2018cow, \cite{2018Smartt}) became the prototype of FELTs. It showed several peculiar characteristics among which there is its high X-ray luminosity ($\sim10^{43}$ \lum) 3 days after the optical discovery, a unique rapid X-ray variability with timescale of a few days in soft X-rays ($0.3-10$~keV) and a hint for a possible periodicity at $\sim$4 days \cite{2019Margutti}, X-ray emission that lasted for several weeks after peak and a non changing soft X-ray spectrum \cite{2018RiveraS}. It also showed a variable X-ray component at energies E $>10$~keV \cite{2019Margutti}, and flare-like variability in IR with duration of 2-3 days \cite{2019Perley}. Besides the Cow, other 2 FELTs have shown X-ray emission: AT2020xnd for which X-rays were also detected at early times ($\sim$21--27 days after explosion), but only in the soft band \cite{2020Matthews}, and CRTS-CSS161010, for which X-rays were detected at late times (99--130 days after explosion, \cite{2020Coppejans}). These observations suggest that early soft X-ray emission is a signature of FELTs. THESEUS will allow the detailed study of these transients by performing simultaneous observations in the hard and soft X-rays, as well as in the NIR band. The sensitivity and fast-slew capabilities of THESEUS will permit to obtain information on the variability of the FELTs in these bands and to investigate whether there is a possible correlation among them, shedding light on the origin of these energetic transients. Based on current optical detection rates (e.g.,\cite{2020Ho}), it is expected that SXI, with its large field of view, will detect $\sim30$ FELTs per year.   

\subsection{Fast X-ray Transients}
\label{sec-FXT}

X-ray surveys often discover new types of objects with characteristics not seen before.
The first Fast X-ray Transient serendipitously discovered in a Chandra pointed observation was announced in 2013 \cite{2013Jonker}: a fast rise ($\sim$10 sec) with a double peak (lasting $\sim100$~s) followed by a fast decline in $\sim100$~s and a subsequent slower gradual decline until the end of the Chandra observation $2\times 10^4$~s later. Several other fast transient events  have been reported since, in both Chandra and XMM-Newton observations 
\cite{2015Glennie,2017Bauer,2019Xue,2020Alp_Larsson}. 
Although they share some similarities, there are also differences that might point to a variety of interpretations. For example, the fast transient CDF-S XT2 showed a plateau in its light curve, consistent with power injection from a millisecond magnetar \cite{2019Xue}.

The strongest constraint on a counterpart to these transients comes from a serendipitous VLT observation where no counterpart was found at $m_R <$ 25.7 within 80 minutes \cite{2017Bauer}. Viable models explaining these events include an orphan GRB, a low luminosity GRB and a jetted TDE. The origin of this class of fast transients remains to be established, requiring wide-field X-ray instruments to find a good number of nearer objects, coupled with the most sensitive telescopes at other wavelengths to complete the picture and especially to provide the much needed transient counterpart that will allow a firm spectroscopic redshift to be determined. \cite{2015Glennie} and \cite{2019Yang} provide a rate estimate (proportional to grasp), which for the SXI amounts to  $\sim50-60$ per year.

\section{Discussion}
\label{sec-disc}

\subsection{THESEUS as a transient-discovery machine}

THESEUS  will provide rapid alerts enabling to trigger observations with other facilities. Such alerts will come not only  from the discovery of new sources, but also from the detection of peculiar states in known sources that will be regularly monitored.  Galactic X-ray binaries containing compact objects   provide a wealth of different phenomena for which the alerts (and in many cases also the subsequent follow-up) with THESEUS will be fundamental. Besides the classical soft and hard X-ray transients containing NS or BH, other particularly interesting cases are outbursts from novae, thermonuclear super-bursts in low mass XRB, transitional accreting millisecond pulsars, and SFXTs.  All these sources reach high luminosities during their outbursts allowing detailed studies. In addition,  THESEUS will have the sensitivity   to detect also the Very faint X-ray transients, a class of sources reaching only lower peak luminosities ($<10^{36}$ erg s$^{-1}$) and of still unknown nature. 

 Other classes of Galactic sources for which THESEUS triggers will be fundamental include stars undergoing super-flares and magnetars. 
THESEUS observations of super-flares, in synergy  with future ground-based observatories, will  allow the application of stellar flare diagnostic techniques that require a high signal-to-noise ratio, providing a unique opportunity  to see whether solar flare physics can be extrapolated in a variety of stellar environments.  Outbursts and short bursts from several   Galactic magnetars will be discovered and/or studied  by THESEUS,  and also giant flares from magnetars in nearby galaxies will be within reach of the SXI and XGIS instruments.

Among extragalactic sources, of particular interest are tidal disruption events, hyperluminous X-ray sources,  supernovae,  changing-look AGNs and flaring episodes in blazars.  THESEUS triggers will enable rapid follow-up observations with the most powerful multiwavelength facilities available in the 30's, such as SKA in radio, the 30-m aperture telescopes in the optical, Athena in X-rays, and CTA at high-energy  $\gamma$-rays.  
Also high-energy neutrino detectors, that will reach better sensitivities thanks to planned upgrades, will be extremely relevant in this context.

\subsection{THESEUS as a survey facility}

Thanks to the large field of view of THESEUS's instruments and its  observing strategy,   a wide energy band coverage and high sensitivity will be combined with a long and high-cadence coverage of individual sources. This will result in a higher overall sensitivity compared to the already operating facilities of this type, and enable variability studies on timescales and flux levels not accessible previously. 
THESEUS will provide spectral information for all detected sources in a wide energy range from 0.3  keV to several MeV. This will enable the simultaneous analysis of broadband spectral variability, currently only feasible with expensive dedicated campaigns, and will yield important information about accretion physics on WDs and NSs in Galactic X-ray binaries, as well as on    black holes on all mass scales from stellar to super massive.
Several classes of sources will be studied thanks to the observations of large samples, enabling systematic studies of their spectral-variability properties.

  The high sensitivity of SXI in the soft X-ray will lead to the detection of a large number of stellar flares from chromospherically active stars. This will permit a systematic study of the occurrence frequency and luminosity function of X-ray flares thanks to the collection of large and unbiased samples from various classes of active stars.

It will be possible to study the outburst cycle of dwarf novae and of LMXBs,  important to obtain insight into the physics of accretion disks and their boundary layers, and to investigate the variability properties on different timescale of particular classes of XRBs, such as the transitional millisecond pulsars and the SFXTs.
The resulting THESEUS  database will be fundamental  to  investigate new theoretical frameworks  for  accretion physics onto WDs and for the interaction between accretion flows and NS magnetospheres.
 SXI will detect about one AGN in each typical pointing and it will be possible to see state evolution, like seen in X-ray binaries,   and to   constrain accretion  duty cycles.
 THESEUS will monitor of order hundreds of thousands of AGN in the low-$z$ Universe ( $z<0.3$), and can catch new changing look AGN events in the act, allowing us to probe the response of AGN structural components to drastic changes in accretion rate.
Long and intensive coverage of radio-loud AGN  will allow for unprecedented monitoring and variability duty cycle investigation.   These will feed new constraints to inhomogeneous jet models, made necessary by recent multi-wavelength observations, and by hitherto few precious UHE neutrino detections.
The SXI and XGIS sensitivity and that of  CTA will enable simultaneous monitoring at soft and hard X-rays and TeV energies on hour- and sub-hour timescales, that match the characteristic light-crossing times of the emitting regions down to a few Schwarzschild radii.  At the same time, plentiful large and sensitive ground-based optical and infrared cameras (and the THESEUS IRT) and extremely sensitive radio facilities will follow the development of the perturbation at the larger scales, making a fully resolved and complete description of the outburst propagation possible.
  
  \subsection{Observatory science with THESEUS}
  \label{sec-GO}

A space-based NIR telescope like the IRT on THESEUS  will be attractive to a wide range of investigators and can be used to 
address important questions in a plethora of scientific areas.  
The chance to use THESEUS to observe substantial samples of interesting sources, both known and newly discovered, to appropriate depths and cadences, while the mission is searching for GRBs, provides
opportunities for additional science that will be implemented through a Guest Observer (GO) program. 
THESEUS will maintain a list of core-programme targets, augmented with targets from a competed GO
programme, with all observations planned and executed by the THESEUS mission operations team, while
THESEUS operates in Survey Mode.  
Some  science objectives based on the use of the IRT, complemented by SXI and XGIS data,  have been discussed in the previous sections. In addition, we mention here a few other examples  and refer to the accompanying paper \cite{WP-Syn} for more details. 

A space-borne IR spectrograph is able to investigate a range of cometary emission and absorption features,
without being restricted to specific atmospheric bands, and with full access to all water and ice features,
impossible to observe from the ground. Several tens of comets per year are likely to be observable as they pass through
the inner solar system, evolving through their approach to and recession from perihelion.
IR spectra of large samples of stars with transiting planets can be obtained by THESEUS. By 2030, tens of
thousands of transiting planets will be known, spread widely over the sky, and with well-determined transit
times, which can be scheduled well in advance to search for potential atmospheric signatures in IR absorption
spectroscopy.  

A prompt spectroscopic IR survey for supernovae that are found taking place out to several tens of Mpc,
unencumbered by atmospheric effects, is likely to remain attractive beyond 2030, and will help to
resolve remaining questions about the impact of environment and metallicity on the nature of supernovae and their
reliability as standard candles. Without sensitive IR spectroscopy, these questions might not be resolved.

The availability of the full spectral window is particularly helpful for observations of emission-line galaxies
and AGN, for which key diagnostic lines are redshifted out of the optical band from the ground at redshifts
$z\sim$0.7.  
IRT will enable H$_{\alpha}$ spectral surveys of interesting classes of the most luminous galaxies, and AGN all
the way to $z\sim$2-3. Furthermore, the THESEUS mission will provide a useful time baseline out to several years,
to see potential changes in the appearance of AGN spectra, and to confirm any changes by revisiting selected
targets.

\section{Conclusions}
\label{sec-conc}

The wide fields of view and high sensitivity of the SXI and XGIS instruments on THESEUS offer great opportunities in the study of practically all classes of Galactic and extra-Galactic variable  X-ray sources. 
The massive grasp (effective area $\times$ field of view) and spectral coverage offered by these instruments will reveal the violent Universe as it happens in real time, bringing three great benefits:

 \begin{itemize}
  
\item the large FOV results in quite unprecedentedly frequent observations. Such frequent long-term monitoring opens up previously unavailable timescales for study, giving access to physical processes which would otherwise be missed.

\item The high-cadence situational awareness of the sky enables the discovery of new rapid phenomena to be broadcast promptly to the world’s greatest astronomical facilities for immediate follow-up, so catching vital source types in revealing new states.

\item The very wide energy bandwidth provides a new opportunity to constrain the emission processes at work in a wide range of source types. The great sensitivity will vastly expand the classes of high energy emitters studied. 

 \end{itemize}

\noindent
Additionally, the use of the IRT for selected sources further broadens the scientific scope by allowing for unique multi-wavelength monitoring as well as to implement a Guest Observer program to study interesting NIR targets, in parallel with the normal operating mode optimized for the discovery of high-redshift GRBs.

\section*{Appendix}

\label{sec-app}

 The SXI instruments is characterized by a low background rate, allowing the significant detection of faint sources providing small numbers of  counts. It is therefore necessary to consider Poisson statistics to define the instrument sensitivity in these cases.
 
For an exposure time of duration  $T$,  we define the  source detection  threshold,  C$_{\rm THR}(T)$,   as the number of counts that has a small probability $P_1$ of being exceeded by a statistical fluctuation of the background.   If the background count rate in the detector region used to extract the source counts is $R_{\rm BKG}$,  the threshold is defined by the following equation\\
 

$P_{\rm POISS} (\geq C_{\rm THR} | R_{\rm BKG} T)= P_1$ 


\bigskip
\noindent
where $P_{\rm POISS} (\geq N | \mu)$ is the integral Poisson probability with average $\mu$ (i.e. the probability of observing $N$ or more counts when $\mu$ counts are expected). When searching for GRBs, the value of $P_1$ must be sufficiently small to reduce the number of false triggers to an acceptable level, also taking into account the large number of independent sky positions in the SXI field of view and of   time intervals that are continuosly monitored (for example, $P_1$=10$^{-10}$ was used in simulations to estimate GRB detection rates). On the other hand,  less conservative thresholds can be used for non-GRB sources.

In addition   to the probability $P_1$ used to define the detection threshold, in order to derive the instrument sensitivity we need to specifiy another probability  value, $P_2$.    This is used to derive the source count rate $R_{\rm SRC}$ that has a reasonable probability of exceeding the threshold:\\

$P_{\rm POISS} (\geq C_{\rm THR} | (R_{\rm SRC} + R_{\rm BKG}) T)= P_2$.

\begin{figure}
\center
\includegraphics[width=8cm]{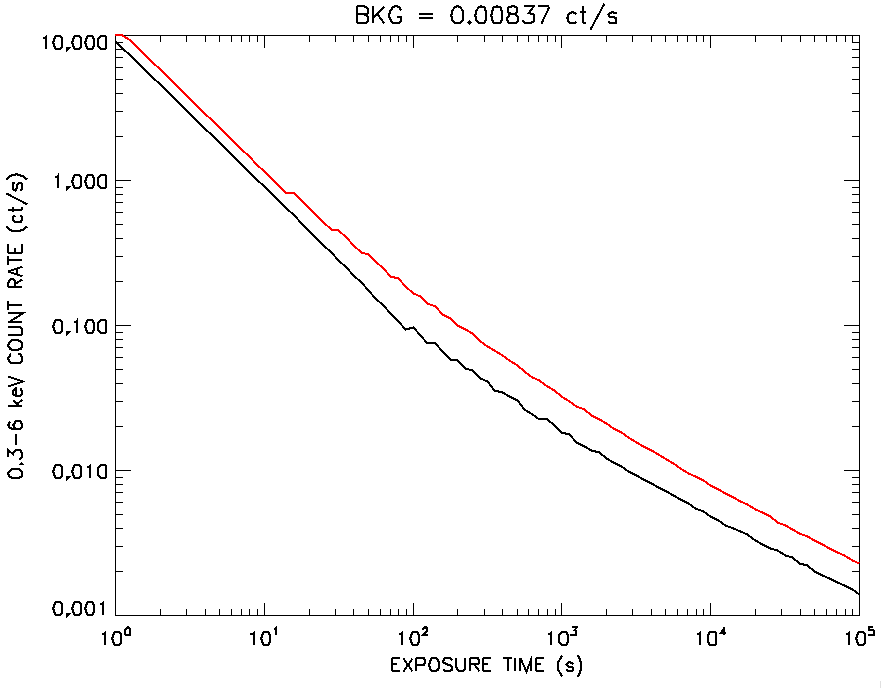}
   \caption{SXI sensitivity as a function of exposure time.  
   The lower curve (black) gives the count rate corresponding to a 95\% probability to exceed a threshold of 3$\sigma$ ($P_1$=1.35$\times10^{-3}$), 
while the upper curve (red) gives the count rate corresponding to a 99\% probability to exceed a threshold of 5$\sigma$ ($P_1$=3$\times10^{-7}$). The assumed background rate of 8.371$\times10^{-3}$ cts s$^{-1}$ is appropriate for a source  extraction region of 675 arcmin$^2$.
}  
  \label{fig-sxi-CRsens}
\end{figure}

\bigskip
\noindent
The SXI sensitivity curves shown in Fig.~\ref{fig-sxi-CRsens} have been computed for two different pairs of $P_1$ and $P_2$.  The black curve refers to $P_2$=0.95 and $P_1$=1.35$\times10^{-3}$, i.e. it gives the source count rates yielding a 95\% probability of exceeding a 3$\sigma$ threshold. A more conservative definition is represented by the red curve, computed for $P_2$=0.99 and $P_1$=3$\times10^{-7}$ (5$\sigma$).

\begin{figure}
\center
\includegraphics[width=9cm]{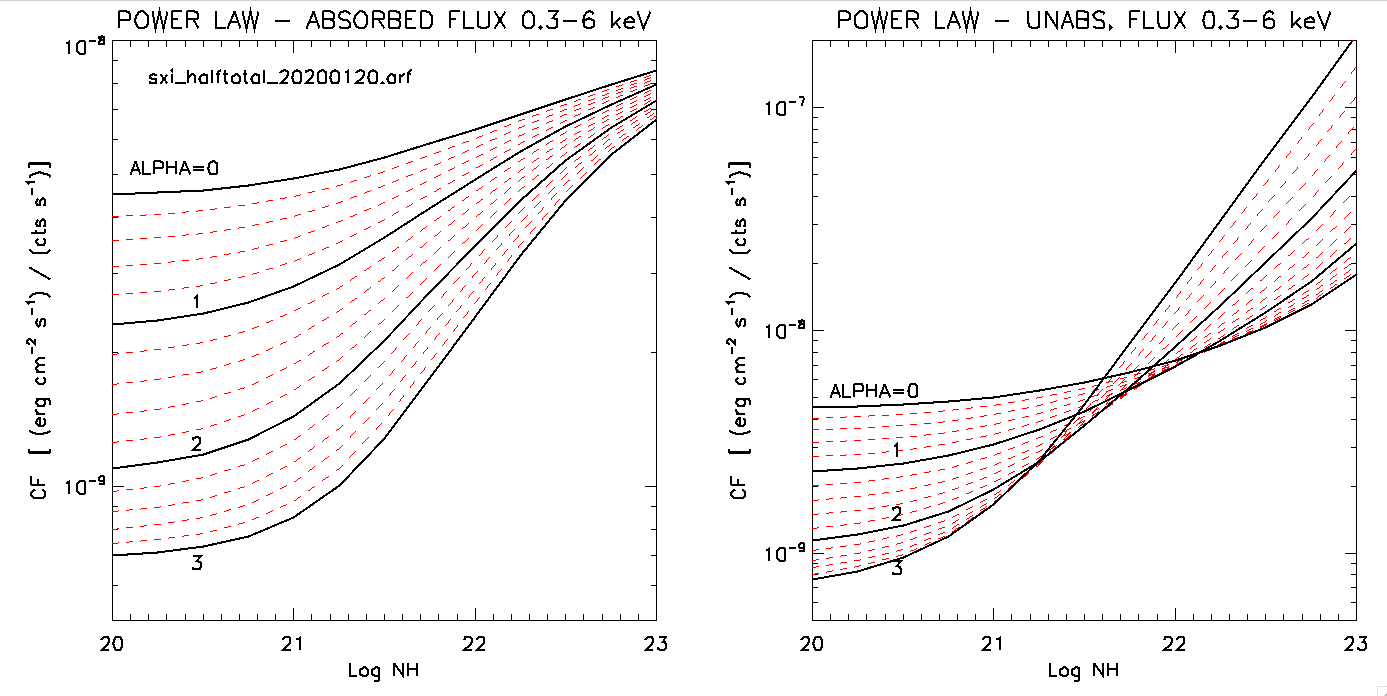}
  \caption{SXI counts to flux conversion rates for a power-law spectrum as a function of absorption. The   lines refer to different values of the photon index. The left figure refers to absorbed fluxes and the rigth one to fluxes corrected for the absorption.}  
  \label{fig-sxi-CF1}
\end{figure}

\begin{figure}
\center
\includegraphics[width=9cm]{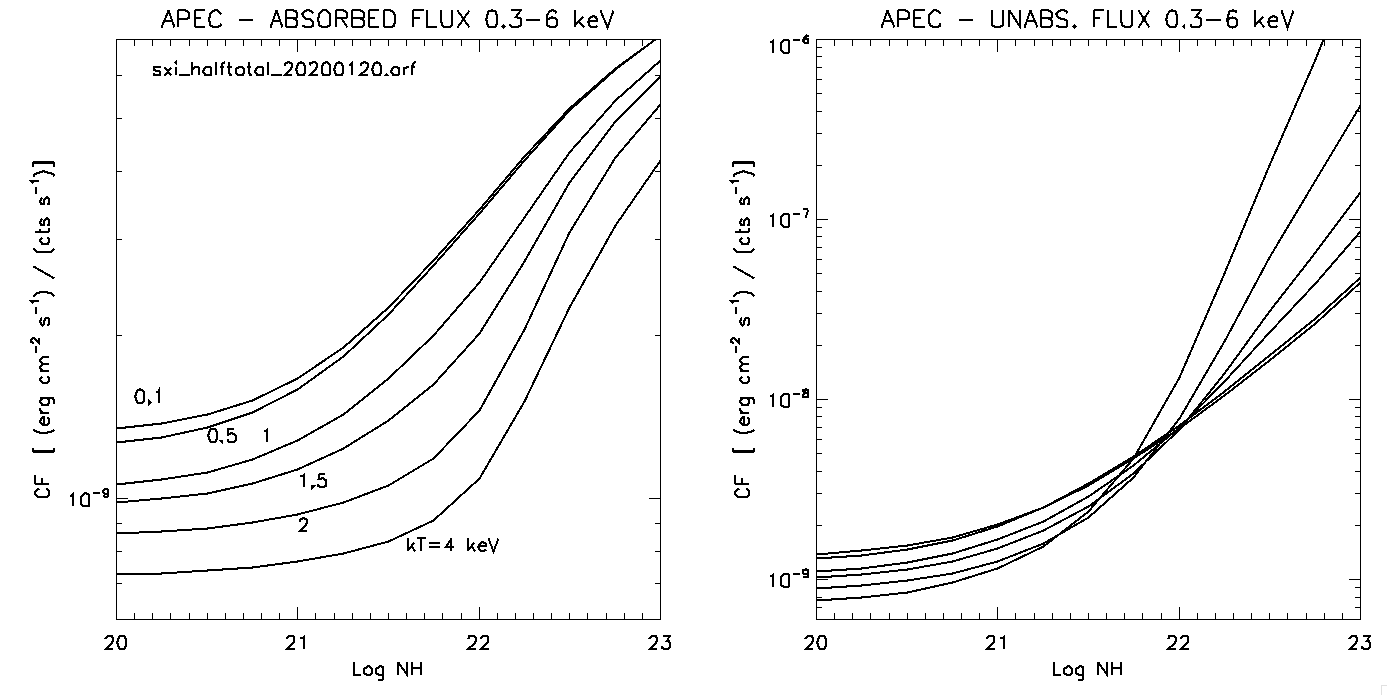}
  \caption{SXI counts to flux conversion rates for a thermal plasma spectrum (APEC in XSPEC)  as a function of absorption. The   lines refer to different temperatures. The left figure refers to absorbed fluxes and the rigth one to fluxes corrected for the absorption. }  
  \label{fig-sxi-CF2}
\end{figure}

The count rate sensitivity curves of   Fig.~\ref{fig-sxi-CRsens} can  be converted in flux units using the conversion factors plotted in  Figs.~\ref{fig-sxi-CF1} and \ref{fig-sxi-CF2} for different source spectral shapes and values of absorption.

\begin{figure*}
\center
\includegraphics[width=8cm]{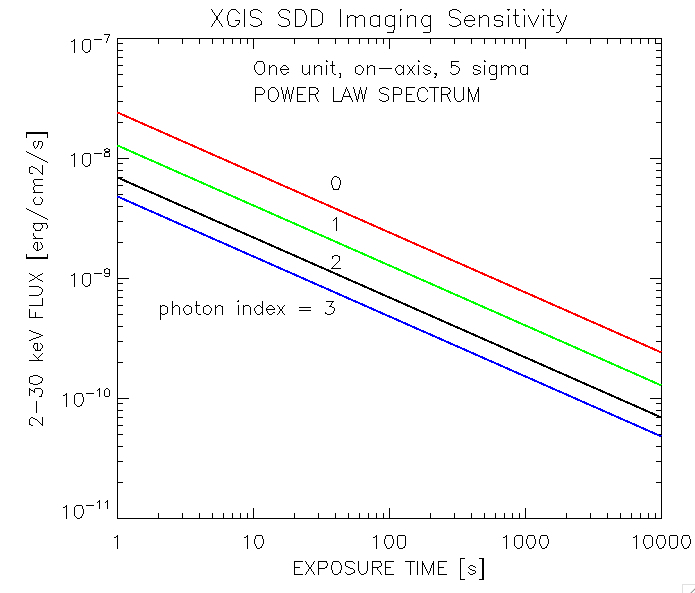}
\includegraphics[width=8cm]{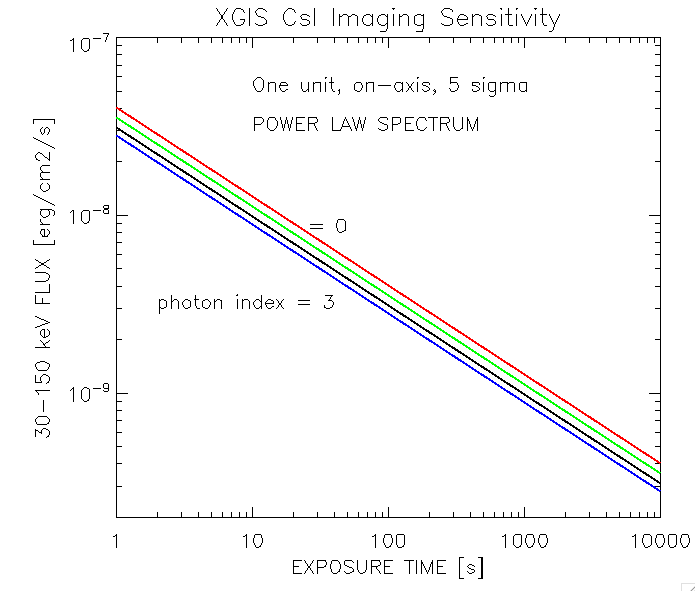}
  \caption{XGIS imaging sensitivity (5$\sigma$ c.l.)  as a function of exposure time  provided by the SDD detectors in the 2-30 keV  range (top panel) and by the CsI detectors in the 30-150 keV (bottom panel). Both figures are for a single XGIS unit and are valid for sources in the fully coded field of view (the central $\sim10\times$10 deg$^2$ of each unit's FoV). The different lines refer to sources with a power law spectrum with the indicated photon index.  By combining the two units, the sensitivity plotted here is achieved over a region of $\sim$60$\times$30 deg$^2$, similar to that of the SXI field of view.}  
  \label{fig-sens-time}
\end{figure*}

In Fig.~\ref{fig-sens-time} we show the sensitivity of the XGIS instrument as a function of exposure time and for sources with power law spectra of different slopes. The plotted sensitivities are for a single XGIS unit in the  fully coded field of view. Note that, by combining the two units, an equivalent   sensitivity   is obtained  over a region of $\sim$60$\times$30 deg$^2$, similar to that of the SXI field of view.

%
%

\bibliographystyle{spphys}       
\bibliography{WPWG3}   

%
%

\end{document}